\documentclass{aastex631}

\usepackage{array}
\newcommand{\head}[2]{\multicolumn{1}{>{\centering\arraybackslash}p{#1}|}{{#2}}}

\shorttitle{SLACS Spatially Resolved Kinematics}
\shortauthors{Knabel et al.}

\graphicspath{{./}{figures/}}

\usepackage{amsmath}
\usepackage{amssymb}
\usepackage{subfigure} 
\usepackage{graphicx}
\usepackage{float}

\begin{document}

\title{Spatially Resolved Kinematics of SLACS Lens Galaxies. I: Data and Kinematic Classification}

\author[0000-0001-5110-6241]{Shawn Knabel}
\affiliation{Department of Physics and Astronomy, University of California, \\
Los Angeles, CA 90095, USA}

\author[0000-0002-8460-0390]{Tommaso Treu}
\affiliation{Department of Physics and Astronomy, University of California, \\
Los Angeles, CA 90095, USA}

\author[0000-0002-1283-8420]{Michele Cappellari}
\affiliation{Sub-Department of Astrophysics, Department of Physics, University of Oxford, \\
Denys Wilkinson Building, Keble Road, Oxford, OX1 3RH, UK}

\author[0000-0002-5558-888X]{Anowar J. Shajib}\thanks{NFHP Einstein Fellow}
\affiliation{Department of Astronomy \& Astrophysics, University of Chicago, \\
Chicago, IL 60637, USA}
\affiliation{Kavli Institute for Cosmological Physics, University of Chicago, \\
Chicago, IL 60637, USA}
\affiliation{Center for Astronomy, Space Science and Astrophysics, Independent University, Bangladesh, Dhaka 1229, Bangladesh}

\author{Chih-Fan Chen}
\affiliation{Department of Physics and Astronomy, University of California, \\
Los Angeles, CA 90095, USA}

\author[0000-0003-3195-5507]{Simon Birrer}
\affiliation{Department of Physics and Astronomy, Stony Brook University, Stony Brook, NY 11794, USA}

\author[0000-0003-2064-0518]{Vardha N. Bennert}
\affiliation{Physics Department, California Polytechnic State University, \\
San Luis Obispo, CA 93407, USA}

\correspondingauthor{Shawn Knabel}
\email{shawnknabel@astro.ucla.edu}

%%%%%%%%%%%%%%%%%%%

\begin{abstract}

We obtain spatially resolved kinematics with the Keck Cosmic Web Imager (KCWI) integral-field spectrograph for a sample of 14 massive ($11<\log_{10} M_*/M_\odot<12$) lensing early-type galaxies at z$\sim$0.15-0.35 from the Sloan Lens ACS  (SLACS) Survey. We integrate kinematic maps within the galaxy effective radius and examine the rotational and dispersion velocities, showing that 11/14 can be classified as slow rotators. The dataset is unprecedented for galaxy-scale strong lenses in terms of signal-to-noise ratio (S/N), sampling, and calibration, vastly superseding previous studies. We find the primary contributions to systematic uncertainties to be the stellar template library and wavelength range of the spectral fit. Systematics are at the $1-1.4\%$ level, and positive covariance is $<1\%$ between sample galaxies and between spatial bins. This enables cosmographic inference with lensing time delays with $<2\%$ uncertainty on H$_0$. We examine the effects of integration of the datacubes within circular apertures of different sizes and compare with SDSS single-aperture velocity dispersions. We conclude that velocity dispersions extracted from SDSS spectra for these 14 SLACS galaxies, which have low S/N ($\sim$9/\AA) relative to the SLACS candidate parent sample, are subject to systematic errors (and covariance) due to stellar template library selection at the level of 3(2)\%, which need to be added to the random errors. Comparison between our KCWI stellar velocity dispersions, our own analysis of SDSS spectra, and previously published measurements based on SDSS spectra shows mean differences within a few percent. However, these differences are not significant given the uncertainties of the SDSS-based stellar velocity dispersions. We find that the correlations between scaling relations using quantities inferred from dynamical, lensing, and stellar population models agree with previous SLACS analysis with no statistically significant change. A follow-up paper will present Jeans modeling in the context of broader studies of galaxy evolution and cosmology.

\end{abstract}

%%%%%%%%%%%%%%%%%%%

\keywords{Galaxies, Galaxy kinematics, Galaxy dynamics}

%%%%%%%%%%%%%%%%%%%

\section{Introduction} \label{sec:intro}

Morphological classification of galaxies began the dialogue between evolutionary theories and observations that led to our current understanding of the variety of galaxy properties in our Universe \citep{Hubble26, Sandage75, Sandage05, Kormendy09, Kormendy12, Graham13a}. In the picture of hierarchical galaxy assembly described in a $\Lambda$ cold dark matter ($\Lambda$CDM) cosmology, early-type galaxies (ETGs) are thought to be the end-products of these evolutionary paths \citep{burkert_naab04, De-Lucia06}. Consequentially, ETGs are important observational tests of our knowledge of the Universe. 

\subsection{Kinematics}

Observations of ETGs have led to powerful empirical scaling relationships like the Fundamental Plane (FP) of ETGs that help us to understand the formation and evolution of these galaxies \citep{Faber76, Kormendy77, Djorgovski87, Dressler87, cappellari06, bolton07_fp, Renzini06}. One of the key insights of the FP is a correlation between the line-of-sight stellar velocity dispersion and several fundamental galaxy properties.
Measurements of elliptical galaxy kinematics with spectroscopy have therefore been very scientifically productive, yielding a wealth of insights into their dynamics, formation, and evolution (see e.g., \citealt{Cappellari2016} for review). 
Integrated line-of-sight velocity dispersions of ETGs from single-aperture spectroscopy have contributed significantly to the characterization of the masses of distant galaxies through dynamical inferences and joint analysis with other mass measurements, e.g., gravitational lens models \citep[henceforth SLACS-IX and SLACS-X]{slacs9, slacs10} and stellar population studies. 

Galaxy kinematics studies gained an additional dimension with the advent and now ubiquity of integral-field spectrographs (IFS) on most major observatories \citep[e.g., TIGER, SAURON, MUSE, KCWI]{bacon95_tiger, bacon01_sauron, bacon10_muse, morrissey12_kcwi}. This technology allows spatially resolved velocity measurements because each spatial pixel (or spaxel) is dispersed into an individual spectrum. 

This gives a 2D projection of the stellar velocity distributions within the galaxy. Dynamical arguments and other physical constraints allow one to deproject and constrain the intrinsic 3D stellar orbits, which trace the gravitational potential of the galaxy. The 3D nature of the stellar orbits encoded in the observable 2D kinematics are described by anisotropy parameters, which quantify how the stellar velocity ellipsoid (in velocity space) at a given spatial coordinate deviates from an isotropic case (i.e., there is a preferred direction of stellar velocities). 

Studies of elliptical galaxy isophotes \citep{kormendy_bender96, faber97, Kormendy12, Kormendy16} and kinematics revealed a dichotomy that prescribes two distinct classes of ETGs. The SAURON survey \citep{bacon01_sauron, dezeeuw02}
used SAURON IFS on the William Herschel Telescope and \textit{HST} images to study the kinematics of galaxies in the local universe \citep{emsellem04, Emsellem07, cappellari06, cappellari07}. This and other studies with IFS \citep[ATLAS$^{\mathrm{3D}}$, SAMI, SDSS MaNGA]{Cappellari11, Krajnovic11, Bryant15, Bundy15} have opened new kinematic-based galaxy-classification schemes and offered insights to the formation mechanisms of ETGs \citep[see][for a full review]{Cappellari2016}. Many galaxies that appear to be elliptical are in fact better described as rotationally-supported, flattened disks with significant bulge components and low star-formation. Their elliptical appearance stems from projection effects. The kinematic properties of these rotating ellipticals challenge an evolutionary scenario where growth and star-formation quenching are dominated by major mergers. Instead, a large contributor to their growth must come from minor ``wet" (i.e., gas rich) mergers and accretion of gas from the surrounding galactic medium, with star formation being quenched by internal mechanisms \citep{cheung12, fang13}. In fact, the parameter space of features that describe nearby younger disk galaxies and their disky elliptical cousins overlaps considerably \citep{falconbarroso15, Querejeta15}. The disks masquerading as ETGs appear to be once-star-forming galaxies that were lucky enough to lead a relatively mergerless life and so retain their disky identity. The cases that resemble the more classic elliptical picture are likely the result of mergers \citep[e.g.,][]{bezanson09, naab09, hopkins10_philip}.
These two classes have been termed ``fast (or regular) rotators" and ``slow (or nonregular) rotators," as they can be distinctly classified with significantly different kinematic attributes \citep{Emsellem07,cappellari07}.

\subsection{Dynamics and Strong Gravitational Lensing}

Dynamical mass models of ETGs with IFS spatially resolved kinematics (e.g., SAURON, ATLAS$^{\mathrm{3D}}$, etc.) have been created using solutions to the two- and three-integral Jeans equations of stellar hydrodynamics in spherical and axisymmetric alignments \citep{emsellem94, cappellari08}, and more generally with \cite{schwarzschild79} models of orbital superposition generalized to fit kinematic data \citep{richstone_tremaine88, rix97, vandermarel98}. These lead to robust measurements of the total mass content (baryonic and dark) and the mass-density profiles of these galaxies, which have been shown to be in excellent agreement with mass measurements obtained with strong gravitational lensing. Studies of lensing ETGs have shown them to be indistinct from other ETGs in the observed mass range \citep{Auger09}, which tends to be relatively large compared to the overall population of ETGs due to strong dependence of lensing optical depth on velocity dispersion \citep{shu15, Knabel20}. In the mid 2000s, the Sloan Lens ACS (SLACS) survey produced a wealth of papers \citep[e.g.,][etc.]{slacs1, slacs2, slacs3} utilizing strong gravitational lensing and other observables to study the properties of a definitive sample of 85 large early-type gravitational lenses (plus another hundred or so in additional follow-up surveys). These galaxies were identified spectroscopically as strong lensing candidates in the Sloan Digital Sky Survey (SDSS) and confirmed with \textit{Hubble Space Telescope} (\textit{HST}) Advanced Camera for Surveys (ACS). The resulting sample combined strong lens models and single-aperture velocity dispersions to jointly constrain the mass profiles and fundamental scaling relations \citep{slacs7, slacs9, slacs10}. SLACS is perhaps the most well-studied set of gravitational lenses, having been followed-up by several additional independent observations and analyses. For example, \cite{czoske08, Czoske12} used VLT/VIMOS-IFU observations to determine spatially resolved stellar kinematics of several SLACS lenses, providing a first comparison to IFS kinematic surveys of more nearby galaxies like SAURON \citep{emsellem04} and ATLAS$^{\mathrm{3D}}$ \citep{Cappellari11}.

\subsection{Time-Delay Cosmography and Overcoming Degeneracies}

A large effort has been undertaken in the past two decades to independently measure H$_0$ through time-delay cosmography of lensed quasars \citep{Refsdal64}; for an up-to-date review, see \cite{birrer22b}, \cite{treu22}; for a historical perspective,
see \cite{Treu16}. The Time-Delay COSMOgraphy (TDCOSMO) collaboration has analyzed seven time-delay lenses to measure the Hubble constant within 2\% error (H$_0 = 74.2\pm1.6$ $\mathrm{km}$ $\mathrm{s}^{-1}$ $\mathrm{Mpc}^{-1}$) assuming simple parametric mass profiles for the lensing galaxies \citep{millon20b}. TDCOSMO encompasses the COSmological MOnitoring of GRAvItational Lenses \citep[COSMOGRAIL,][]{courbin05, millon20a}, the H0 Lenses in COSMOGRAIL’s Wellspring \citep[H0LiCOW,][]{Suyu17, Bonvin17, Birrer19, rusu20, wong20}, the Strong-lensing High Angular Resolution Programme \citep[SHARP,][]{Chen19}, and the STRong-lensing Insights into the Dark Energy Survey \citep[STRIDES,][]{Treu18, shajib21, schmidt23} collaborations. 

An important and delicate caveat in the application of lens modeling methods is called the ``mass-sheet degeneracy" (MSD), which can be succinctly described as the non-uniqueness of mass profile solutions that can result in the observed lensed image features  \citep{falco85, schneider_sluse13, schneider_sluse14}. 
The TDCOSMO collaboration has tested several potential sources of uncertainty and concluded that the MSD is the most significant source of uncertainty in the measurement of H$_0$ with time-delay cosmography of lensed quasars \citep{birrer20_tdcosmo_iv, millon20b, gilman20, vandevyvere22a, shajib22, vandevyvere22b}. A non-lensing tracer of gravitational potential (e.g., kinematics) is required to break the degeneracy \citep{treu_koopmans02, slacs10, shajib18, yildirim20, yildirim21}. Using single-aperture velocity dispersions, while allowing for more flexible mass models, \cite{birrer20_tdcosmo_iv} (henceforth TDCOSMO-IV) measured H$_0$ to within 8\% precision ($\mathrm{H_0} = 74.5^{+5.6}_{-6.1}$ $\mathrm{km}$ $\mathrm{s}^{-1}$ $\mathrm{Mpc}^{-1}$) using just 7 time-delay lenses.

Line-of-sight velocity dispersions from single-aperture spectroscopy can constrain galaxy dynamics, but there exists another degeneracy, the mass-anisotropy degeneracy (MAD, \citealt{Binney1982}), which cannot be broken without spatially resolved kinematics \citep{cappellari08, barnabe09, Barnabe11, Collett18, shajib18, Birrer20, birrer_treu21, shajib23}. MAD arises because we cannot observe all components of the stellar velocities, but only their projection along the line of sight. The anisotropy parameter $\beta_{\rm \rm ani}$ that encodes this information is the crucial unknown component. Studies of nearby ETGs like the SAURON survey \citep{Emsellem07, cappellari07} show that this anisotropy is not a universal parameter, nor is it radially uniform within a single galaxy. Anisotropy profiles can be illuminated with the study of spatially resolved kinematics. Breaking either the MSD or MAD on its own is a daunting task, but the union of lensing and dynamics can put joint constraints on both. This is the strategy outlined by \cite{shajib18} and \cite{birrer_treu21}, and demonstrated by \cite{shajib23} (henceforth TDCOSMO-XII) using KCWI resolved kinematics of quasar time-delay lens RXJ1131-1231 to constrain H$_0$ within 9\% error for the single lens (comparable to the sample of 7 lenses in TDCOSMO-IV). Additional constraints can be placed on these cosmography measurements by auxilliary observations of non-time-delay (i.e., the lensed source is not variable in time) lensing galaxies as in TDCOSMO-IV. 
Through Bayesian hierarchical inference, constraints from the population level increase the precision of time-delay lens analysis. TDCOSMO-IV used single-aperture SDSS \citep[SLACS-XII,][]{shu15} and VLT/VIMOS-IFU \citep{czoske08, Czoske12} kinematics from 33 SLACS lenses to increase the precision of their sample from 8\% to 5\%.

However, the kinematic data need to have sufficient S/N and spectral resolution to enable measurements of the Hubble constant with the precision and accuracy required to help settle the 8\% difference between the early and late universe probes \citep[e.g.,][and references therein]{abdalla22_cosmology}. The relative error on the Hubble constant is to first order equal to double the relative error on stellar velocity dispersion \citep[$\delta {\rm H}_0/{\rm H}_0\approx 2\delta \sigma/\sigma$, e.g.,][]{Chen21}. Whereas random errors can be constrained with larger samples, it is crucial to mitigate the systematic bias and covariance within and between datasets, ideally to the sub-percent level. Reaching this goal is possible with data of sufficient resolution and S/N, if the data and stellar templates are clean of defects and careful methods are applied for extracting the kinematics \citep{TDCOSMO19}.

Unfortunately, the VLT/VIMOS-IFU data had insufficient spatial sampling (0.67$''$ per spaxel) and spectral resolution (R$\sim$2500) to reach the precision and accuracy required to overcome the MAD and measure H$_0$ to our desired levels. Furthermore, VIMOS as a fiber-based first-generation IFS on VLT did not have the stability and accuracy of sky subtraction and wavelength and resolution calibration of modern day IFS such as KCWI and MUSE. TDCOSMO-IV found VIMOS data to require a significant error boost to account for systematic uncertainties. They did not utilize the absolute measurements of the velocity dispersion and instead used only the relative shape to constrain the MAD. Likewise, the SDSS data of the SLACS lenses had limited S/N \citep{TDCOSMO19}, and stellar velocity dispersion had been measured for galaxy formation and evolution purposes and not with precision cosmography in mind. Systematic errors were thought to be at the 5\% level \citep{Bolton08,shu15}.

We present spatially-resolved kinematics of a selection of 14 strong-lensing ETGs selected from the SLACS catalog. The SLACS survey is an ideal parent sample from which to select our sample for the wealth of auxiliary data and measurements that exist and their observability in terms of size and brightness. The resolution (R $\sim$ 3600), spatial sampling (spaxel size $ 0.1457'' \times 0.1457''$), and sub-arcsecond seeing (average $0.9''$ across the observing dates) afforded by Keck and KCWI offer a more detailed description of the 2D velocity distribution than any previous kinematics study of these galaxies.  In order to minimize systematic errors and take full advantage of the high quality of our spectroscopic data, we adopt the recently developed methods and clean stellar libraries described by \citet{TDCOSMO19} (henceforth TDCOSMO-XIX). Lensing models have been conducted in a uniform manner for all objects in our sample by \cite{tan23}.

New dynamical models for each galaxy are being computed via Jeans Anisotropic Modeling \citep[JAM,][]{cappellari08, cappellari20_jeans_axis} and will be presented in a follow-up paper (Paper II). 

Observations were taken with the Keck Cosmic Web Imager (KCWI) and are presented in Section \ref{sect:observations}. Kinematic analysis methods are described in Section \ref{sect:analysis}, including systematic and random error estimates, estimated on the basis of the methods recently developed by TDCOSMO-XIX. 
Kinematic maps and initial results, including kinematic classification and aperture-integrated kinematics, are introduced in Section \ref{sect:results}. Correlations between kinematics and other observables are discussed in Section \ref{sect:correlations}. Additional points of discussion are considered in Section \ref{sect:discussion}, and finally we summarize our results in Section \ref{sect:conclusions}. Where necessary, we assume a standard $\Lambda$CDM cosmology with $\Omega_m=0.3$ and H$_0=70$ km $\mathrm{s}^{-1}$ $\mathrm{Mpc}^{-1}$. We note of course that stellar velocity dispersions are independent of cosmology.

%%%%%%%%%%%%%%%%%%%

\section{Observations}\label{sect:observations}

We select 14 objects from the Sloan Lens ACS survey that have been uniformly modeled by \cite{tan23}. The sample is representative of the parent sample, being selected mostly based on observability, declination and right ascension, according to when telescope time was scheduled. Spectra were observed with the Keck Cosmic Webb Imager \citep[KCWI,][]{morrissey12_kcwi, morrissey18_kcwi} on Keck 2 during both 2021 semesters and the first semester of 2022 (nights of 2021 May 6 and 15, June 7, August 12, November 26; 2022 April 7). See Table \ref{tab:seeing_table} for average seeing estimates obtained at the telescope on the dates of observation. At redshifts around z$\sim$0.15-0.35, a S/N of $\sim$10 can be achieved with 2-3 hours of exposure. As in TDCOSMO-XII we take a series of 30 minute exposures with the low-resolution blue grating centered at 4600$\mathrm{\AA}$ with the small slicer and 1$\times$1 binning. The spectral resolution of R $\sim$ 3600 corresponds to an instrumental dispersion $\sigma_{\mathrm{inst}} = 35$ km $\mathrm{s}^{-1}$. The reciprocal dispersion is 0.5$\mathrm{\AA}$ per pixel. We dither by around 8 arcseconds along the long axis of the $8.4''\times20.4''$ field of view to obtain background sky simultaneously with the on-source exposure for sky subtraction. Standard stars are observed periodically through the night for calibration. 

\begin{table}[]
    \centering
    \begin{tabular}{c|c}
         Observing date & Average seeing ($''$) \\
         \hline
         7 May, 2021 & 1.08 \\
         16 May, 2021 & 1.2 \\
         8 June, 2021 & 0.5 \\
         13 August, 2021 & 1.2 \\
         27 November, 2021 & 0.8 \\
         8 April, 2022 & 0.9
    \end{tabular}
    \caption{Average seeing estimate from telescope focus on dates of observation with Keck Cosmic Web Imager (KCWI).}
    \label{tab:seeing_table}
\end{table}

The resulting 4-5 exposures for each object are reduced with the KCWI official Python-based data reduction pipeline (DRP; developed by Luca Rizzi, Don Neill, Max Brodheim; \url{https://kcwi-drp.readthedocs.io/}). The pipeline translates the 2D information from the detector into a 3D datacube, performing geometry correction, differential atmospheric refraction correction, wavelength calibration, and standard-star calibration. The calibration with the standard star corrects for instrumental response and scales the data to flux units \citep{morrissey18_kcwi}. We use the final output files with the suffix “icubes” and create mosaics from the frames by drizzling \citep{fruchter_hook02}.  Drizzling is performed using a modification of the routine utilized in the official Keck OSIRIS IFU pipeline, using recommended settings \citep{avila15}. Throughout the drizzling process, the rectangular ($ 0.1457'' \times 0.3395''$) geometry of KCWI pixels is maintained. We transform to square pixels of dimensions $ 0.1457'' \times 0.1457''$ after drizzling by resampling to conserve flux. 

\begin{figure}[t]
    \centering
    \includegraphics[width=0.8\linewidth]{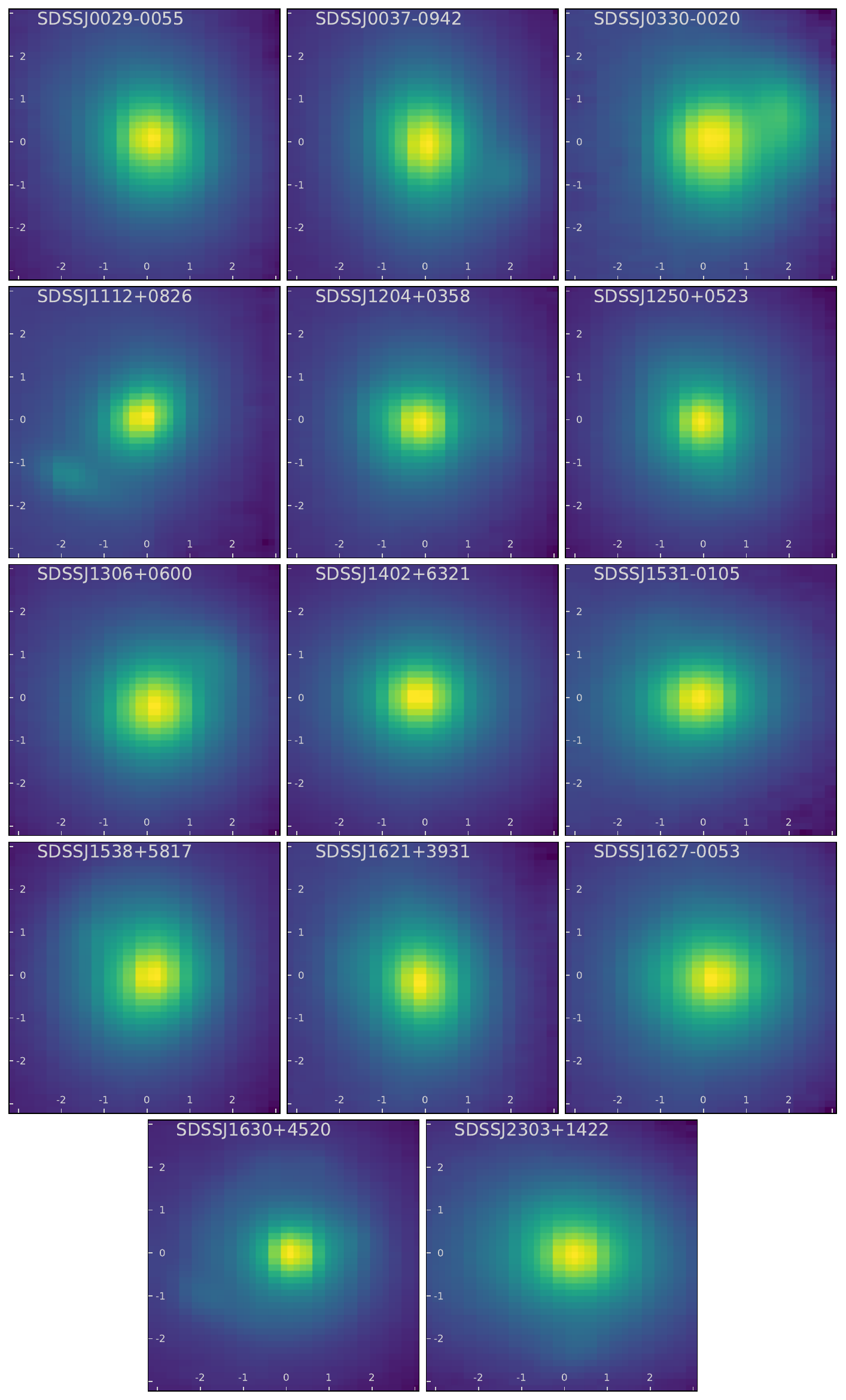}
    \caption{Datacube for each object summed across all wavelength bins in the fitted range 3600-4500$\rm \AA$ and cropped to 43$\times$43 pixels ($6.235''\times6.235''$).}
    \label{fig:datacube}
\end{figure}

In order to measure the point spread function (PSF), we sum the datacube across wavelength bins to create a 2D image, which we show in Figure \ref{fig:datacube}. 
\textit{HST} imaging is available for each of these objects (ACS/WFC F814W and F435W, WFPC2 F606W; \textit{HST}-SNAP 10174; \textit{HST}-GO 10494, 10886, 10798, 11202; PIs Koopmans and Bolton). We convolve the high-resolution HST image with a Gaussian PSF kernel and fit to the KCWI summed datacube for optimal values of the PSF width $\sigma$, with two additional parameters to correct for offset of the two images.  
Surface brightness profiles from \textit{HST} also allow us to examine the kinematic structure in relation to the morphology, especially the alignment (or misalignment) of the kinematic and photometric major axes. The kinematic maps and photometry are aligned with the brightest pixels in the central foreground deflecting galaxy. We take B-spline models and PSF models from \cite{slacs5} for these \textit{HST} images. We use Multi-Gaussian Expansion \cite[MGE]{emsellem94, cappellari02} to translate the B-spline models and PSF models (for both KCWI and \textit{HST} data) to a parametrization that is ideal for future dynamical fitting.

%%%%%%%%%%%%%%%%%%%

\section{Kinematics Measurements} \label{sect:analysis}

The spatially-resolved kinematics analysis is done with Penalized Pixel-Fitting method\footnote{We used v9.0.1 of the Python package from \url{https://pypi.org/project/ppxf}.} \citep[{\sc pPXF},][]{Cappellari04, cappellari17, cappellari23_ppxf}, which extracts stellar kinematics by fitting the observed spectrum with a linear combination of template stellar spectra convolved with (broadened by) a line of sight velocity distribution (LOSVD) constructed with a Gauss--Hermite parametrization. Observed and template spectra are all logarithmically rebinned before fitting. Additive Legendre polynomials give greater fitting power by adjusting the strength of individual absorption lines \citep[see][for a full description]{cappellari17}, and multiplicative polynomials account for inaccuracies in the relative spectral flux calibration (see TDCOSMO-XIX for further discussion).

Traditionally, the singly-ionized Calcium (Ca II) H and K stellar absorption lines (3934 \& 3969\AA, hereafter CaHK) have been used to measure stellar kinematics \citep[see e.g.,][]{Dressler79}. CaHK are very strong and intrinsically broad features, which makes them ideal for detections with low brightness and low S/N. Discussion of the consistency of CaHK kinematics \cite[e.g.,][]{Kormendy82, Murphy11} have become insignificant since {\sc pPXF} standardized the practice of using linear combinations of multiple templates from a flexible stellar library.
Alternative features throughout all wavelength ranges have also been used. 
The redshifted CaHK lines for all objects are present well within the observed wavelength range (3500-5600\AA). NaD and Mg I lines will be redshifted beyond 5600$\mathrm{\AA}$ for all objects (z$\sim0.15-0.35$). G-band absorption due to CH and neutral Fe (restframe 4304\AA ) is observable for all but one target (SDSSJ0330$-$0020 at $\mathrm{z=0.351}$). 

Our extracted kinematics are unbiased when excluding the G-band, and we take full advantage of all the available line features for each individual object in the sample. Our baseline extraction wavelength range is 3600-4500$\mathrm{\AA}$. The blue cutoff is set by the quality requirements on the template stellar spectra, which have low S/N and imperfect calibration at shorter wavelength \citep{TDCOSMO19}. The red cutoff is set by the need to have a uniform wavelength range and capture the G-band.
We examine the effects of the selection of wavelength range, as well as other systematics, in Section~\ref{sect:uncertainties}.

\begin{figure}
    \centering
    \includegraphics[width=0.49\textwidth, trim={0.25cm 0.25cm 0.5cm 0.cm}, clip]{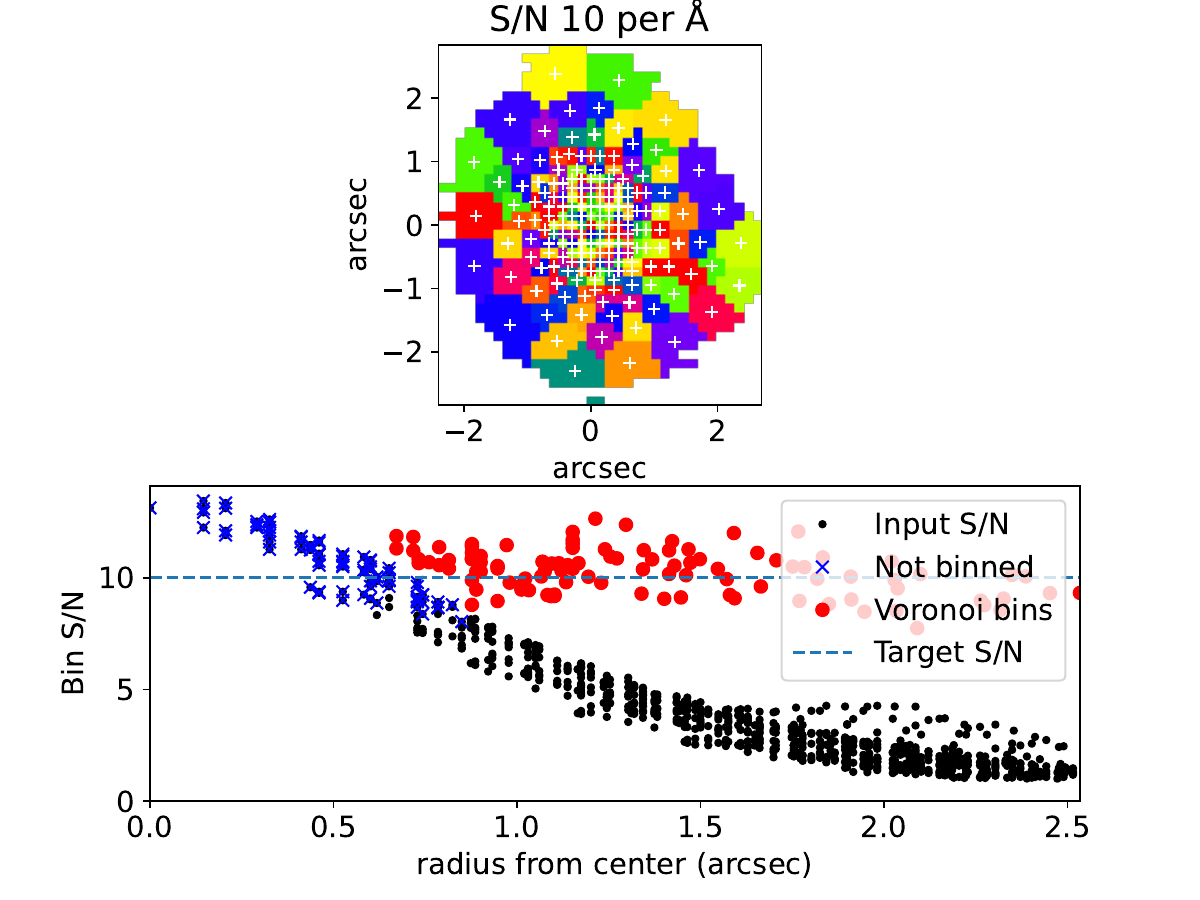}
    \includegraphics[width=0.49\textwidth, trim={0.5cm 0.25cm 0.25cm 0.cm}, clip]{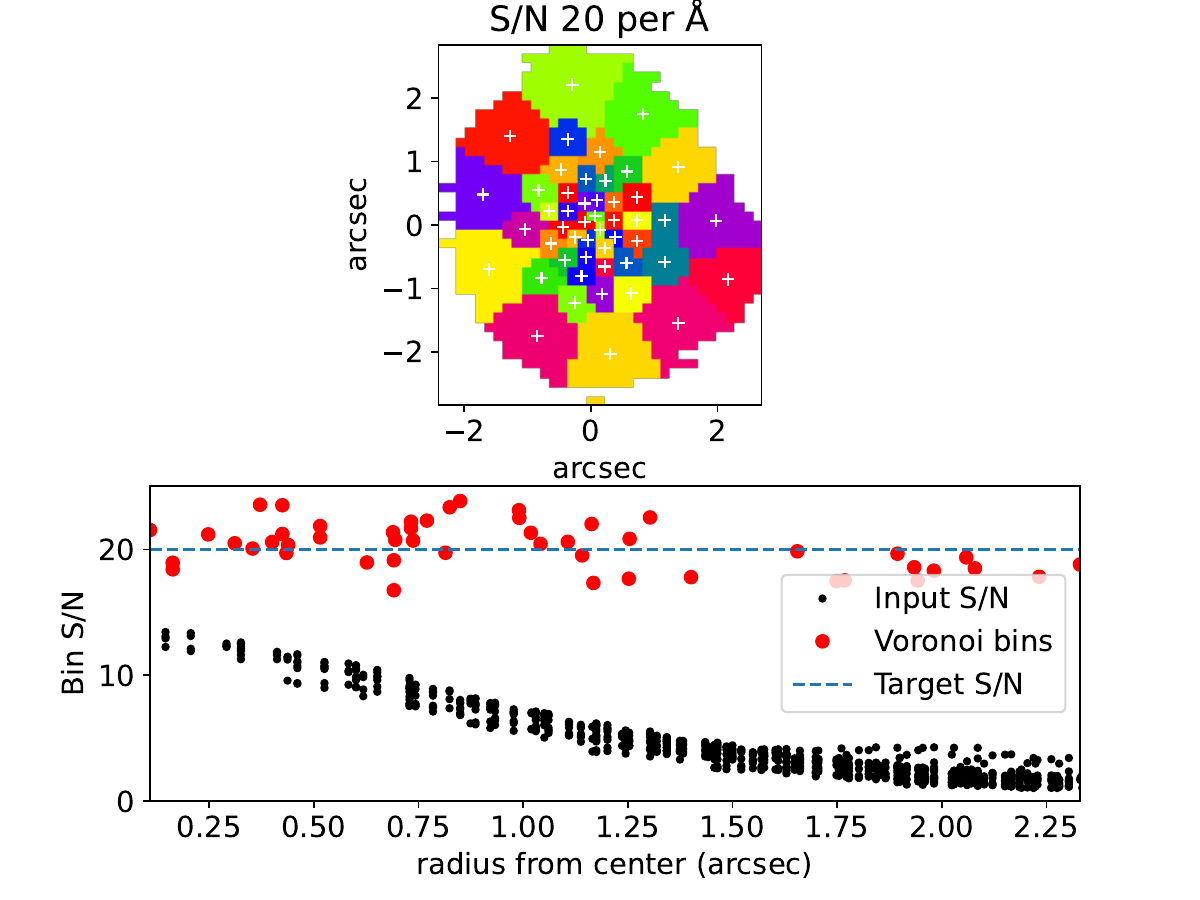}
    \caption{Voronoi binning of SDSSJ0037$-$0942 with target bin S/N ratios of 10 (\textit{left}) and 20 (\textit{right}). The upper plot for each is a map showing the spatial boundaries of bins, with bin centers marked with ``+". Beneath, we show the binning. The vertical axis is the S/N per $\mathrm{\AA}$, and the horizontal axis is radius from the center. Black dots are the spaxels that are binned, and blue crosses indicate spaxels with a high enough S/N to be a bin on their own. Red markers are the Voronoi bins. Taking a lower target S/N offers more spatial information at the cost of increased uncertainty in the kinematics of individual bins.}
    \label{fig:voronoi_binning}
\end{figure}

In principle, kinematics can be measured for each spaxel in the datacube. Spaxels at larger radii from the center of the galaxy will suffer from lower S/N than the central spaxels, so we bin them in order to measure kinematics at a consistent S/N and avoid introducing potential biases. We use {\sc VorBin} \footnote{We use v3.1.5 of the Python package from \url{https://pypi.org/project/vorbin}.} \citep{capp03_vorbin} Voronoi binning software to select only spaxels with S/N (per $\mathrm{\AA}$) greater than 1 and group those spaxels to achieve close to a designated target S/N within each bin. An example of this selection for target S/N values of 10 and 20 is shown in Figure~\ref{fig:voronoi_binning}. For our baseline, we use a S/N target of 15, which results in roughly 50-100 spatial bins depending on the brightness of the object.

In order to avoid overfitting and introducing biases, we do not fit each Voronoi bin independently with freely weighted stellar templates. Instead we first create a global template from a high-S/N (typically $\sim100$ per $\mathrm \AA$) aperture-integrated spectrum extracted from the datacube within half the effective radius of the galaxy. 

\begin{figure}
    \centering
    \includegraphics[trim={0.25cm 0.25cm 0.25cm 0.25cm},width=0.8\textwidth,]{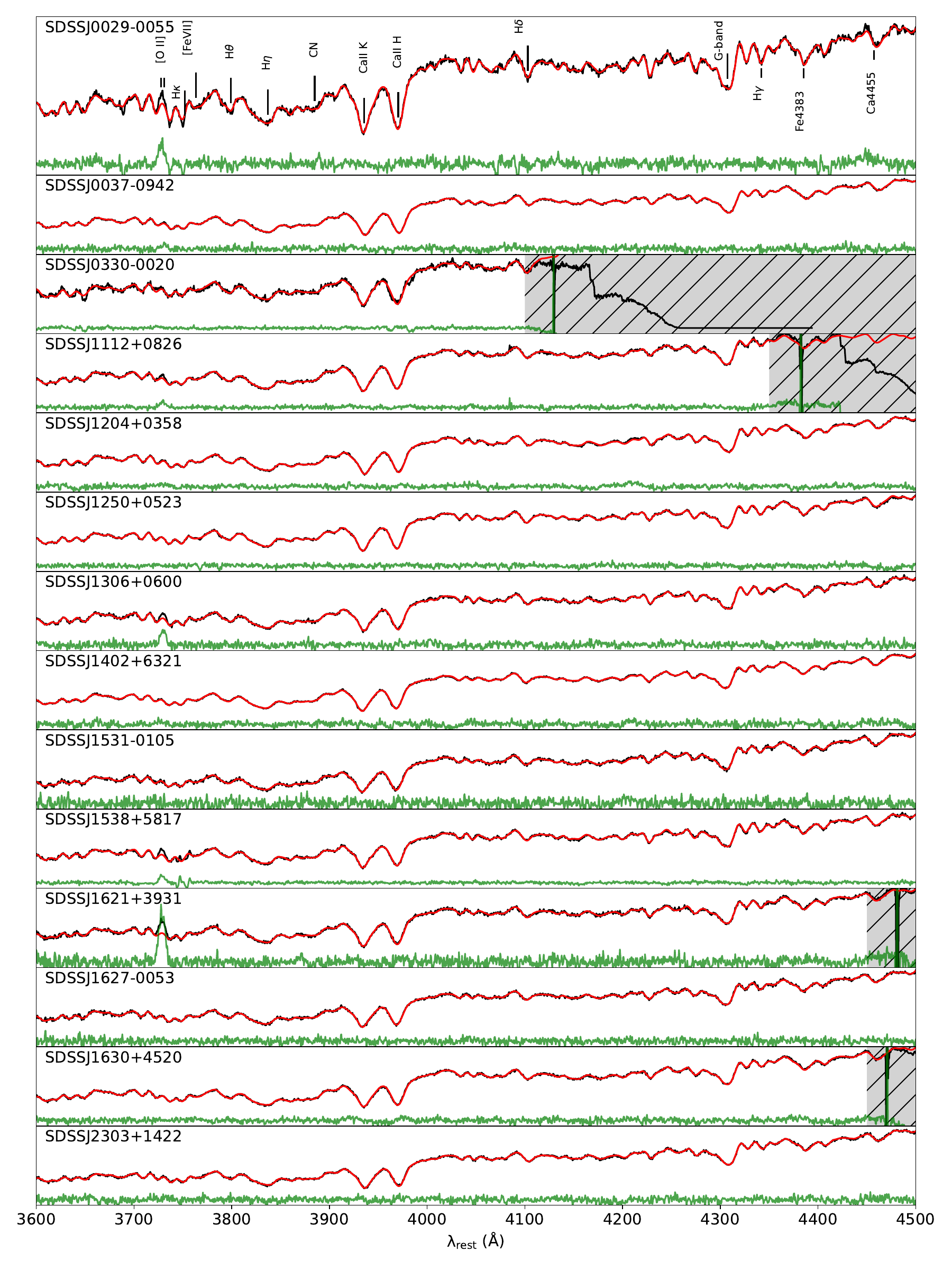}
    \caption{Restframe spectrum of each deflector galaxy in our sample extracted from an aperture of half the galaxy effective radius, fit with stellar template spectra from the X-shooter Spectral Library (XSL). Black lines show the data. Red shows the model. Green shows rescaled residuals. Gray regions show masked wavelengths where the redshifted observed wavelengths are redder than the instrument range. Notable line features are highlighted in the first spectrum for demonstration. During fitting, sigma-clipping removes some contaminating emission lines, especially $\rm \left[OII\right]\lambda$3727, and spurious single wavelength elements.
    }
    \label{fig:central_spectrum}
\end{figure}

Following TDCOSMO-XIX, for our baseline model, we fit the aperture-integrated spectrum with linear combinations of stellar templates from the Indo-US library of empirical stellar templates\footnote{\url{https://noirlab.edu/science/observing-noirlab/observing-kitt-peak/telescope-and-instrument-documentation/cflib}} \citep{valdes04_indo_us} in a wavelength range of 3600-4500$\mathrm{\AA}$. We include corrective additive and multiplicative polynomials of degree 6 and 2, respectively, and convolve with a Gaussian (\texttt{moments=2} in pPXF) LOSVD kernel. This linear combinations of typically $\sim20-30$ stellar templates is the global template for fitting the spatial Voronoi bins of the datacube. Example global template spectra extracted from the centermost spaxels for each of the 14 objects are shown in Figure~\ref{fig:central_spectrum}. The composite spectrum in each Voronoi bin is then fit with the appropriate annular global template with fixed template weights by scaling the overall normalization, introducing new corrective polynomials as described above, and convolving with a LOSVD. The extracted mean velocity is a small correction to the galaxy cosmological recession velocity. Due to how the velocity is defined in pPXF, the peculiar velocity in the galaxy rest frame can be obtained by simply subtracting the velocity of the barycenter from the fitted velocity \citep[eq.~3]{cappellari23_ppxf}. This gives the galaxy's rotation with positive values showing receding (redshifted) bins. The resulting kinematic maps describe the 2D projection of stellar motion within the galaxy.

This method assumes that the kinematic tracer population is adequately described by a single spatially-uniform stellar population and is standard practice for extracting the kinematics of ETGs \citep[see, e.g.,][]{zhu23}. Following the methods outlined by TDCOSMO-XIX for evaluating the goodness of fit across several models by comparing the Bayesian information criterion (BIC, see TDCOSMO-XIX Section 5 and Appendix for full derivations) for each model, we investigate the possibility of a radial stellar population gradient by fitting radial annuli instead of a single central aperture. We take annuli of radial width $0.5''$ centered at the brightest spaxel in the center of the deflector galaxy. We ensure the aperture-integrated spectra from annular bins have S/N $>$ 30, but most are closer to 60-70. Voronoi bins are fit with the appropriate annular global template according to the radius of the luminosity-weighted center of the Voronoi bin. In this way, we allow for gradients in the stellar population while controlling for spurious individual bin fits. At the Voronoi bin level, compared with our baseline bin measurements (with a single global template), the flexible stellar population results in random variations to the extracted velocity dispersions at the $1\%$ level that are highly correlated with the bin's membership in the annular bins. The resulting BIC evidence shows no indication of improved goodness of fit for the added flexibility, and there is a clear risk of introducing a systematic bias to the radial profiles of the velocity dispersions. Therefore, we utilize a single global template for each galaxy. We go into further detail regarding systematic effects in Section~\ref{sect:uncertainties}.

\subsection{Systematics, Biases, and Error Budget}\label{sect:uncertainties}

Formal statistical uncertainty estimates in the {\sc pPXF} model are computed from the standard errors in the fit and are corrected by multiplying by $\sqrt{\chi^2}$, in order to account for potential error under/overestimates. $\chi^2$ is calculated as the square sum of the residuals between model spectrum and data divided by the degrees of freedom (approximately the number of spectral bins). In order to test the robustness of the statistical uncertainties on the extracted bin kinematics, we use wild bootstrapping with a Rademacher distribution \citep[see][eq.~7]{davidson_flachaire08_wild_bootstrap} of residuals for each bin fit for each object in our sample. Across all bins, the uncertainties on mean velocity (velocity dispersion) from bootstrapping are on average 2.6\% (3.1\%) smaller than the {\sc pPXF} formal uncertainties, with a standard deviation of 5.6\% (5.3\%). Similar tests were done with MaNGA data \citep[see right panel of Figure 20,][]{Westfall19}, finding that the {\sc pPXF} formal uncertainties are very close to the true ones, except at unreasonably low S/N. We therefore utilize and report the formal statistical uncertainties directly from {\sc pPXF}.

%%%%%%%%%%%%%
% Systematics

With our state-of-the art methods and the high quality of KCWI data, statistical errors at the Voronoi bin level are on average $3-4\%$, which easily achieves the precision required for population-level inferences. However, potential sources of systematic error may bias individual measurements or introduce covariance at the bin or sample level, introducing an effective noise floor that is independent of sample size or S/N. Some well-studied sources of systematics include instrumental resolution, S/N, wavelength range, mismatch between templates and galaxy spectra, flux calibration, and interstellar absorption. Many excellent studies on this topic have been published in the past \citep[e.g.,][]{Dressler84,kelson00,treu01,Barth02,Spiniello15,Spiniello21a,Spiniello21b,Mehrgan2023,DAGO23}. We experimented with a variety of additional potential sources of systematic uncertainty by testing small adjustments to virtually every step in our fitting procedure. Systematic choices that we tested and found to result in negligible differences include the degrees of corrective polynomials 0.6\%, target S/N thresholds for Voronoi binning, subsets of stellar templates from within the same library, and masks over individual absorption features present in the wavelength range of fitting. Ultimately, we found that only two aspects made a difference beyond the 1$\%$ level in bin velocity dispersions: the library of stellar templates from which the galaxy spectra are fit and the wavelength range used.

% Template library

To address the first, the TDCOSMO collaboration implemented a rigorous study of template libraries and our methods for extracting stellar kinematics (TDCOSMO-XIX). We demonstrated with four independent datasets (including the KCWI data presented here) and three independent code implementations that the most popularly-used stellar template libraries are unbiased in velocity dispersion to a sub-percent level when they are properly cleaned of contaminants and applied to appropriate wavelength ranges for data of high S/N. The libraries that were cleaned and examined were Indo-US, MILES\footnote{\url{https://miles.iac.es/}} \citep{ SanchezBlazquez2006,Falcon-Barroso11}, and  X-shooter Spectral Library\footnote{\url{http://xsl.u-strasbg.fr/}} \citep[XSL,][]{verro22_xshooter}. For that analysis, the KCWI datacubes were integrated over a 1-arcsecond radius from the center of the galaxy in order to achieve an average S/N of 160 per \AA, which reduces the statistical uncertainty well below the effects of systematics. We introduced a procedure to evaluate the evidence for each library based on the Bayesian information criterion (BIC) measurement for each dataset. The SLACS KCWI dataset revealed a strong preference for the Indo-US stellar template library, with a negligible contribution to the systematic uncertainty when accounting for the BIC preference. We therefore take the Indo-US extractions as the fiducial models. When averaged over the sample, velocity dispersions measured with the MILES library are $0.7\pm0.3\%$ higher, and XSL results in velocity dispersions $1.0\pm0.2\%$ lower. Even when averaged with equal weight over the stellar template libraries, the systematic uncertainty is still sub-percent at $0.79\%$. The stellar template libraries were constructed from observations with different instruments and different stars, so across a sample of galaxies, the measured velocity dispersions from one template library compared with another are correlated. Sample-level covariances are $0.47\%$ when BIC-weighted and $0.86\%$ when averaged with equal weight over the stellar template libraries. We also included single stellar population (SSP) models and found they lack the flexibility provided by individual stellar spectra. We do not recommend their use for sub-percent accuracy. We also tested subsets of libraries selected at random or by stellar temperature and metallicity and found them to consistently recover the same sets of stellar templates and extracted kinematics, indicating that {\sc pPXF} is able to efficiently select the most appropriate templates from a given library without a pre-selection based on assumptions about the stellar population. Given well-calibrated and appropriately representative empirically-derived template libraries, and with careful selection of the wavelength range of the fit, the systematic effects from the template selection are well-controlled.

% Wavelength range

The wavelength range specified for spectral fitting can introduce systematic effects in the selection of stellar templates and in the extracted bin kinematics. For example, the truncation of any line feature by wavelength boundaries (especially at the longer wavelength end) can affect the fit. To mitigate this effect, we select bounds where the continuum is relatively flat and there are no contaminating spectral features. In addition, we find that fits suffer when including much of the rest-UV spectrum below 3600$\mathrm{\AA}$. This is due to low S/N and poor flux calibration of the stellar libraries below this blue cutoff.  Extending fits to bluer regions will require better stellar templates than the ones currently available. However, this is not an issue, since the complex spectrum there does not contain absorption features that are clearly informative for the stellar kinematics, and fits that include those wavelengths include large contributions from hot F, A, and B stars that may be thermally broadened and bias the velocity dispersion measurements. 

We test three regions of equal wavelength range enclosed within our baseline extraction range: $\rm 3600-4250\AA$, $\rm 3700-4350\AA$, and $\rm 3800-4450\AA$. Two of our galaxies (SDSSJ0330-0020 and SDSSJ1112+0826) are higher redshift than the others, and their restframe spectra do not span the full range of wavelengths needed for this test. We remove those two and test the systematics with the other 12 objects. Unlike the analysis presented by TDCOSMO-XIX, the wavelength effect on the extracted velocity dispersion cannot be assessed in terms of BIC goodness-of-fit. This is because BIC deals with the likelihood of the data given a systematic model choice. Comparing kinematic extractions of different ranges of spectra is by definition utilizing different data with different noise realizations. In order to get the closest possible comparison, we ensure that the variations in wavelength ranges span the same number of sampled data points, which depends on the galaxy's redshift. Because the data is log-rebinned before fitting, the wavelength ranges centered at longer wavelengths will have fewer sampled data points. We calculate the number of data points in the range $\mathrm 3600-4250\AA$ and extend the blue end of the other two ranges to include the same number of total sampled points. Because the noise realization is different for each range, the systematic effect is muddled by random noise variations, so we need to use the highest-possible S/N. Following TDCOSMO-XIX, we integrate each datacube within 1 arcsec. At the sample level, the effect of one wavelength selection compared with another averages to 0. We create a $12 \times 12$  covariance matrix to quantify the covariance between objects. Matrix element (\textit{n, m}) is calculated as

\begin{equation}
    \Delta_{n,m} = \frac{ \langle \Delta\sigma_{n,i} \Delta\sigma_{m,i} \rangle}
    { \overline{\sigma}_n \ \overline{\sigma}_m }
\end{equation}

\noindent where

\begin{equation}
    \Delta\sigma_{n,i} = \overline{\sigma}_n - \sigma_{n,i}
\end{equation}

\noindent and $\sigma_{n,i}$ is the velocity dispersion from the $i$th variation of object $n$.

The average of the diagonal is $0.8 \%$, and the average off-diagonal is consistent with 0 within the error on the mean. We test for spatial covariance by repeating the test at the bin level, creating an $N\times N$ covariance matrix for each object where $N$ is the number of spatial bins. We then average over each object and then over the sample to quantify the average contribution to the bin variance. We test increasing S/N of the spatial Voronoi bins and find the average diagonal and off-diagonal terms are consistently at the noise level, approaching $1\%$ for both the diagonal and off-diagonal terms for bins with a S/N of 60 per $\mathrm \AA$. This suggests that the contribution of the wavelength effect to the error budget is $<1\%$ and the bin covariance is mostly due to the residual noise.

The sub-percent covariance from both the template and wavelength contributions results in a total systematic component per object in the range 1-1.4\% and a subpercent covariance. It is thus sufficient to improve our hierarchical inference of the Hubble constant to 2$\%$ with the addition of more SLACS lenses to the external population of non-time-delay lenses alone (given equivalent data to this sample). Improved kinematic datasets reaching further in the red, with reduced sensitity to templates \citep{Barth2002}, e.g. with the new KCRM red arm of KCWI, or with NIRSPec on JWST, are needed to further reduce these covariances and their effects on cosmological inferences.

%%%%%%%%%%%%%%%%%%%%

\subsection{Contamination from Background Source}

For 12 of 14 objects, the lensed background source arcs are faint enough that we are able to extract from spatial bins even with some contamination from the lensed arcs. We note that several of these bins have higher uncertainty in extracted kinematics than bins that are farther from the lensed arcs. We retain those bins and flag those with uncertainty in the velocity dispersion greater than 20 km $\mathrm{s^{-1}}$. In two cases, SDSSJ0330$-$0020 and SDSSJ1112+0826, the background source at higher redshift is bright enough to significantly affect the Voronoi binning and extraction of kinematics from bins that include contaminated spaxels. For these objects, we mask these pixels by eye and fit as above. We tested two methods for iteratively fitting the foreground and background galaxy spectra. We fit composite spectra taken from spaxels at the center of the object and from the region most contaminated by the background source arc. Initially, we used a technique similar to the method used by \cite{shajib23} to remove contaminating background quasar continuum spectra using KCWI data. We take the extracted background source spectrum from the data as a template for fitting the contamination in other spatial bins. The background source galaxy arcs in our data are not bright enough relative to the foreground deflector for a clean separation of the light; the deflector CaHK absorption features are clearly present in the template constructed from the extracted background source arc. To improve on this, we simultaneously fit both spectra extracted from the foreground and background components with the same stellar template library at the restframe of the foreground deflector. We redshift the templates to the appropriate redshift for the background galaxy relative to the restframe of the foreground deflector. Then we alternately fit the extracted central spectrum and background source spectrum to iteratively improve both fits. The result still does not cleanly separate the foreground and background source light, so we opt to mask the most affected spaxels and extract kinematics where there is no contamination. This method will be useful for future IFS datasets of strong lenses, e.g., with JWST, where the sensitivity and resolution will allow cleaner separation of the background source spectral features.

%%%%%%%%%%%%%%%%%%%

\section{Results}\label{sect:results}

We present the final kinematic maps and \textit{HST} photometry in Figure~\ref{fig:kinematics_maps}. 
The following subsections discuss kinematic classification and integrated kinematics. We present some select details in Table \ref{tab:results}.

% table

\begin{table}[h]
    \centering
    \begin{tabular}{c|c|c|c|c|c|c|c|c|c|c|c|c}
        Object & z$_{\mathrm{lens}}$ 
        & 
        \head{1.0cm}{$\mathrm{\sigma_{1/2}^{KCWI}}$ $\mathrm{km \ s ^{-1}}$} 
        &
        \head{1.0cm}{$\mathrm{\sigma_{eff}^{KCWI}}$ $\mathrm{km \ s ^{-1}}$} 
        & 
        \head{1.0cm}{$\mathrm{\sigma_{1.5}^{KCWI}}$ $\mathrm{km \ s ^{-1}}$} 
        & 
        \head{1.0cm}{$\mathrm{\sigma_{SDSS}^{this \ work}}$ $\mathrm{km \ s ^{-1}}$} 
        & 
        \head{1.0cm}{$\mathrm{\sigma_{XII}^{SLACS}}$ $\mathrm{km \ s ^{-1}}$} 
        & 
        \head{1.0cm}{$\mathrm{\sigma_{IX}^{SLACS}}$ $\mathrm{km \ s ^{-1}}$} 
        & \head{0.8cm}{$r_{\mathrm{eff}}$ $ \ '' \ $} 
        & $\epsilon_{\mathrm{obs}}$ 
        & $V/\sigma$ 
        & $\lambda_R$ 
        & Class \\
    \hline
SDSSJ002907.77-005550.5 & 0.227 & 207±7 & 200±9 & 206±2 & 197±18 & 216±15 & 229±18 & 2.30 & 0.16 & 0.070 & 0.056 & slow \\
SDSSJ003753.21-094220.1 & 0.195 & 279±7 & 265±8 & 274±1 & 309±14 & 265±8 & 279±10 & 2.30 & 0.27 & 0.162 & 0.151 & slow \\
SDSSJ033012.14-002051.9 & 0.351 & 258±12 & 247±13 & 250±5 & 288±26 & 273±23 & 212±21 & 1.26 & 0.23 & 0.106 & 0.094 & slow \\
SDSSJ111250.60+082610.4 & 0.273 & 274±7 & 264±9 & 270±2 & 271±22 & 260±15 & 320±20 & 1.55 & 0.23 & 0.236 & 0.208 & fast \\
SDSSJ120444.07+035806.4 & 0.164 & 262±5 & 249±7 & 246±2 & 254±18 & 251±12 & 267±17 & 1.63 & 0.03 & 0.071 & 0.059 & slow \\
SDSSJ125028.26+052349.1 & 0.232 & 243±6 & 231±7 & 235±1 & 220±12 & 242±10 & 252±14 & 1.86 & 0.03 & 0.129 & 0.113 & fast \\
SDSSJ130613.65+060022.1 & 0.173 & 230±7 & 221±9 & 223±2 & 242±18 & 248±14 & 237±17 & 2.08 & 0.09 & 0.110 & 0.101 & slow \\
SDSSJ140228.21+632133.5 & 0.205 & 285±7 & 276±8 & 279±1 & 249±14 & 274±11 & 267±17 & 2.65 & 0.23 & 0.084 & 0.073 & slow \\
SDSSJ153150.07-010545.7 & 0.160 & 273±9 & 267±10 & 272±2 & 262±13 & 261±10 & 279±12 & 2.73 & 0.32 & 0.052 & 0.045 & slow \\
SDSSJ153812.92+581709.8 & 0.143 & 236±5 & 225±7 & 216±2 & * & 177±9 & 189±12 & 1.45 & 0.18 & 0.055 & 0.048 & slow \\
SDSSJ162132.99+393144.6 & 0.245 & 255±10 & 242±12 & 256±2 & 238±21 & 234±15 & 236±20 & 2.30 & 0.27 & 0.224 & 0.200 & fast \\
SDSSJ162746.45-005357.6 & 0.208 & 262±10 & 261±12 & 261±2 & 254±13 & 274±11 & 290±14 & 2.02 & 0.15 & 0.083 & 0.070 & slow \\
SDSSJ163028.16+452036.3 & 0.248 & 280±7 & 267±9 & 271±2 & 269±16 & 283±13 & 276±16 & 2.01 & 0.16 & 0.095 & 0.088 & slow \\
SDSSJ230321.72+142217.9 & 0.155 & 259±9 & \textdagger{258±9} & 257±2 & 267±21 & 251±13 & 255±16 & 3.46 & 0.36 & 0.062 & 0.058 & slow \\
    \end{tabular}
    \caption{Kinematic details for the 14 SLACS lenses. 
    KCWI velocity dispersions have all been extracted with the Indo-US library. $\mathrm{\sigma_{1/2}^{KCWI}}$ and $\mathrm{\sigma_{eff}^{KCWI}}$ are the KCWI velocity dispersion within 1/2 and 1 $\mathrm{R}_{\mathrm{eff}}$, respectively, measured by luminosity-weighted integration over the 2D velocity dispersion map. $\mathrm{\sigma_{1.5}^{KCWI}}$ is the KCWI velocity dispersion extracted from the datacube integrated over an aperture of 1.5$''$ with mock SDSS seeing. $\sigma^{\rm this \ work}_\mathrm{{SDSS}}$ is the velocity dispersion measured from SDSS spectra using XSL stellar templates in this work. $\mathrm{\sigma_{XII}^{SLACS}}$ is measured from SDSS spectra by \cite{shu15}. $\mathrm{\sigma_{IX}^{SLACS}}$ is measured from SDSS spectra in \cite{slacs9}. Errors quoted here are statistical errors. The effective radius $r_{eff}$ is reported from SLACS-X. Observed ellipticity $\epsilon_{\mathrm{obs}}$ is measured from MGE models at the isophote enclosing half the total luminosity. $V/\sigma$, $\lambda_R$, and kinematic classification are defined in Section \ref{sect:classification}. \\
    *{\footnotesize SDSSJ1538+5817 has two SDSS observations in the SDSS Science Archive Server (SAS). We measured both with {\sc pPXF} and report both. One yields $180\pm12$ km $\mathrm{s}^{-1}$, which closely agrees with SLACS-IX and -XII values. The other yields $253\pm15$ km $\mathrm{s}^{-1}$. The value we measure over the same aperture with KCWI data and mock seeing $216\pm2$ is at the midpoint between these two values}. \\
    \textdagger {\footnotesize SDSSJ2303+1422's large effective radius is beyond the size of the kinematic map. We report the value integrated over the whole map.}}
    
    \label{tab:results}
\end{table}

% kinematic maps figure
% this plots the kinematic maps in three columns

\def\plotht{0.27\textwidth}

% maps, etc
\begin{figure}[ht]
    \centering
    % SDSSJ0029-0055
    \includegraphics[height=\plotht]{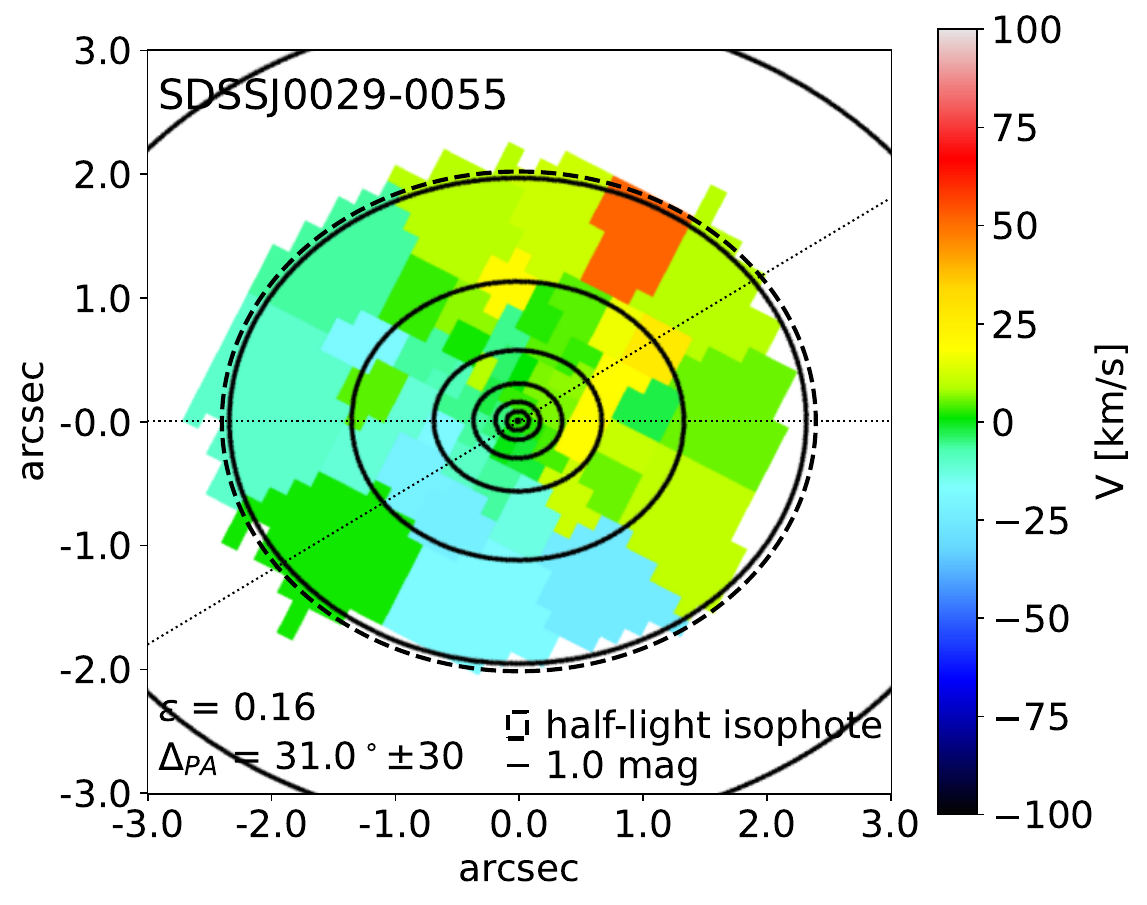}
    \includegraphics[height=\plotht]{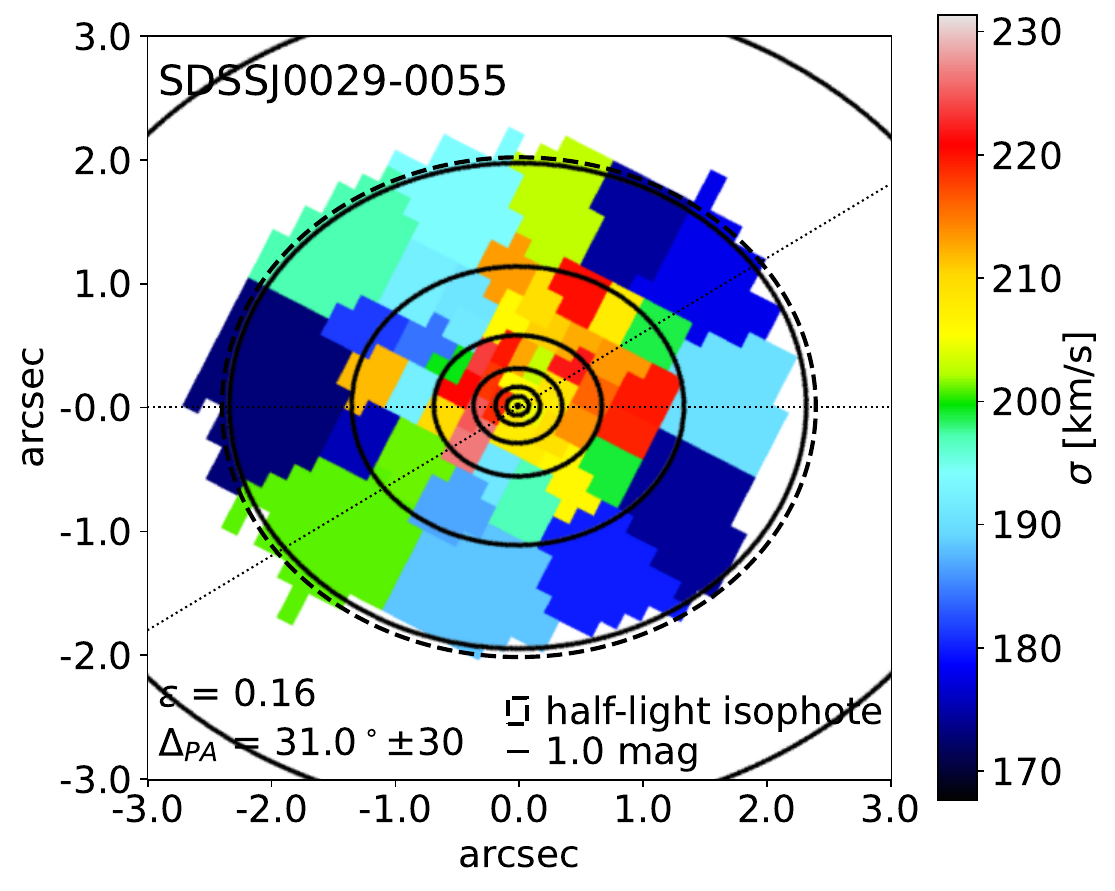}
    \includegraphics[height=\plotht]{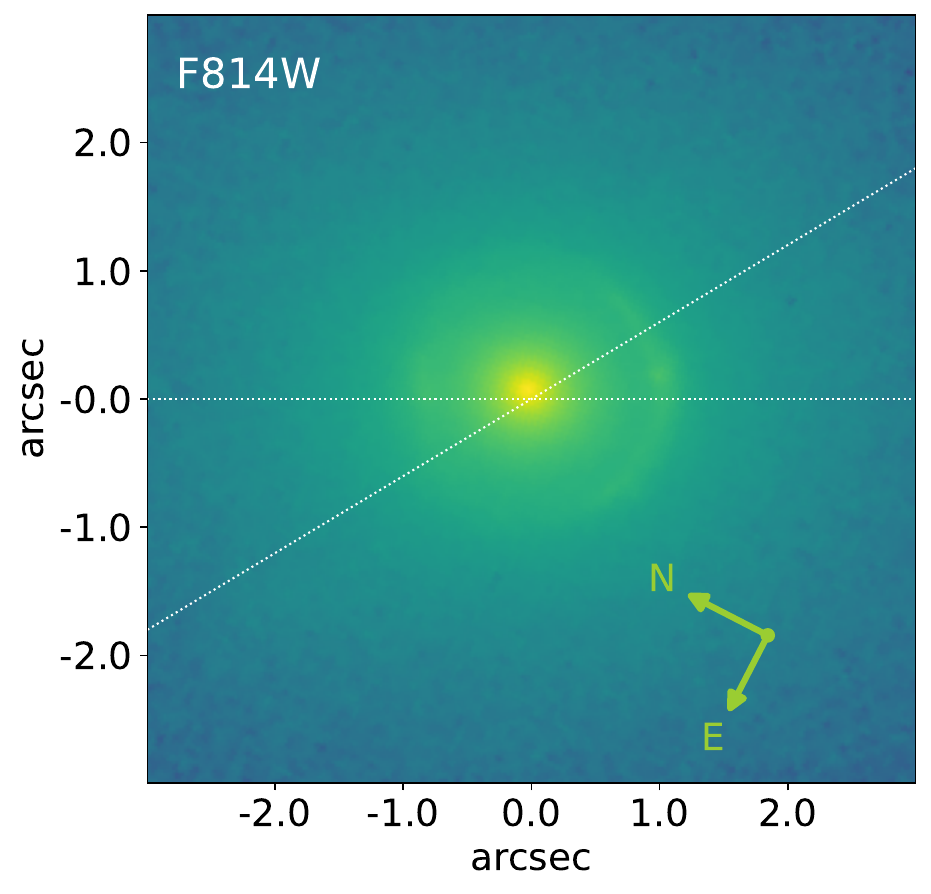}
\end{figure}
\begin{figure}[h]
    % SDSSJ0037-0942
    \centering
    \includegraphics[height=\plotht]{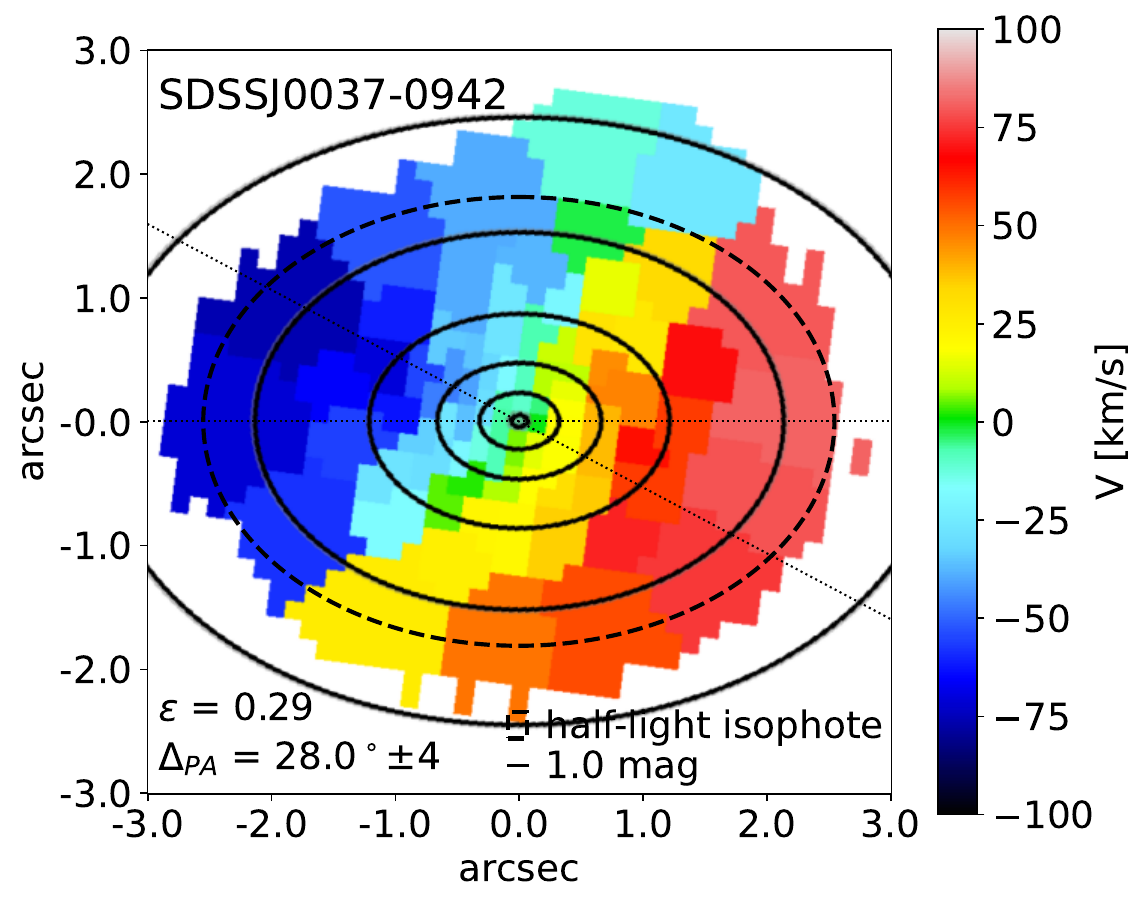}
    \includegraphics[height=\plotht]{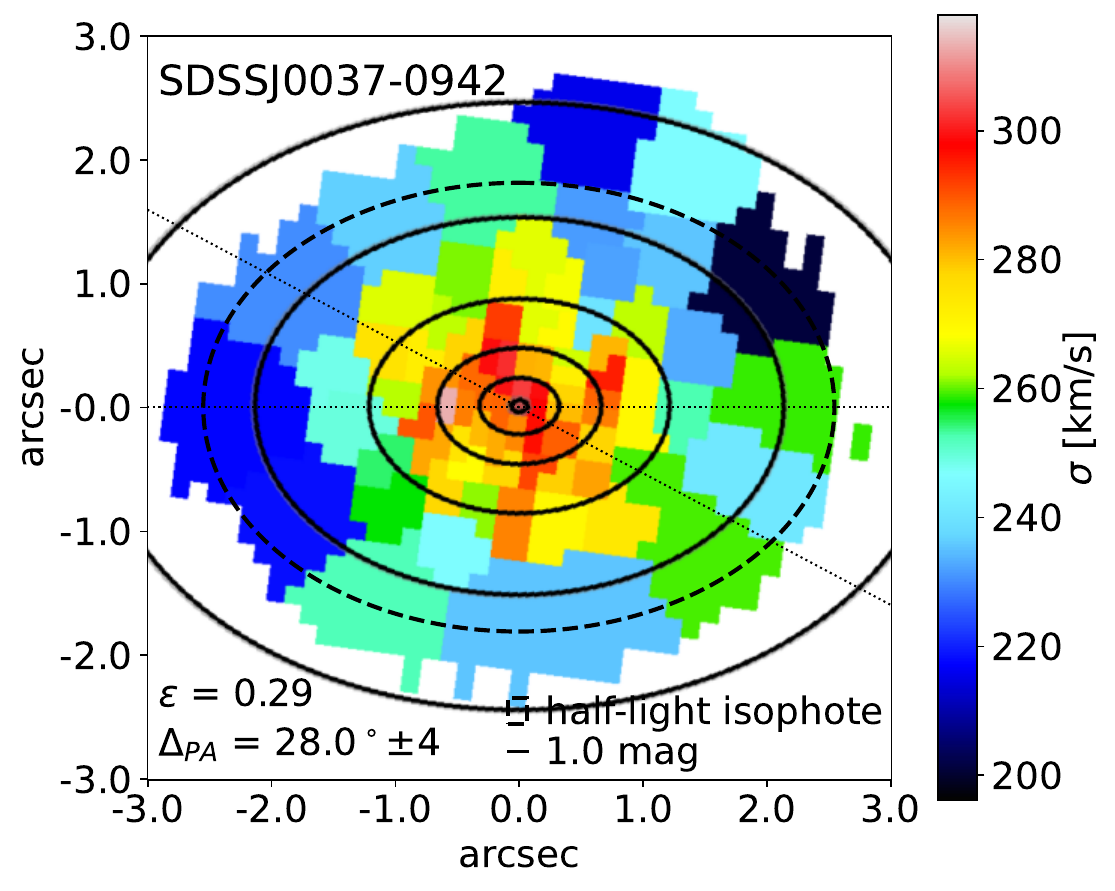}
    \includegraphics[height=\plotht]{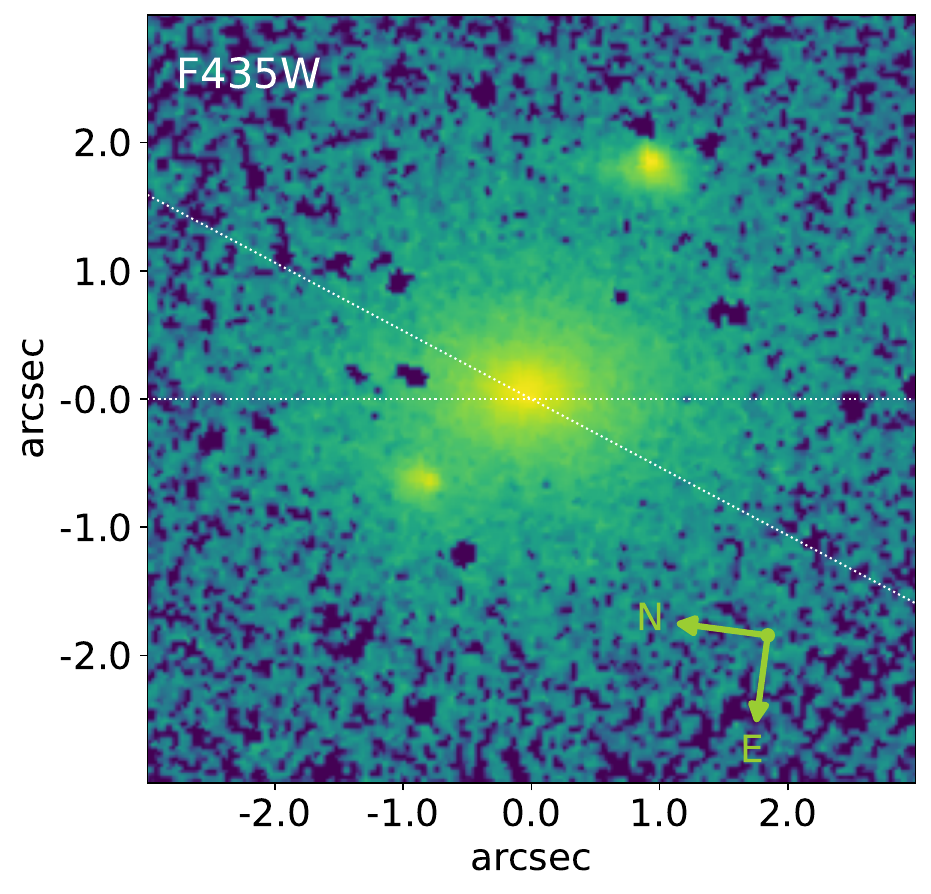}
\end{figure}
\begin{figure}[h]
    % SDSSJ0330-0020
    \centering
    \includegraphics[height=\plotht]{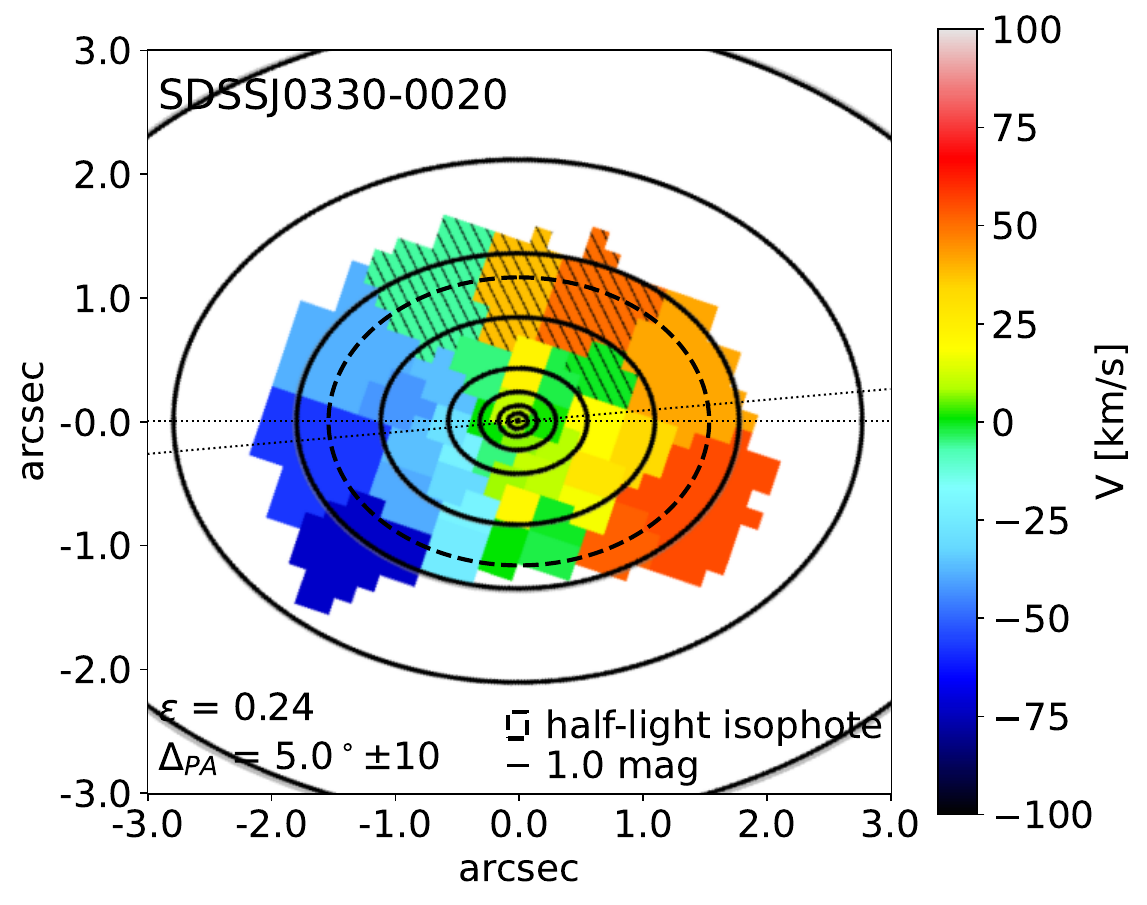}
    \includegraphics[height=\plotht]{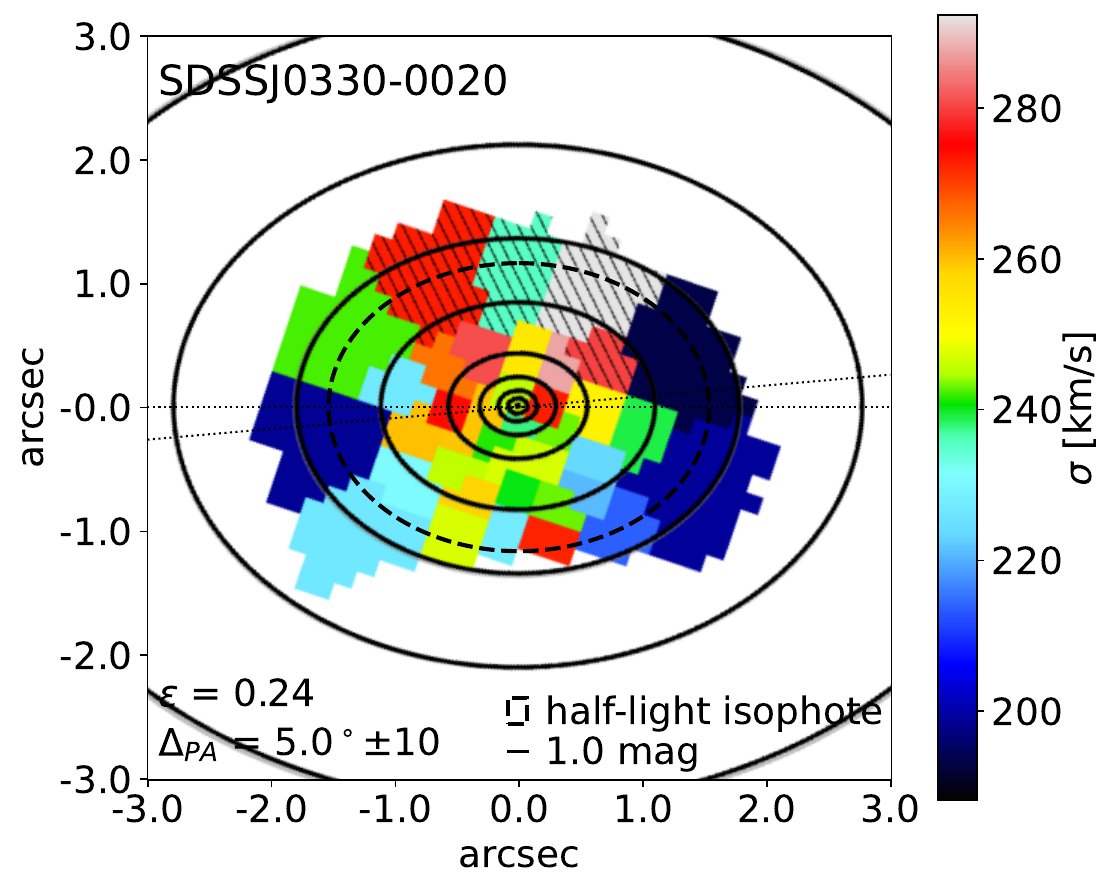}
    \includegraphics[height=\plotht]{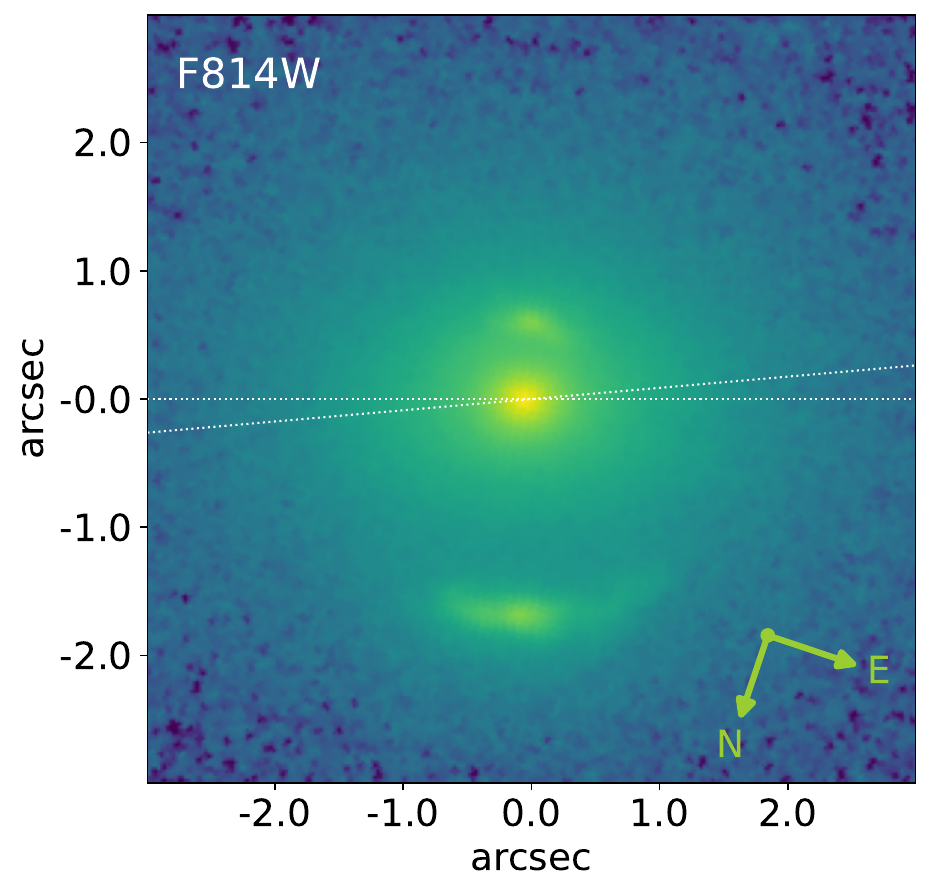}
\end{figure}
\begin{figure}[h]
    % SDSSJ1112+0826

    \centering
    \includegraphics[height=\plotht]{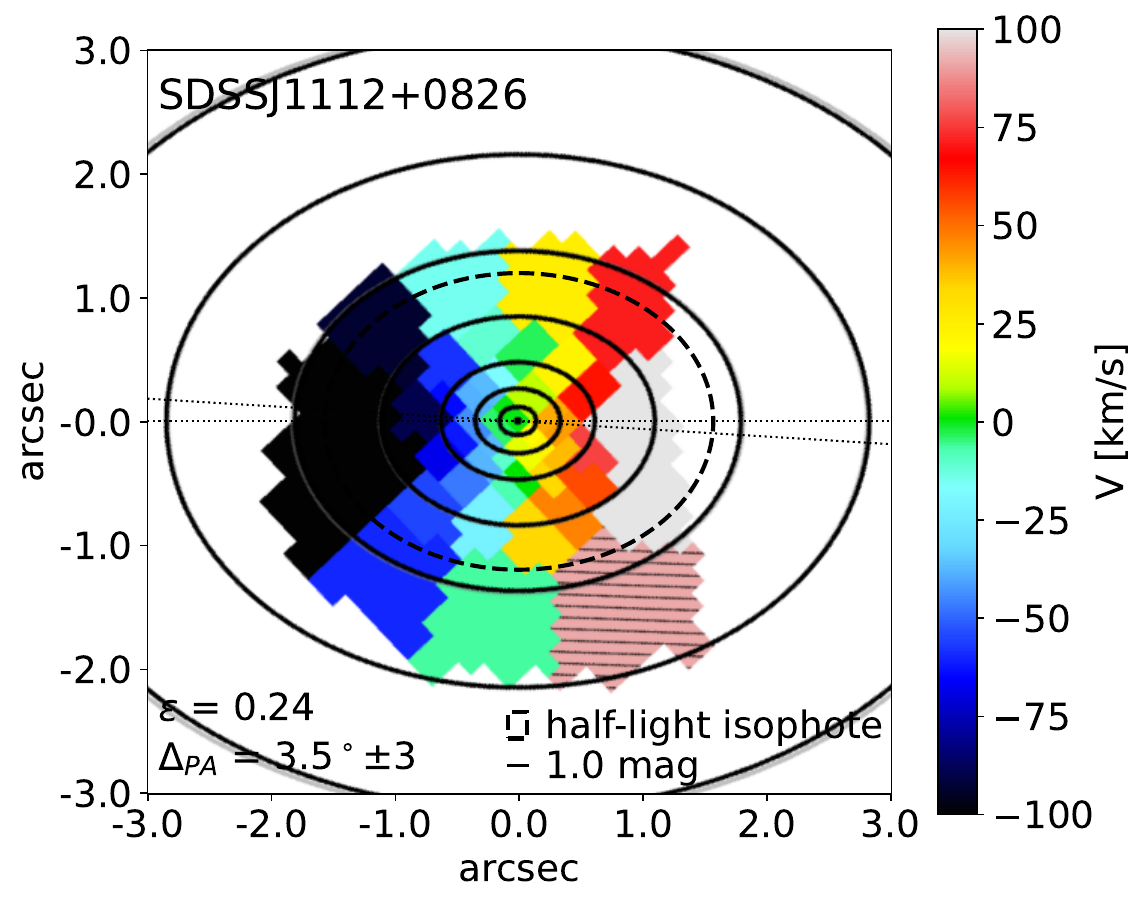}
    \includegraphics[height=\plotht]{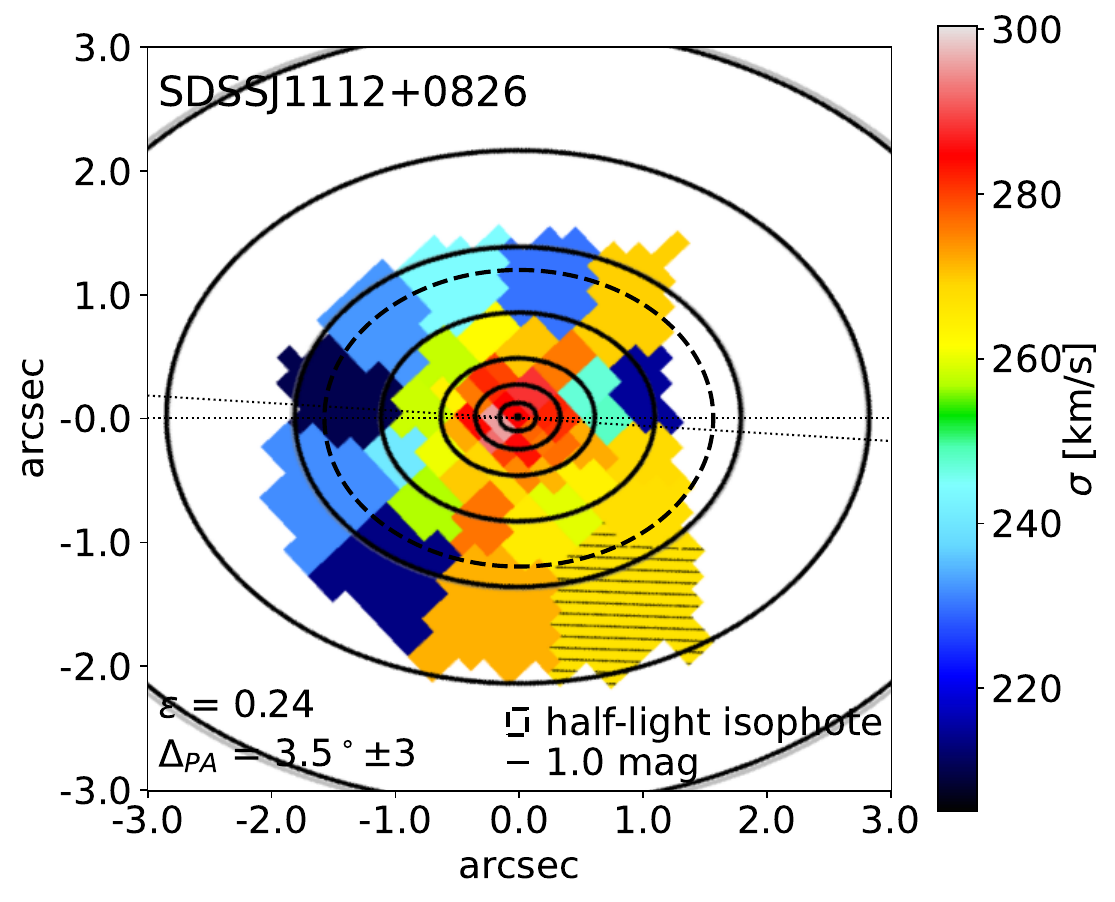}
    \includegraphics[height=\plotht]{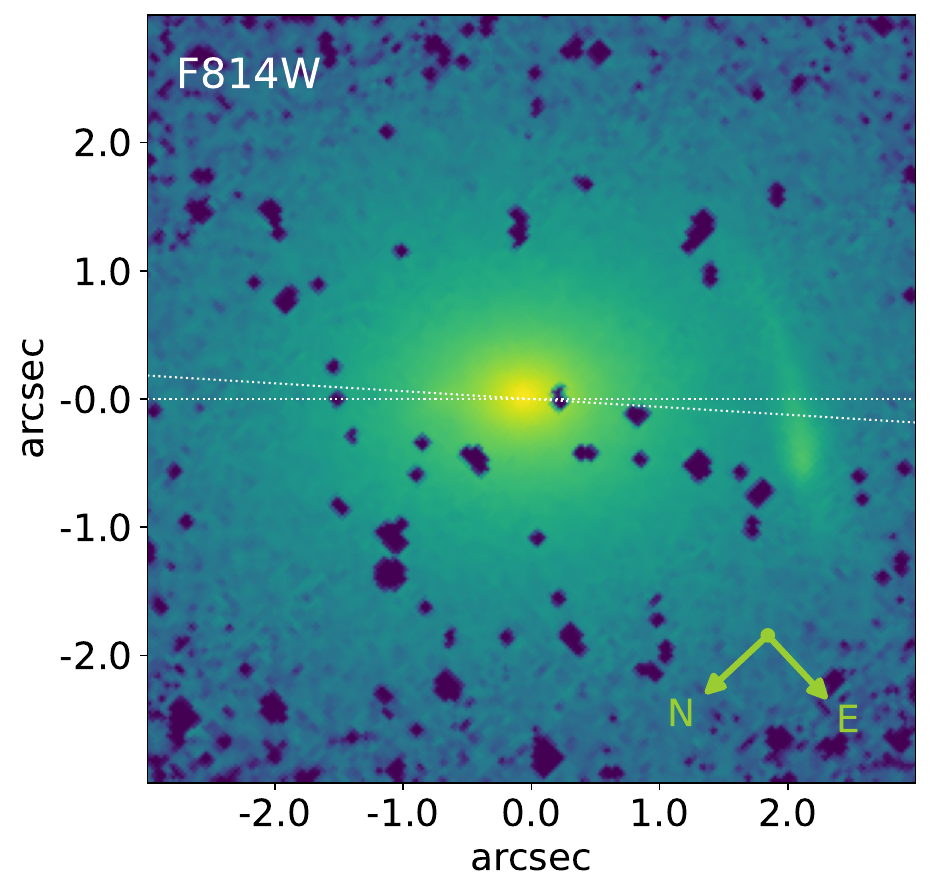}
\end{figure}
\begin{figure}[h]
    % SDSSJ1204+0358
    \centering
    \includegraphics[height=\plotht]{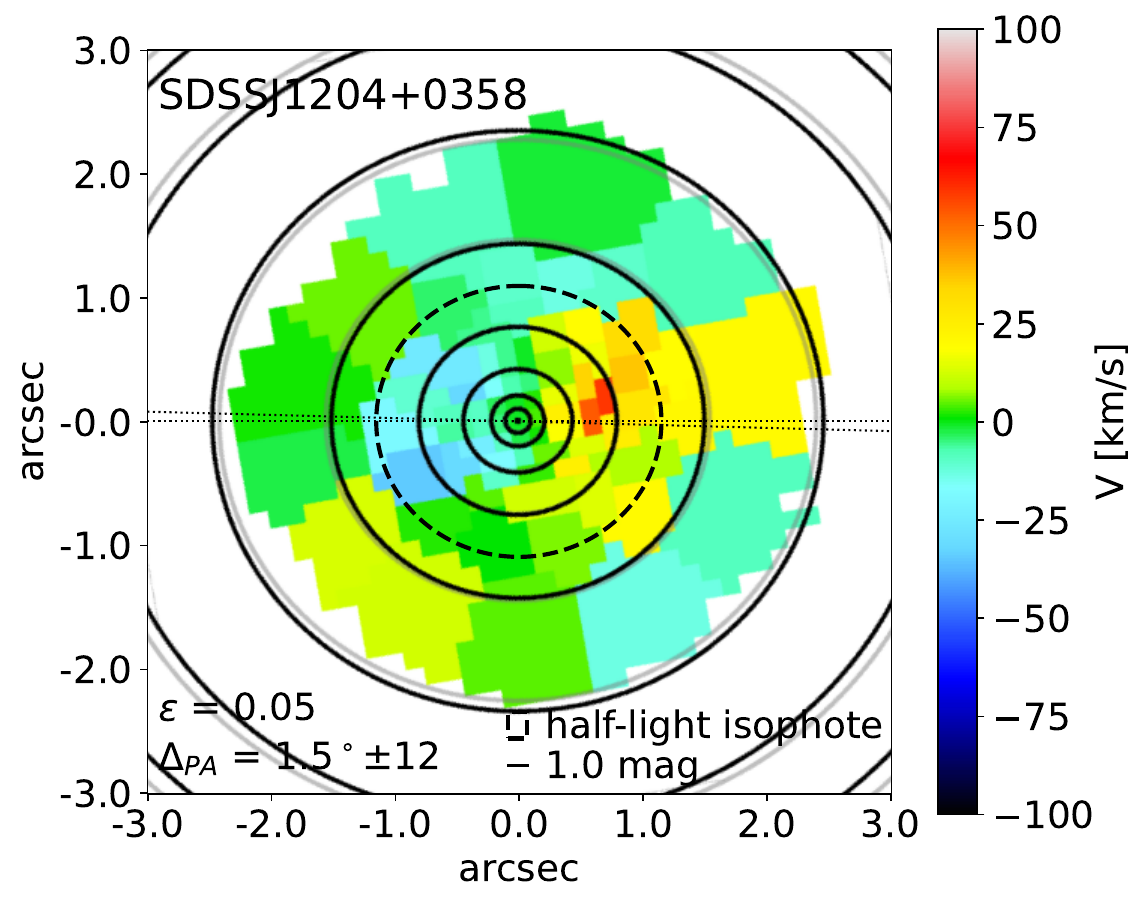}
    \includegraphics[height=\plotht]{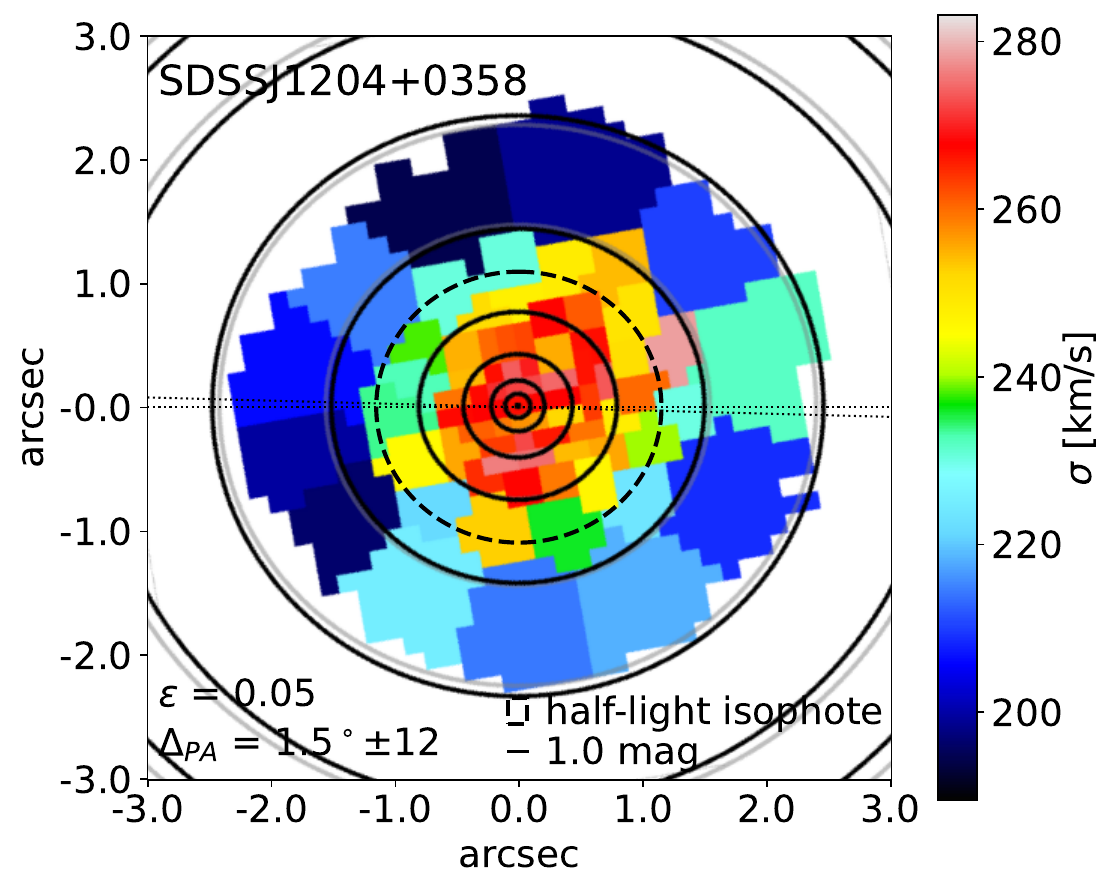}
    \includegraphics[height=\plotht]{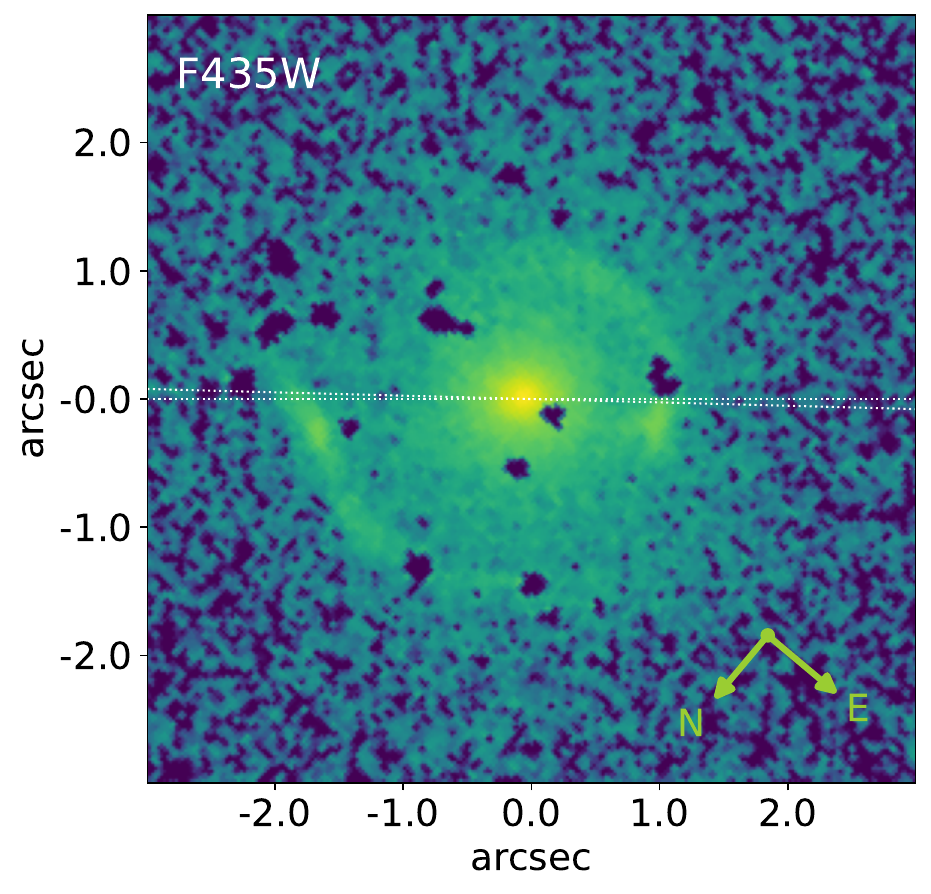}
\end{figure}
\begin{figure}[h]
    % SDSSJ1250+0523
    \centering
    \includegraphics[height=\plotht]{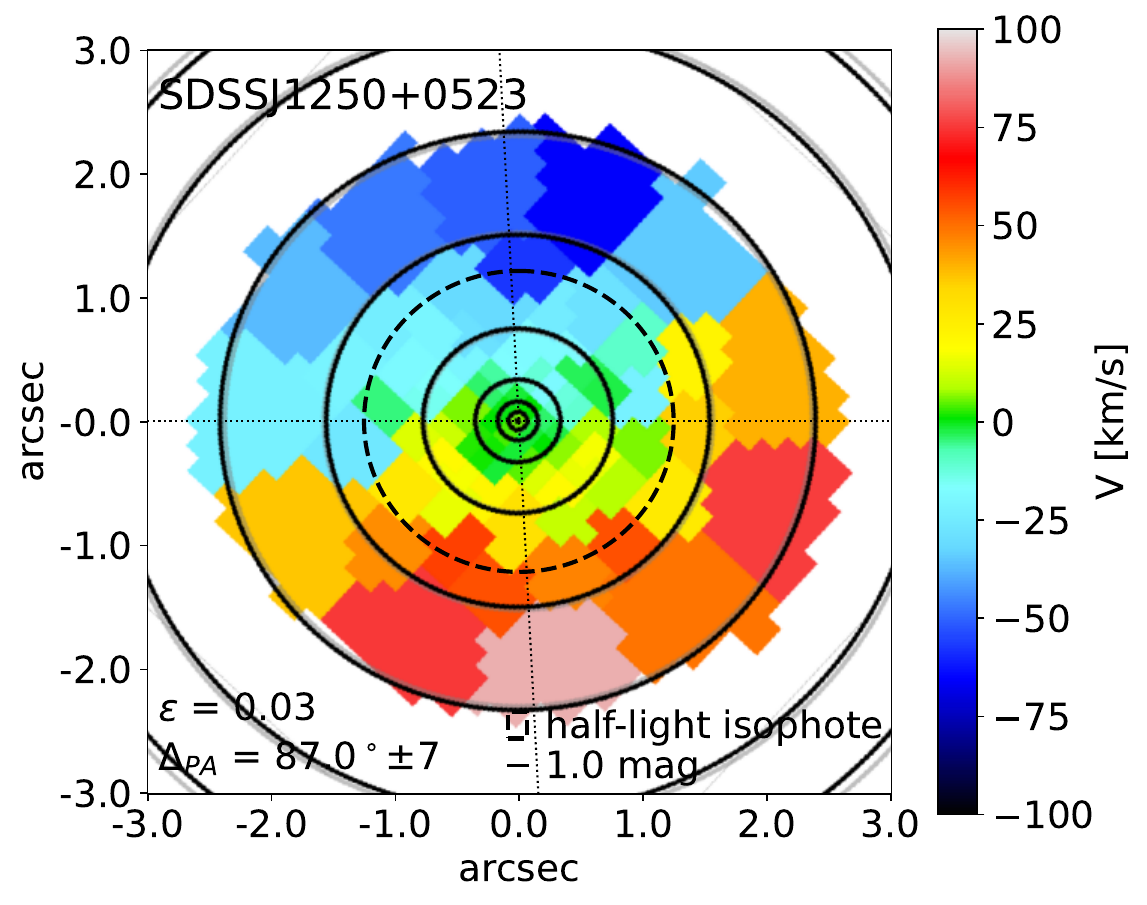}
    \includegraphics[height=\plotht]{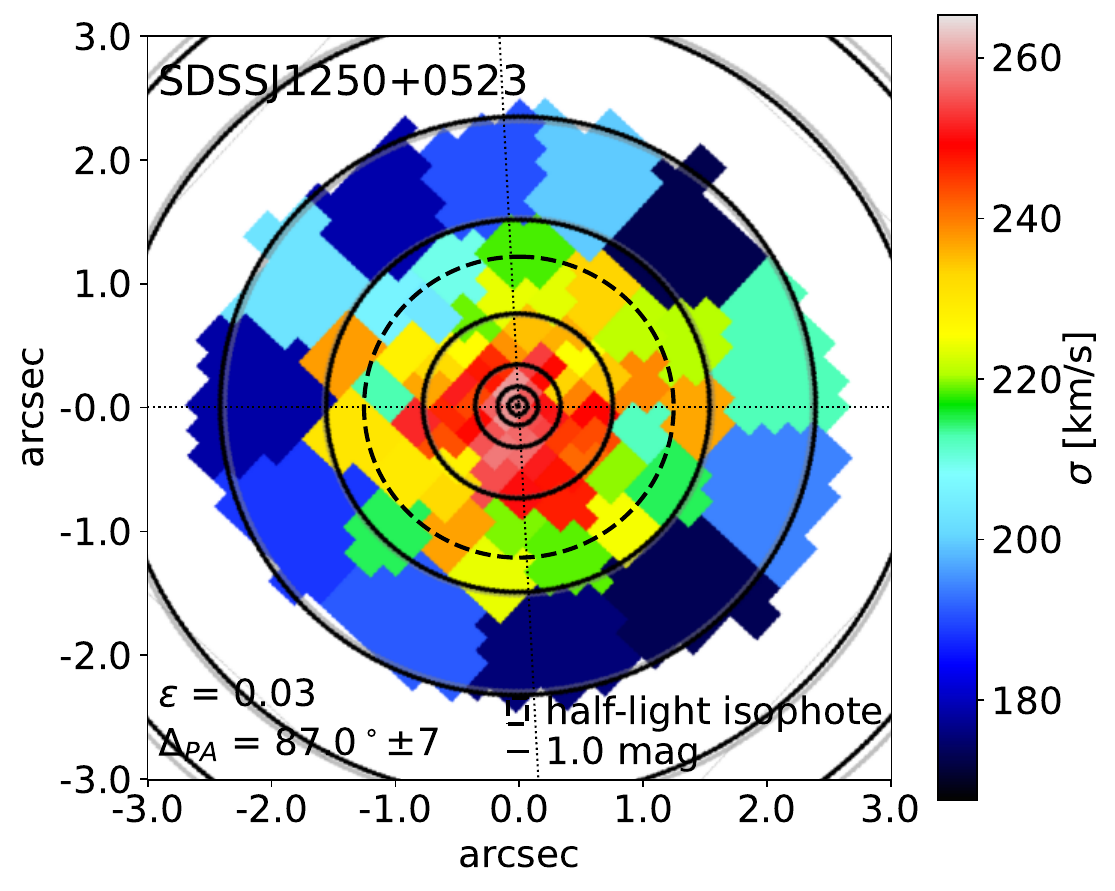}
    \includegraphics[height=\plotht]{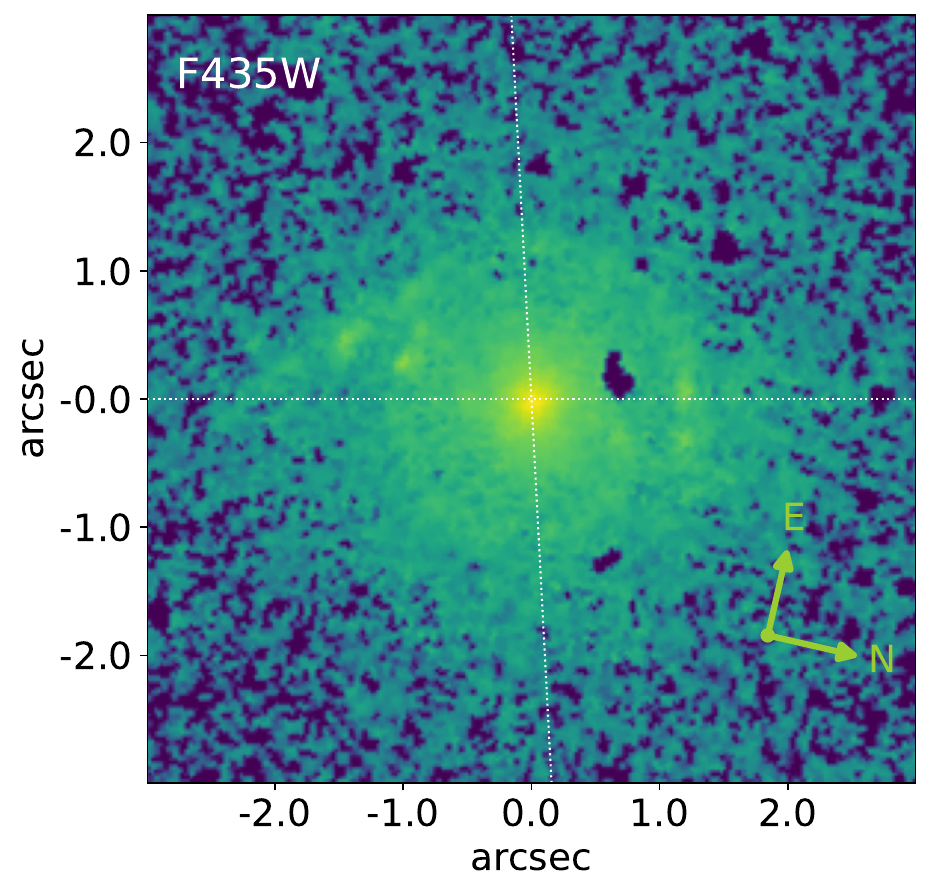}
\end{figure}
\begin{figure}[h]
    % SDSSJ1306+0600
    \centering
    \includegraphics[height=\plotht]{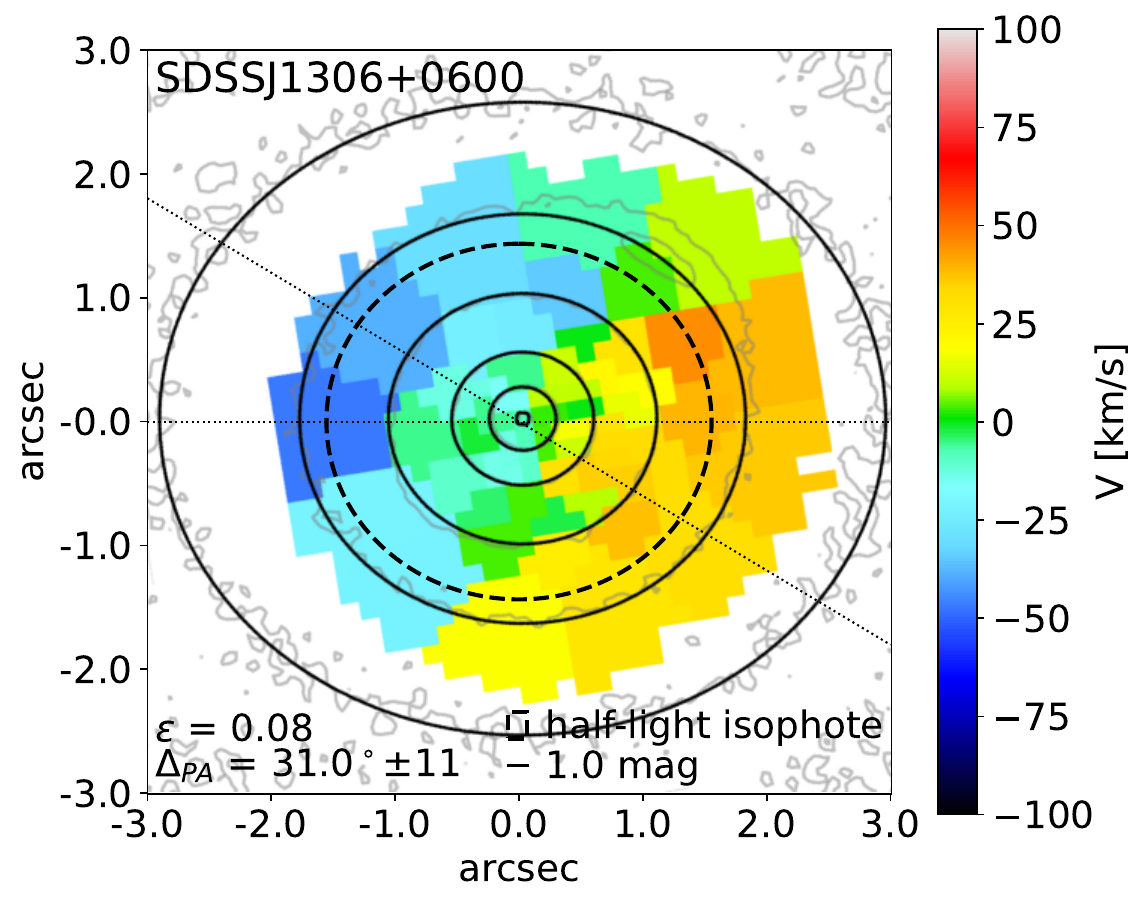}
    \includegraphics[height=\plotht]{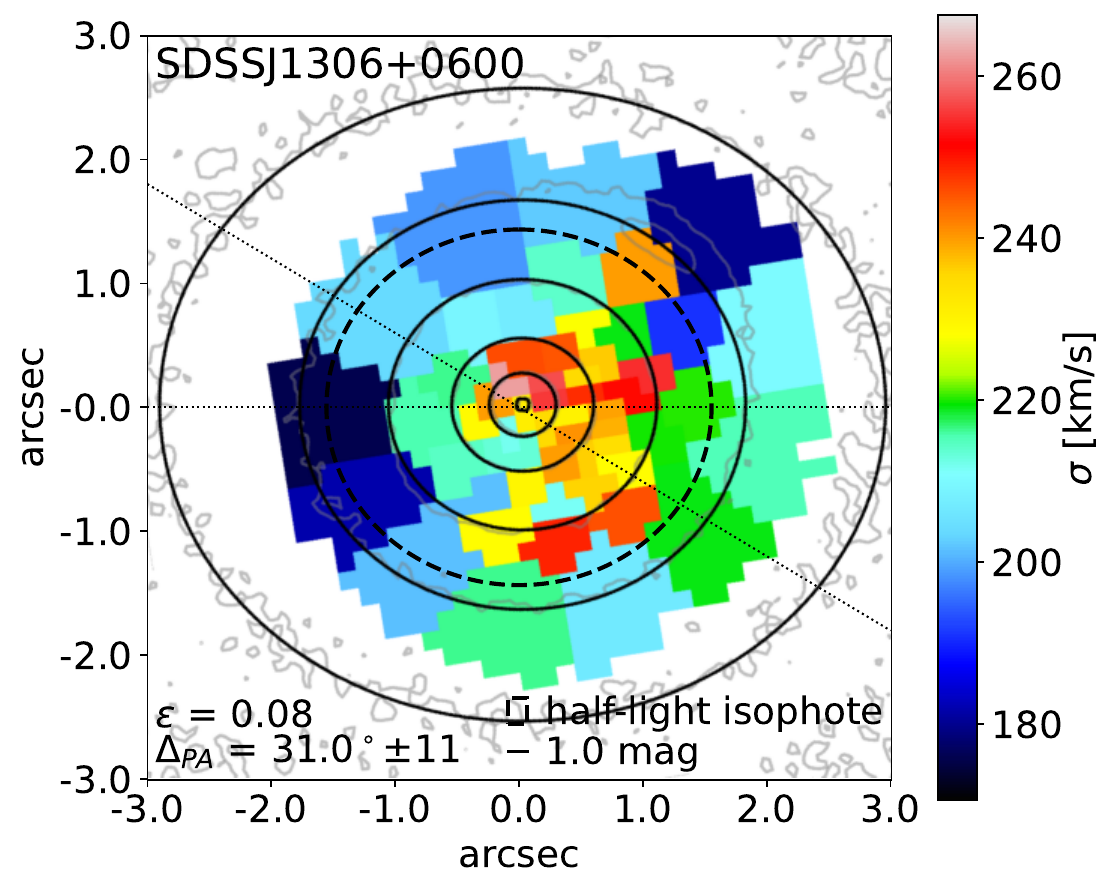}
    \includegraphics[height=\plotht]{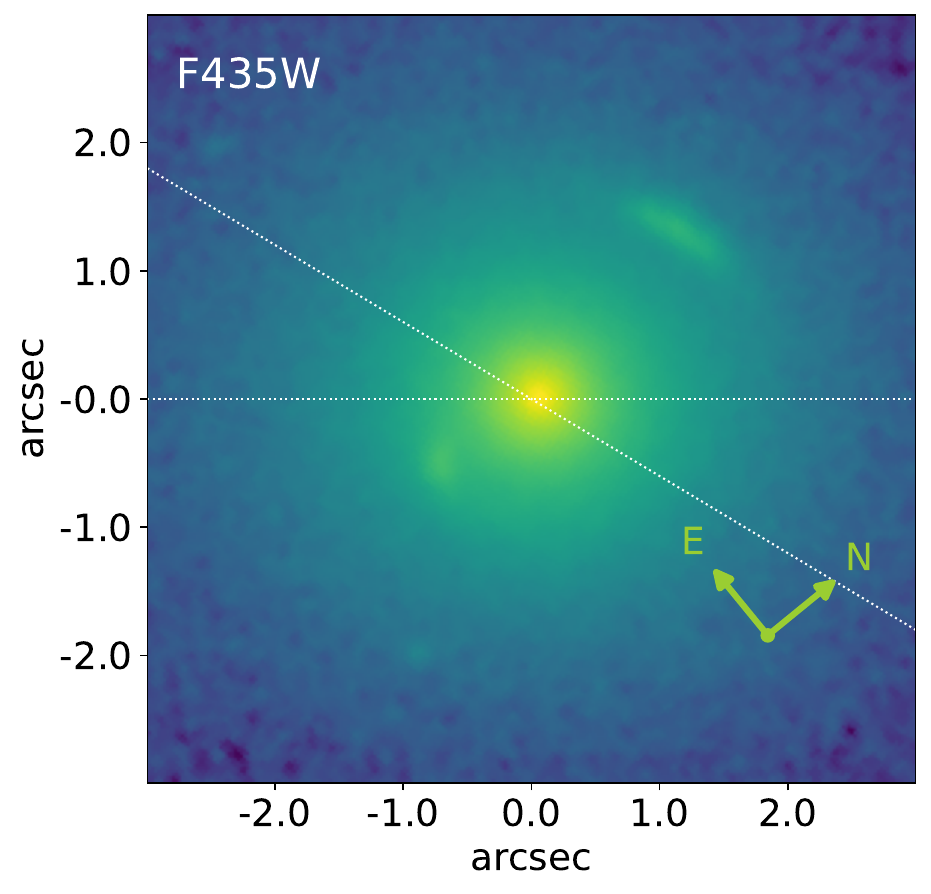}
\end{figure}
\begin{figure}[h]
    % SDSSJ1402+6321
    \centering
    \includegraphics[height=\plotht]{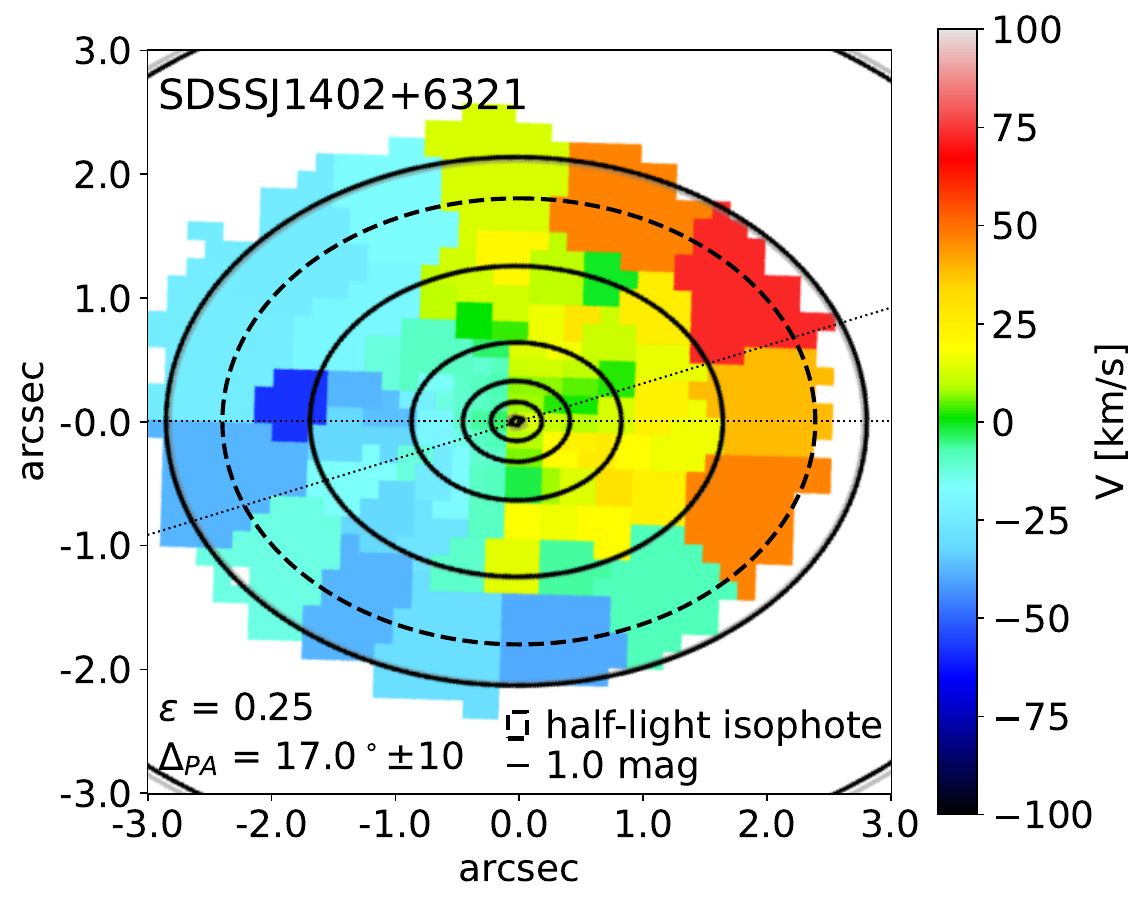}
    \includegraphics[height=\plotht]{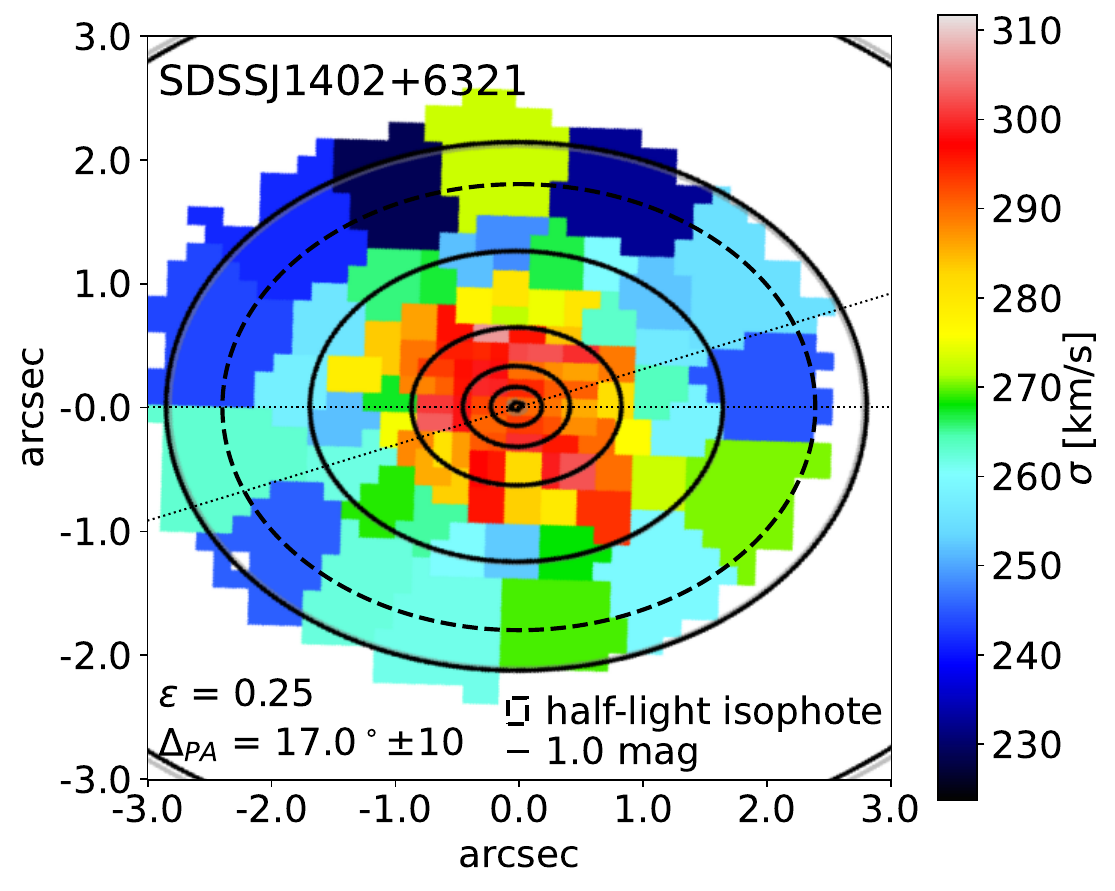}
    \includegraphics[height=\plotht]{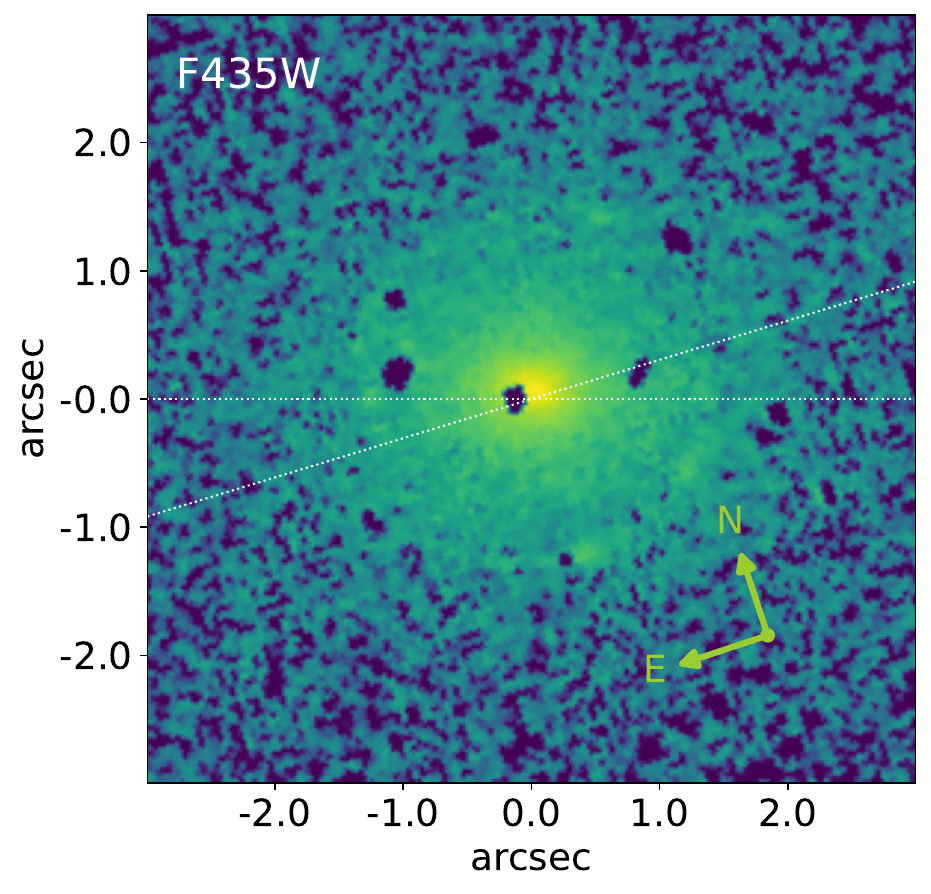}
\end{figure}
\begin{figure}[h]
    % SDSSJ1531-0105
    \centering
    \includegraphics[height=\plotht]{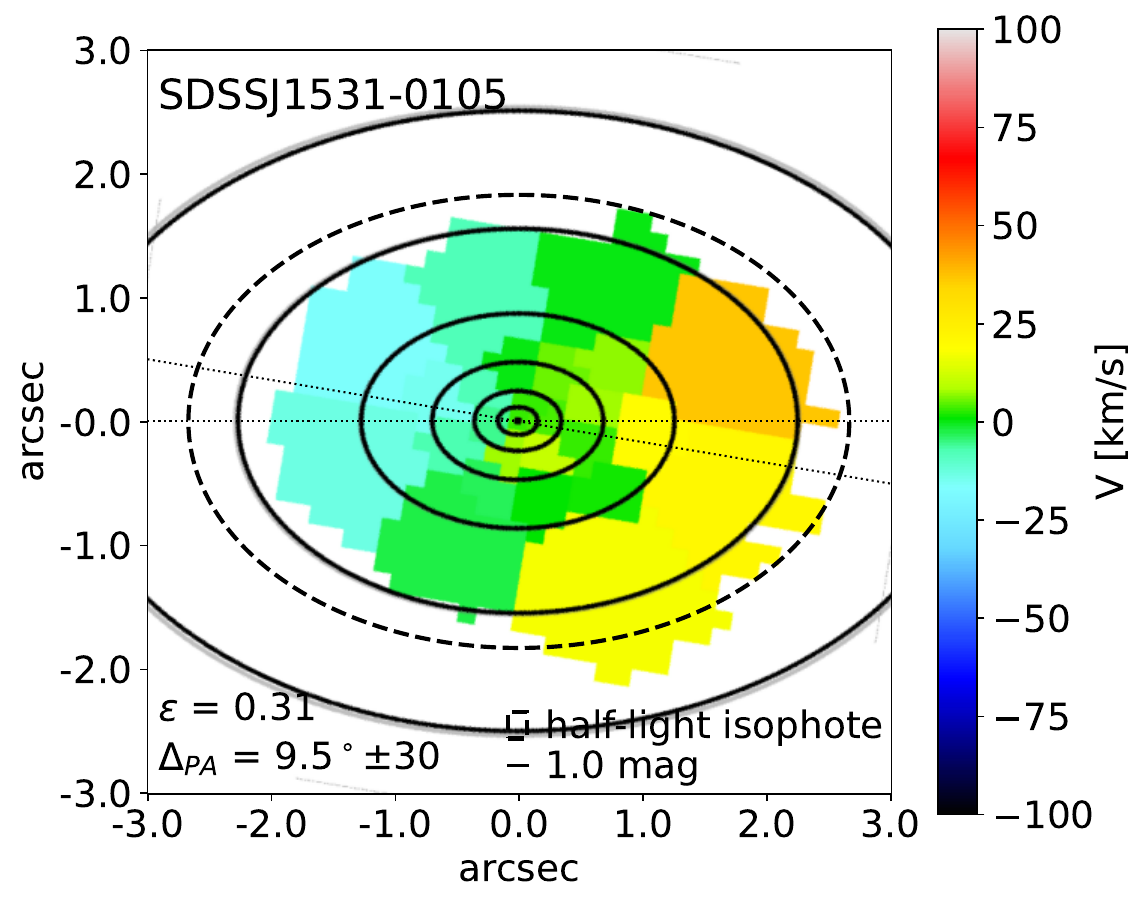}
    \includegraphics[height=\plotht]{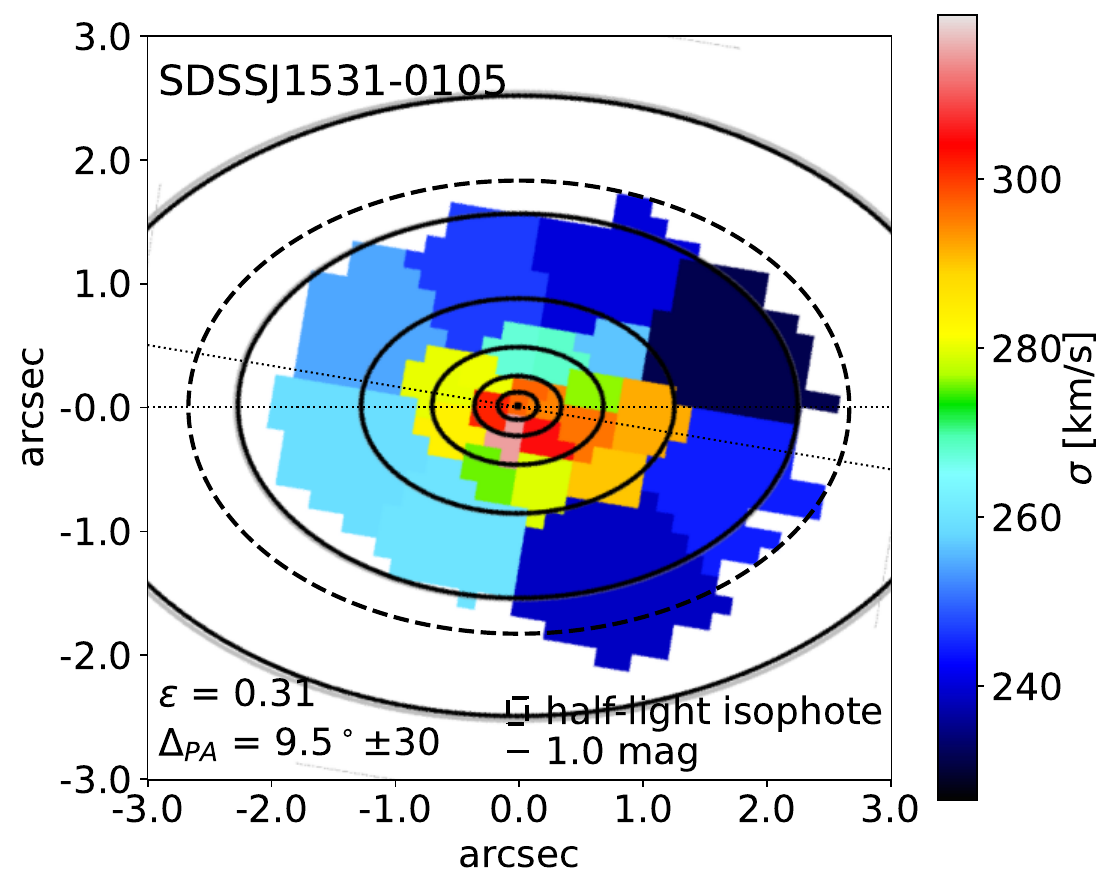}
    \includegraphics[height=\plotht]{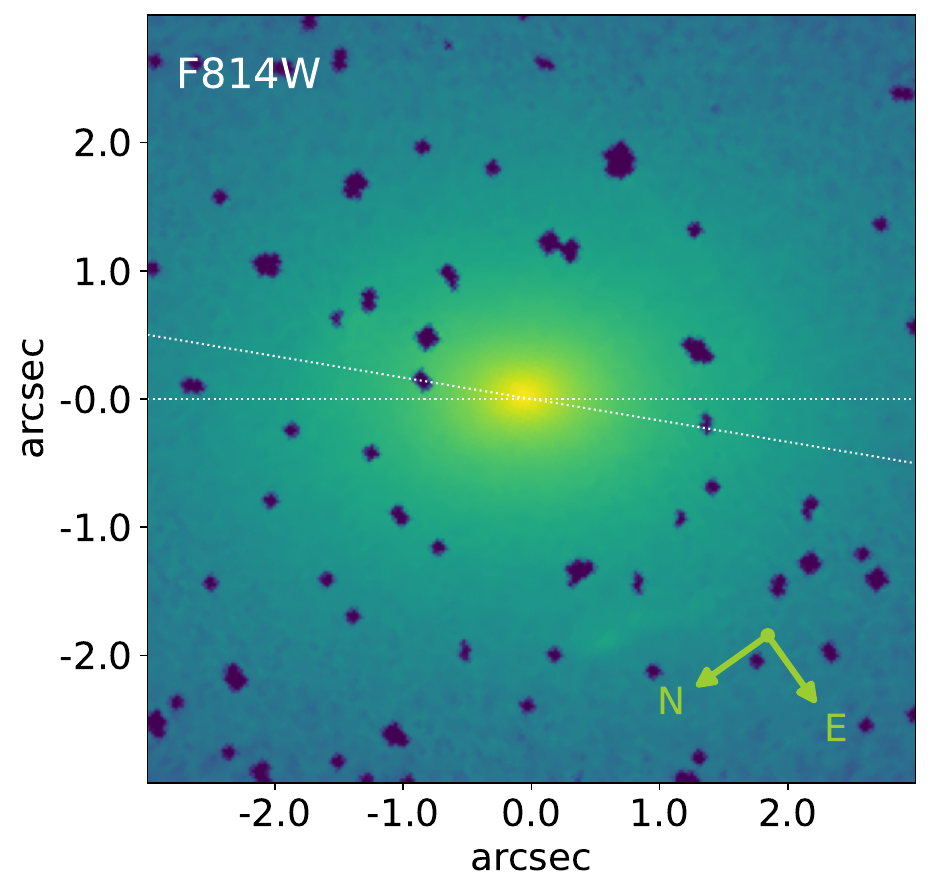}
\end{figure}
\begin{figure}[h]
    % SDSSJ1538+5817
    \centering
    \includegraphics[height=\plotht]{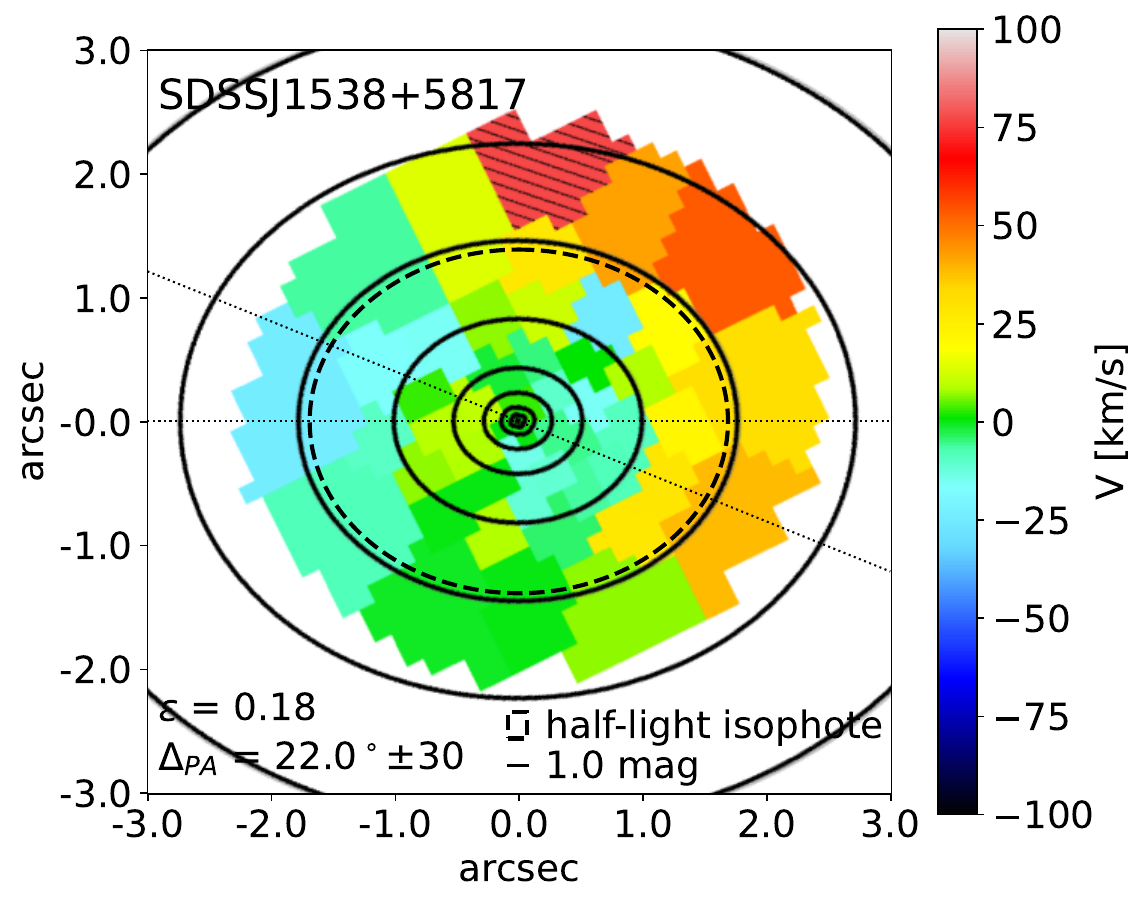}
    \includegraphics[height=\plotht]{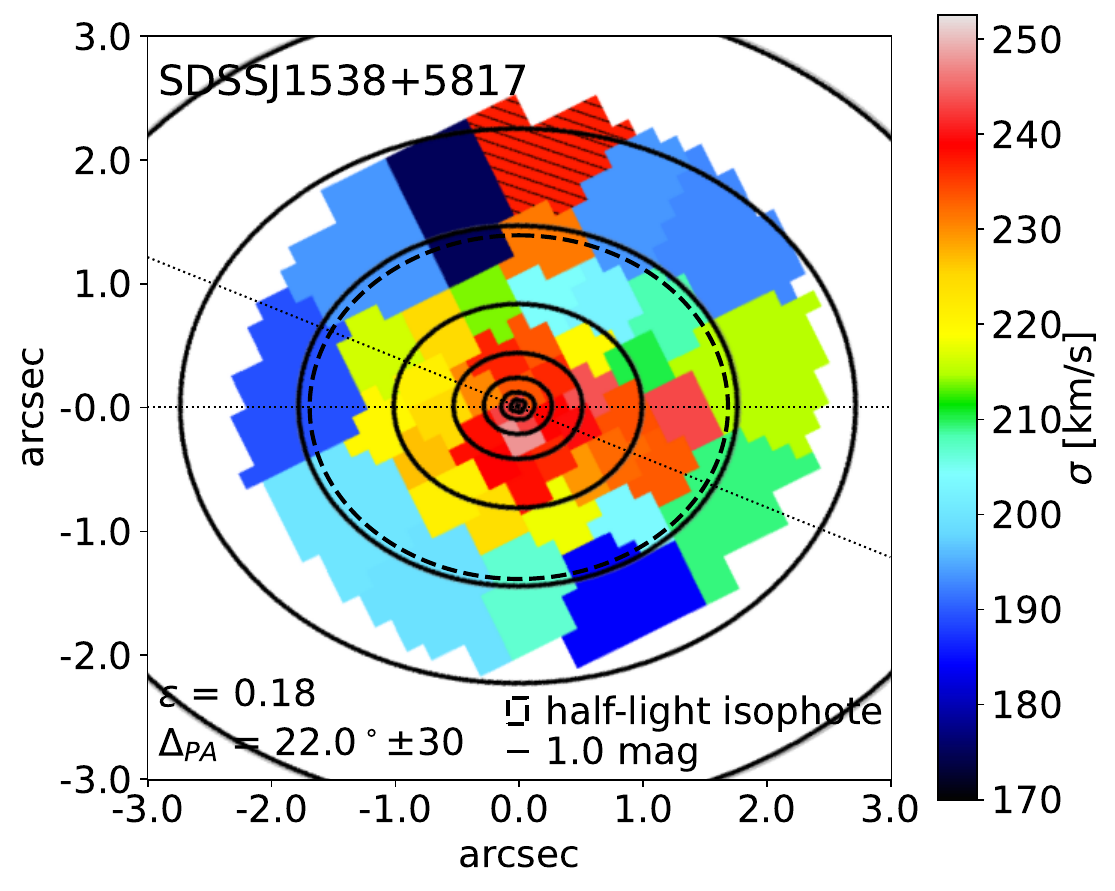}
    \includegraphics[height=\plotht]{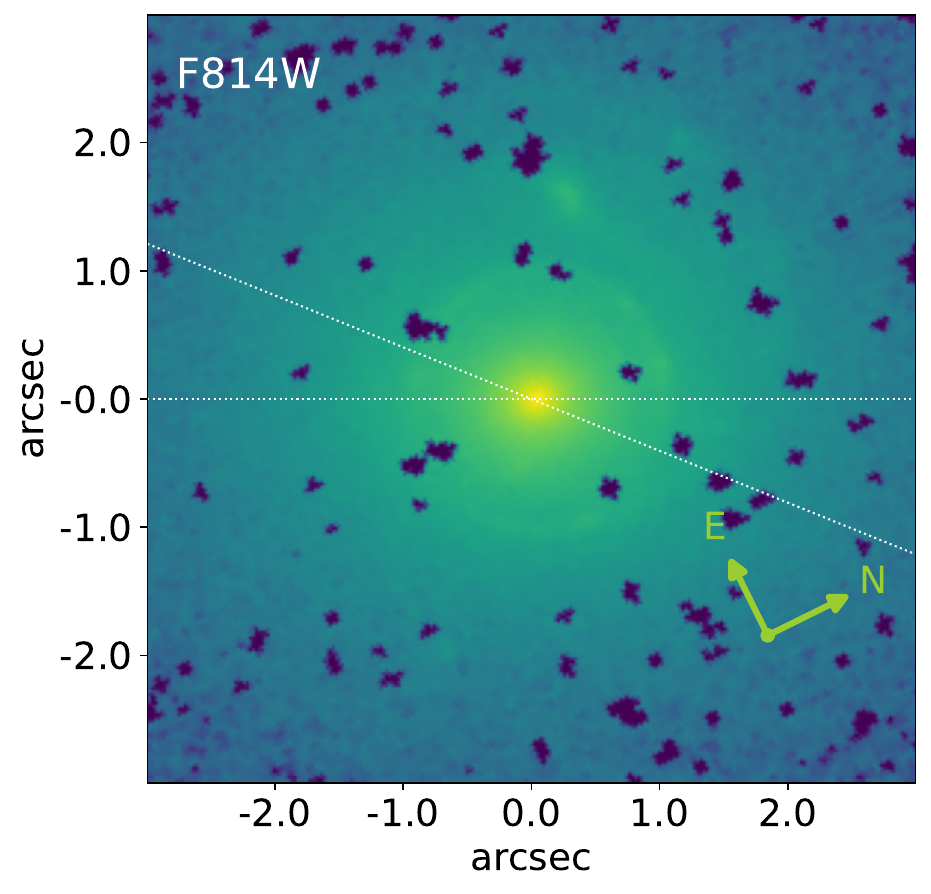}
\end{figure}
\begin{figure}[h]
    % SDSSJ1621+3931
    \centering
    \includegraphics[height=\plotht]{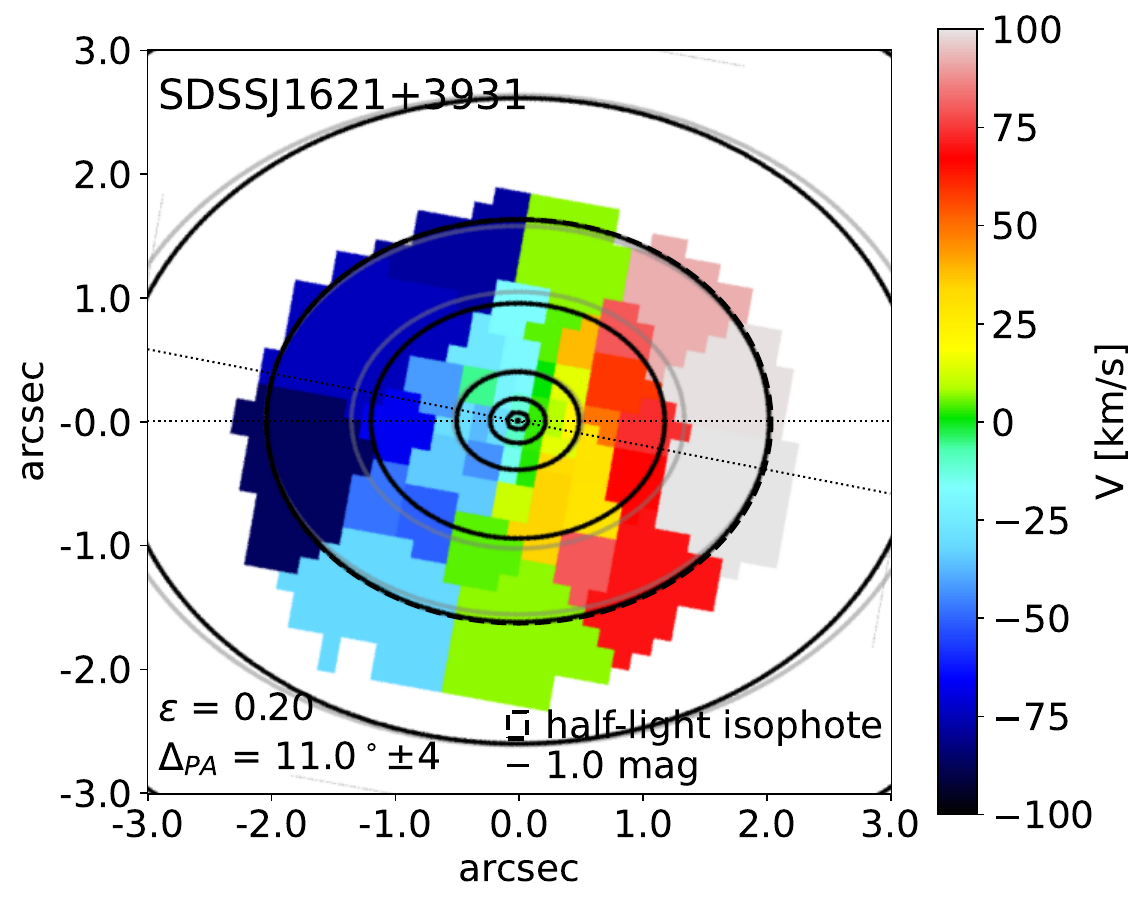}
    \includegraphics[height=\plotht]{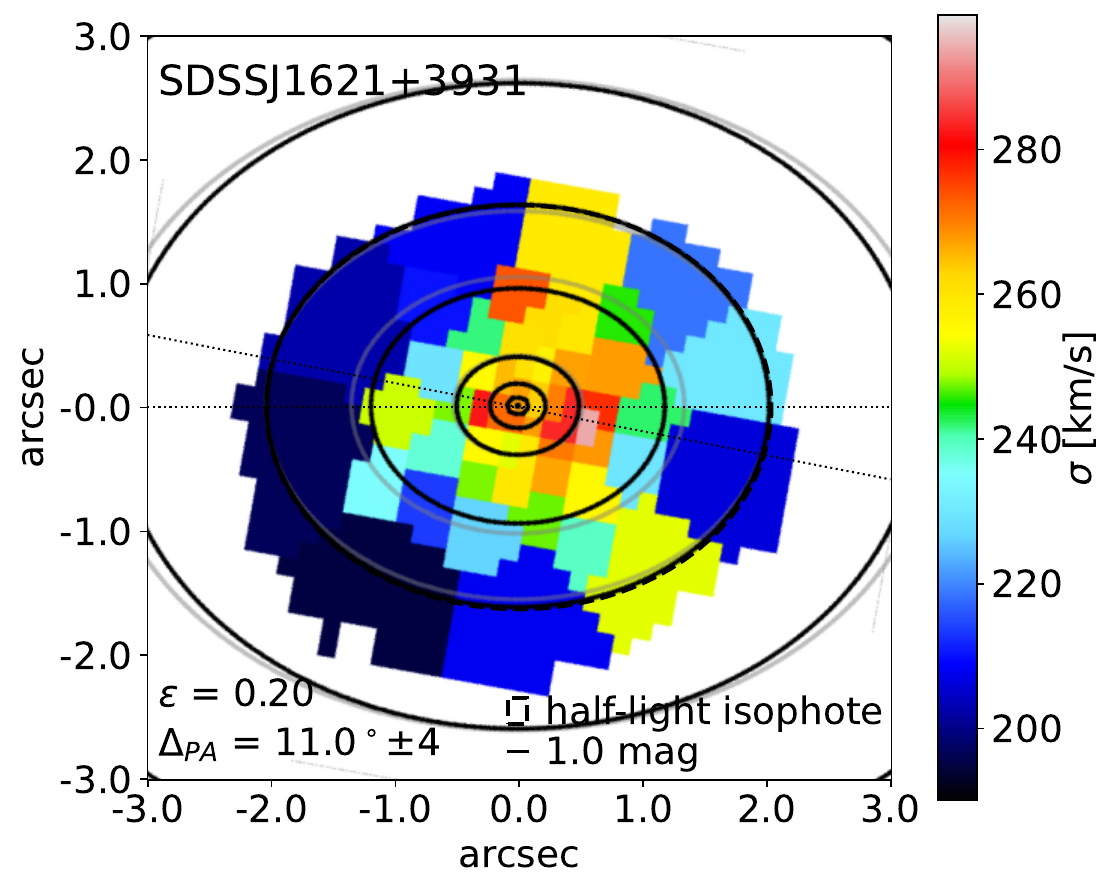}
    \includegraphics[height=\plotht]{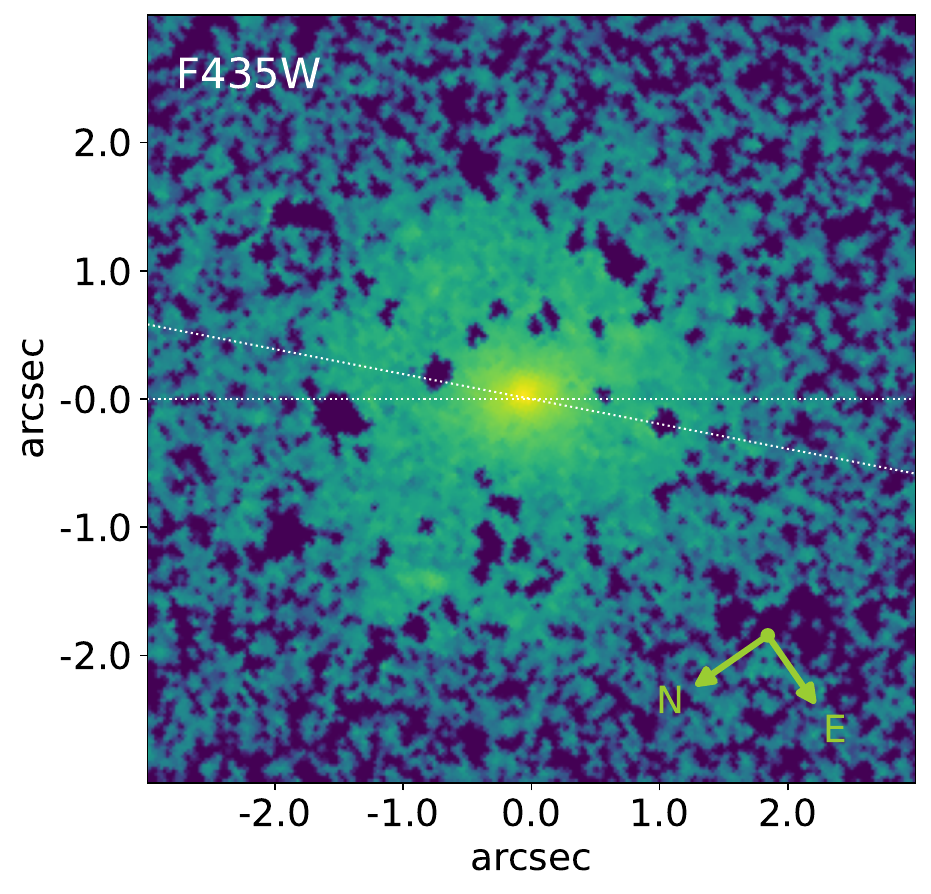}
\end{figure}
\begin{figure}[h]
    % SDSSJ1627-0053
    \centering
    \includegraphics[height=\plotht]{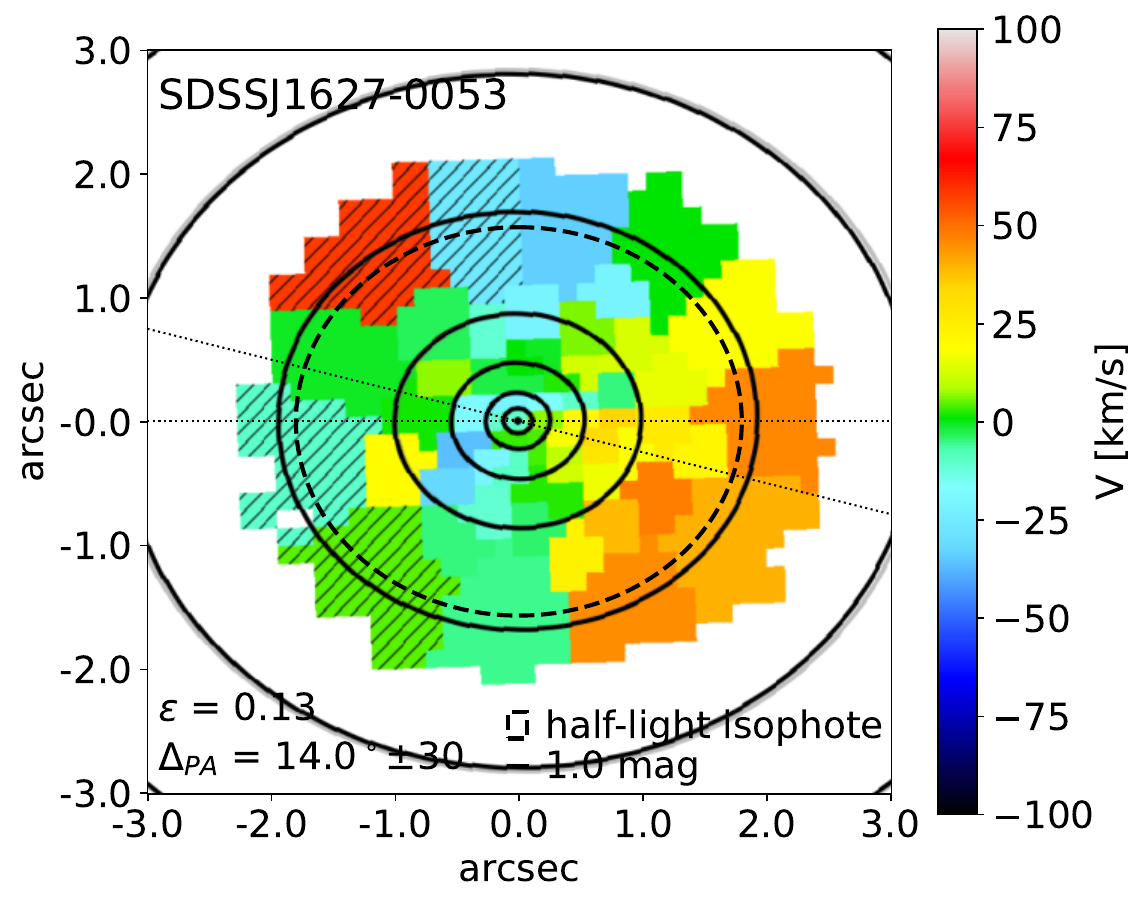}
    \includegraphics[height=\plotht]{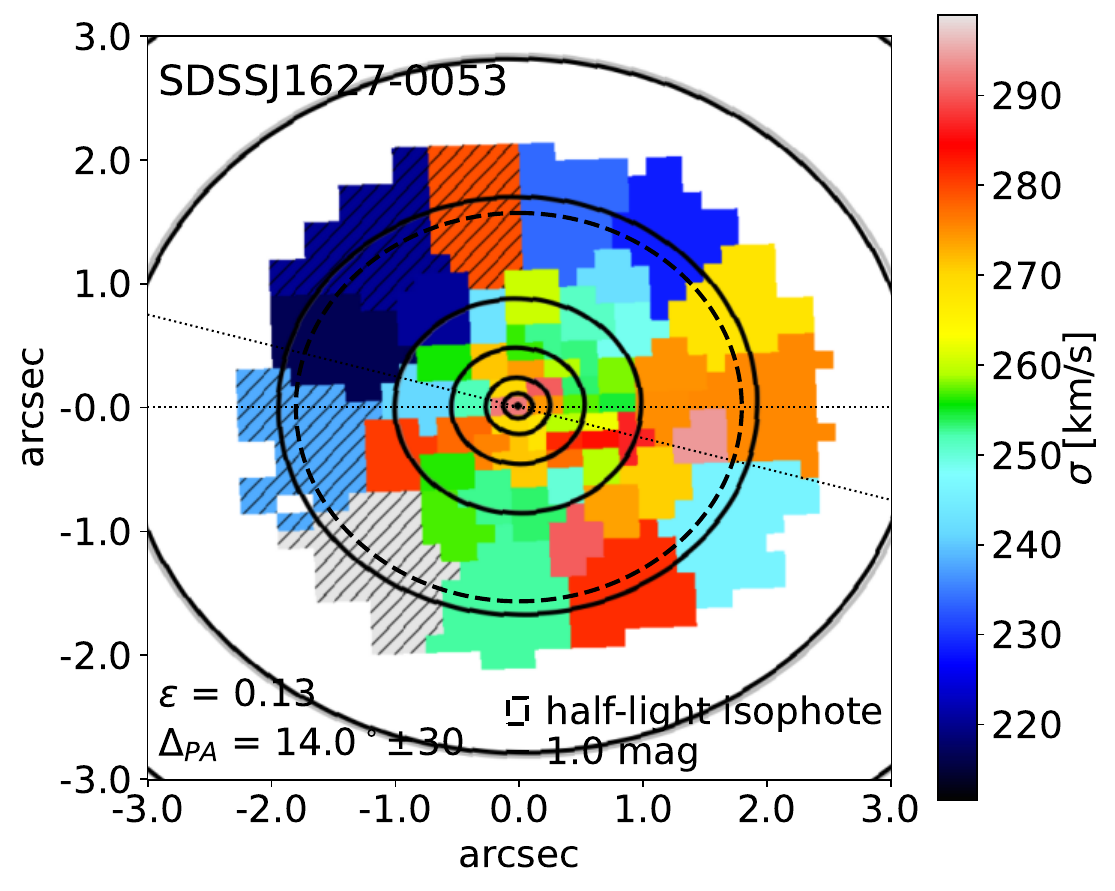}
    \includegraphics[height=\plotht]{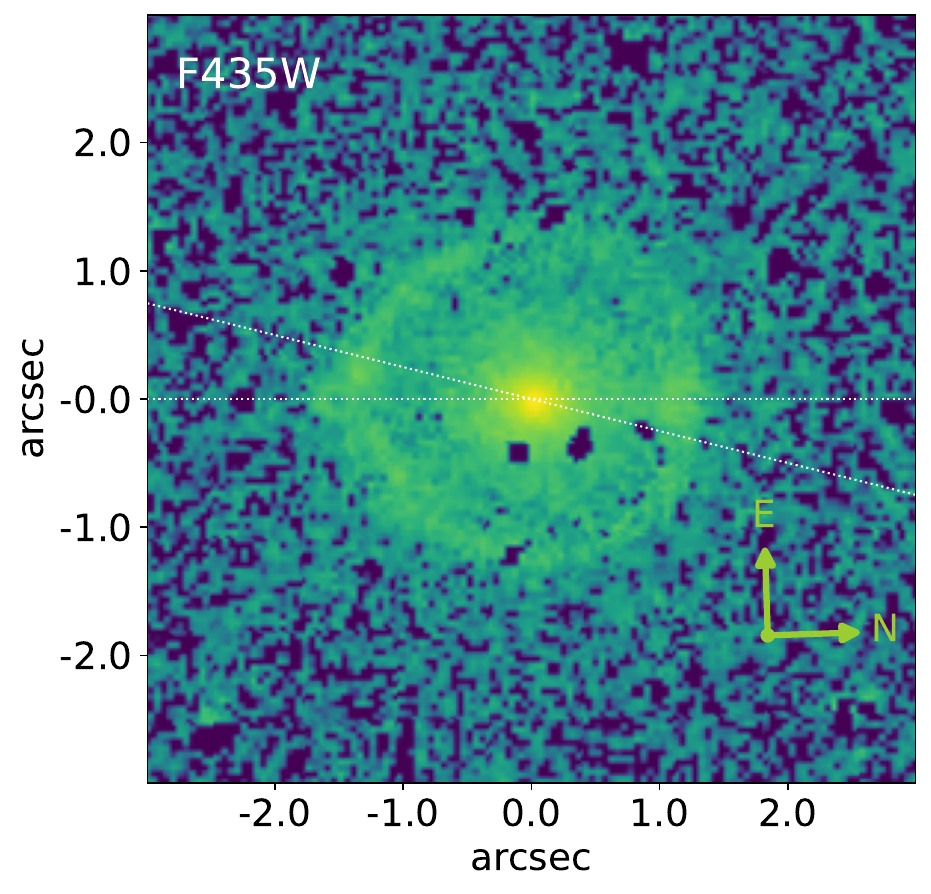}
\end{figure}
\begin{figure}[h]
    % SDSSJ1630+4520
    \centering
    \includegraphics[height=\plotht]{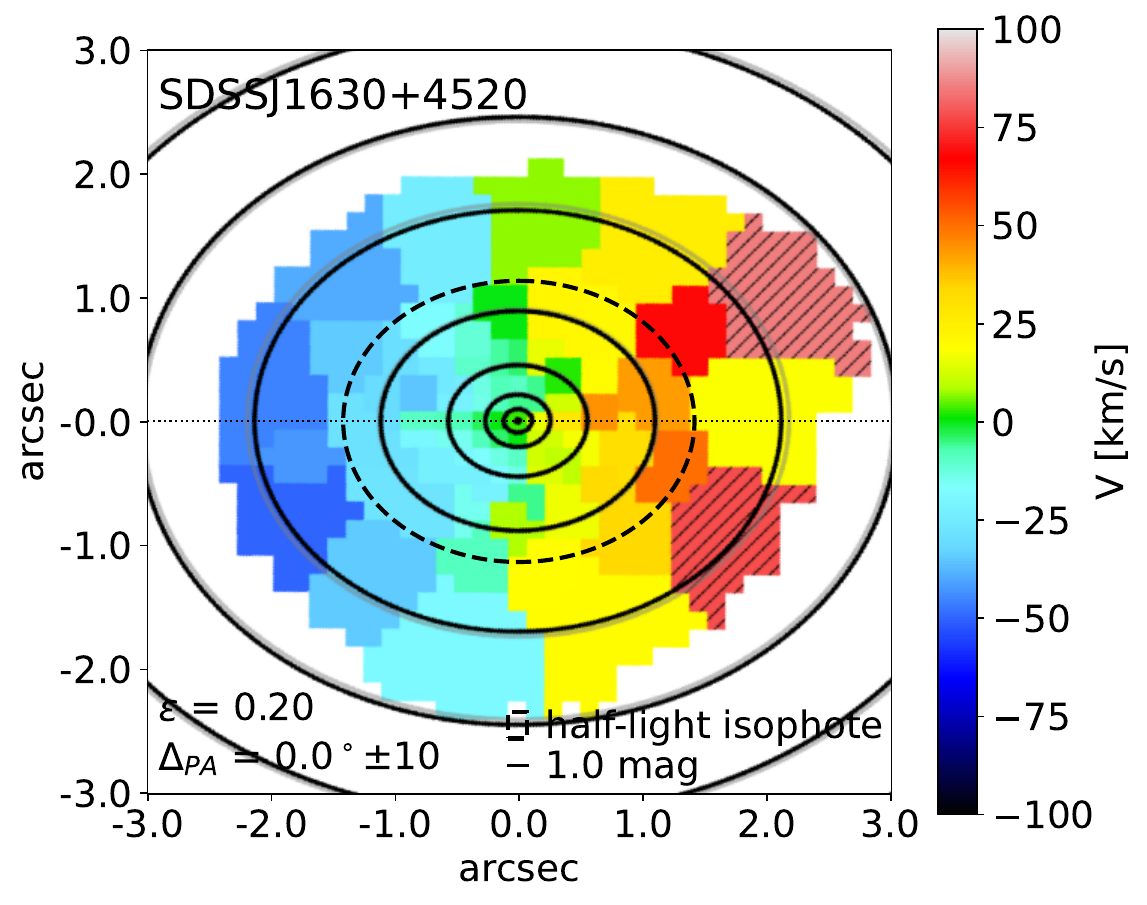}
    \includegraphics[height=\plotht]{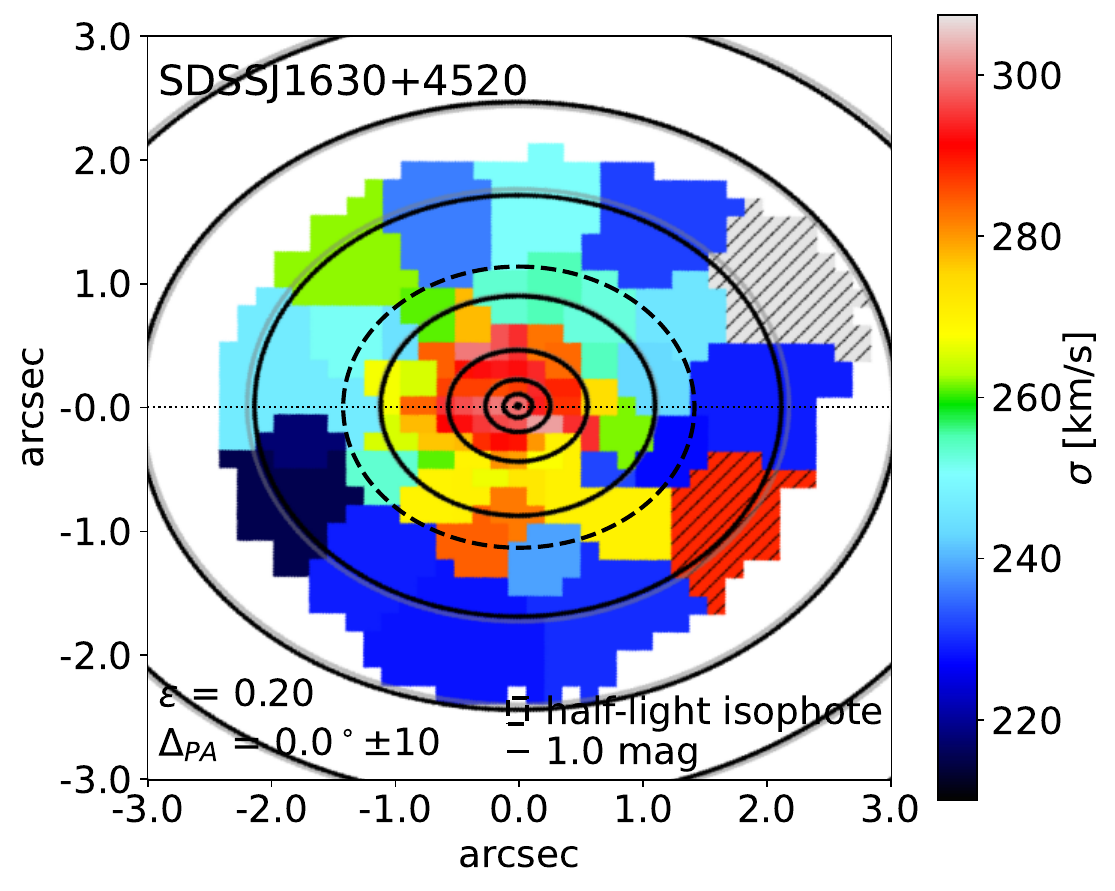}
    \includegraphics[height=\plotht]{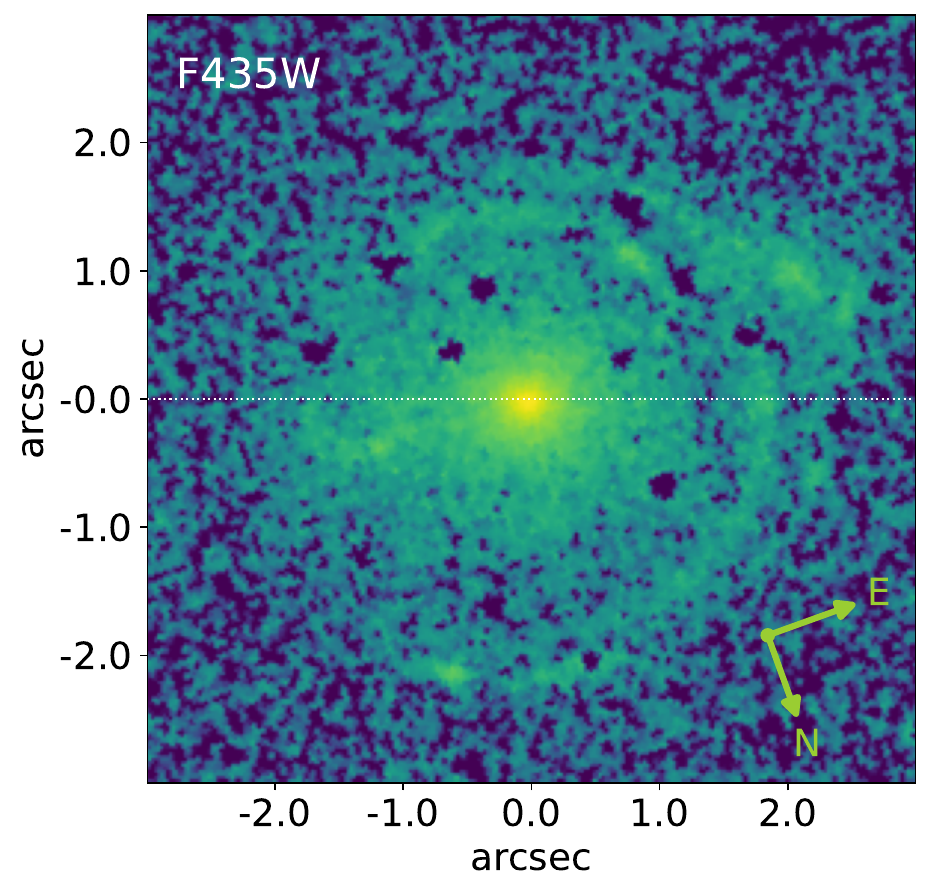}
\end{figure}
\begin{figure}[h]
    % SDSSJ2303+1422
    \centering
    \includegraphics[height=\plotht]{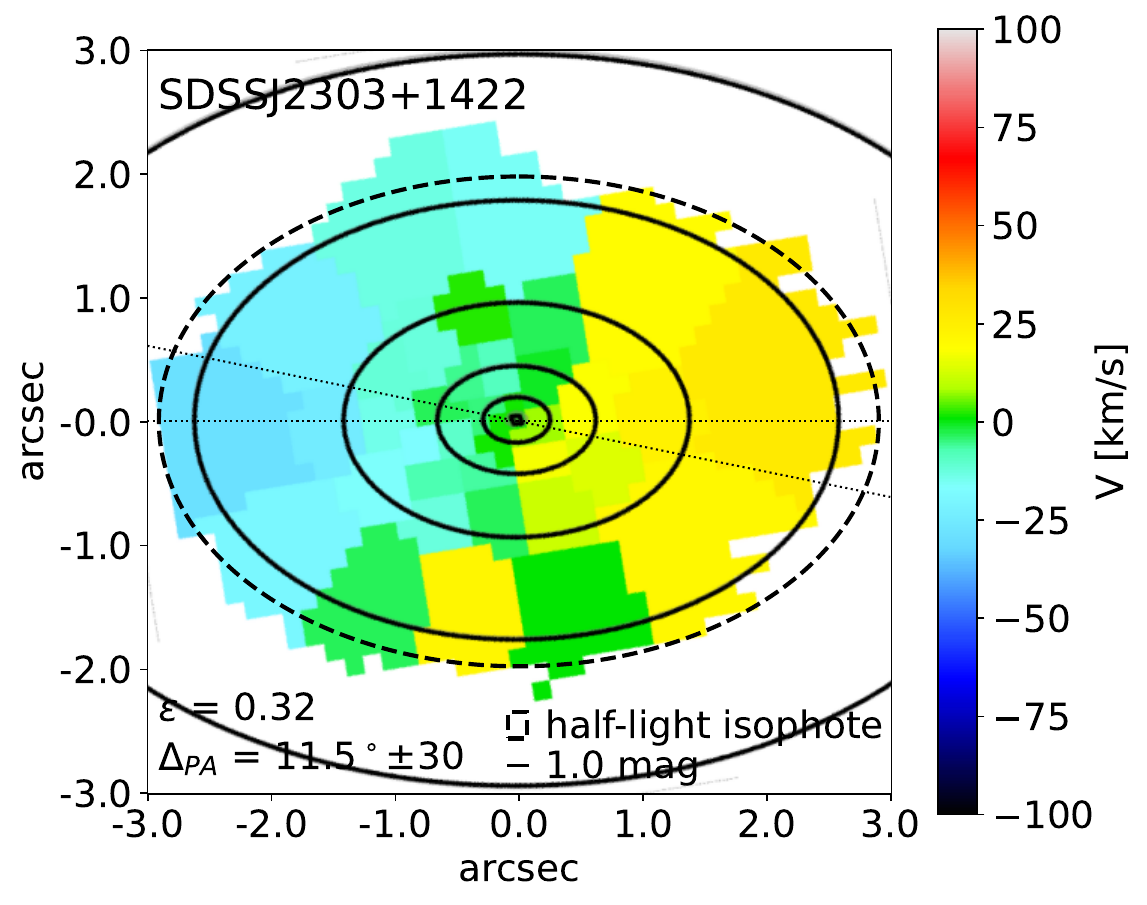}
    \includegraphics[height=\plotht]{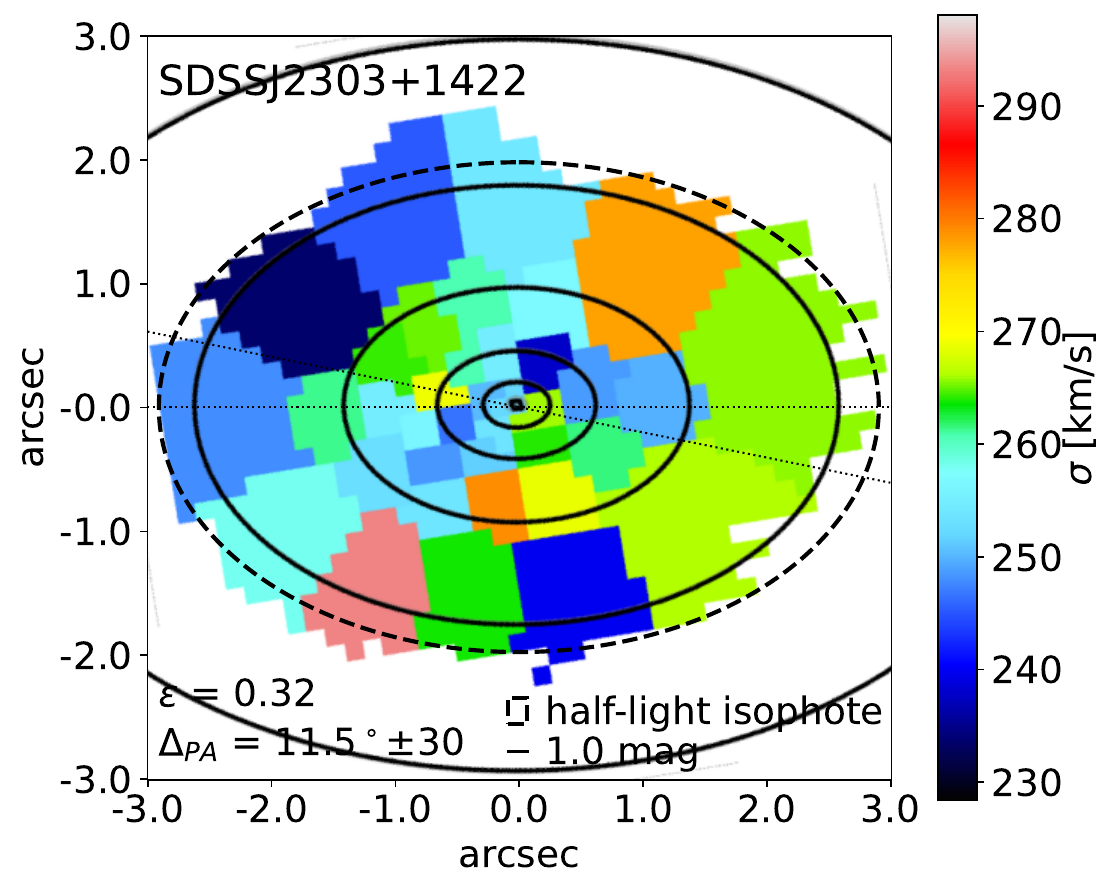}
    \includegraphics[height=\plotht]{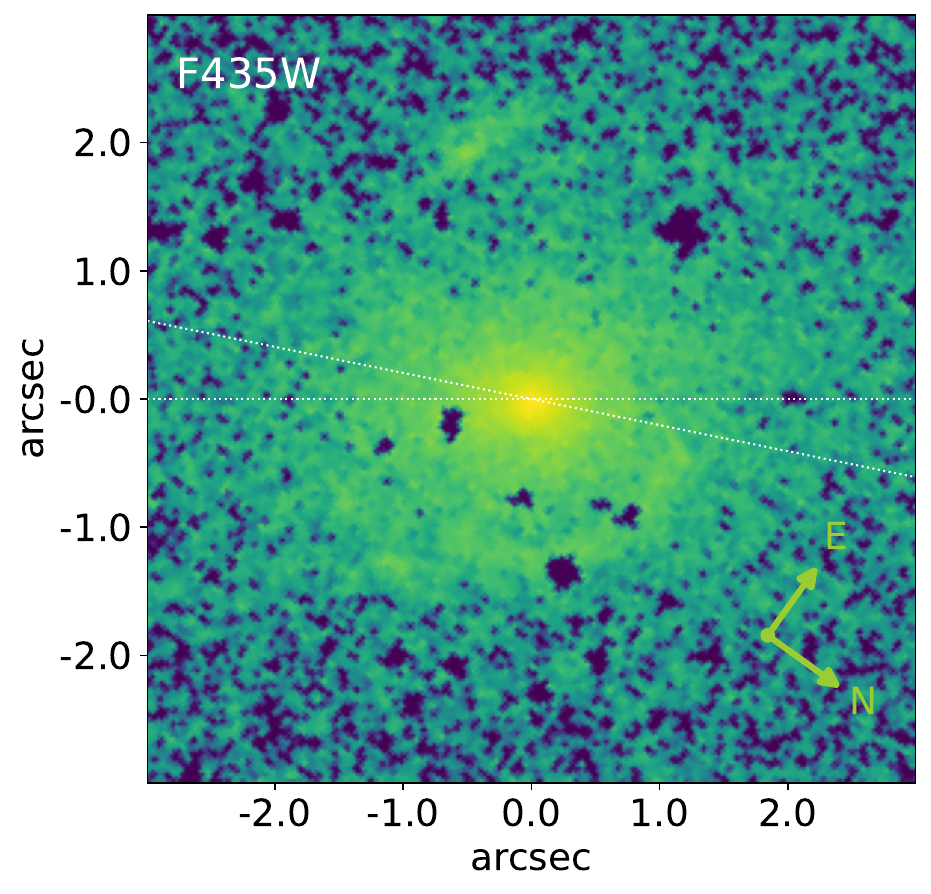}
    \caption{We show the kinematic maps of each object: mean velocity (\textit{left column}) and velocity dispersion (\textit{middle column}). We also show the available HST image closest in wavelength to KCWI (\textit{right column}, F435W, F606W, or F814W). We scale mean velocities to 100 $\mathrm{km \ s^{-1}}$. Velocity dispersions are shown in the range of the minimum and maximum value ($\pm5$ $\mathrm{km \ s^{-1}}$) shown in the colorbar to the right of each kinematic map. Red-to-white indicates the highest values (or receding velocity) and blue-to-black indicates the lowest values (or approaching velocity), with green indicating ``zero" velocity or midpoint. Bins with random uncerainties $> 20 \mathrm{km \ s^{-1}}$ are hashed. These bins often correspond to regions where the background source is significant, but we chose not to mask before fitting. We use {\sc MGEfit} to translate B-spline surface brightness models of the foreground lens galaxy (with background source features subtracted) into multiple Gaussian components for use in dynamical modeling. Surface brightness contours are overlaid on the kinematic maps. Dark contours indicate the best-fit MGE solution to the gray B-spline model surface brightness. 
    %The legend in the upper right describes the step in magnitude (normalized to the peak brightness) for each subsequent contour. 
    Images are oriented so that the photometric major axis is aligned horizontally. This axis is shown with a dotted line, in addition to a second dotted line showing the kinematic axis. The lower left details the best-fit ellipticity from the MGE translation of the B-spline models and the misalignment of the kinematic and photometric axes. For some objects, misalignment shown here is insignificant because of high uncertainty in the determination of the kinematic axis or very round isophotes. Arrows indicating north and east are placed in the HST images in the right column.}
    \label{fig:kinematics_maps}
\end{figure}

\subsection{Kinematic Classification}\label{sect:classification}

Hints of the distinction between two main classes of ETGs were evident even before spatially resolved kinematics were possible, when velocities much lower than theoretical predictions were measured in the first single-aperture stellar kinematics observations of ETGs \citep{bertola75, illingworth77, binney78}. This challenged the assumption that ETGs are universally homogeneous isotropic systems. To quantify this, \cite{binney78} introduced the ($\mathrm{V}/\sigma$, $\epsilon$) diagram. $\mathrm{V}/\sigma$ is the ratio of rotational to ``random" velocity integrated spatially across the galaxy, and $\epsilon$ is the observed ellipticity of the galaxy. This can be extended to spatially resolved kinematics by taking the flux-weighted ($F_p$) average of velocities $V_p$ and $\sigma_p$ of the \textit{p}th pixel replicated from the kinematics of the bin to which pixel \textit{p} belongs \citep{cappellari07, Graham18} 

$$\frac{\langle V^2\rangle}{\langle\sigma^2\rangle} \approx \left(\frac{V}{\sigma}\right)_e^2 \equiv \frac{\sum_p F_p V_p^2}{\sum_p F_p \sigma_p^2} $$

\noindent summing over all pixels inside the effective radius of the galaxy. $\mathrm{V}/\sigma$ contains no spatial information, so \cite{Emsellem07} introduced another quantitative kinematic description with a proxy for angular momentum $\textbf{L}=\langle \textbf{R}|\textbf{V}|\rangle$, normalizing it with the rms velocity $V_{\rm \rm rms} = \sqrt{V^2 + \sigma^2}$. With $R_p$ the position of pixel $p$, we get

$$ \lambda_{\rm \rm R} \equiv \frac{\langle R|V| \rangle}{\langle R\sqrt{V^2+\sigma^2}\rangle} 
= \frac{\sum_p F_p R_p |V_p|}{\sum_p F_p R_p \sqrt{V_p^2 + \sigma_p^2}} $$

\begin{figure}
    \includegraphics[width=0.49\textwidth]{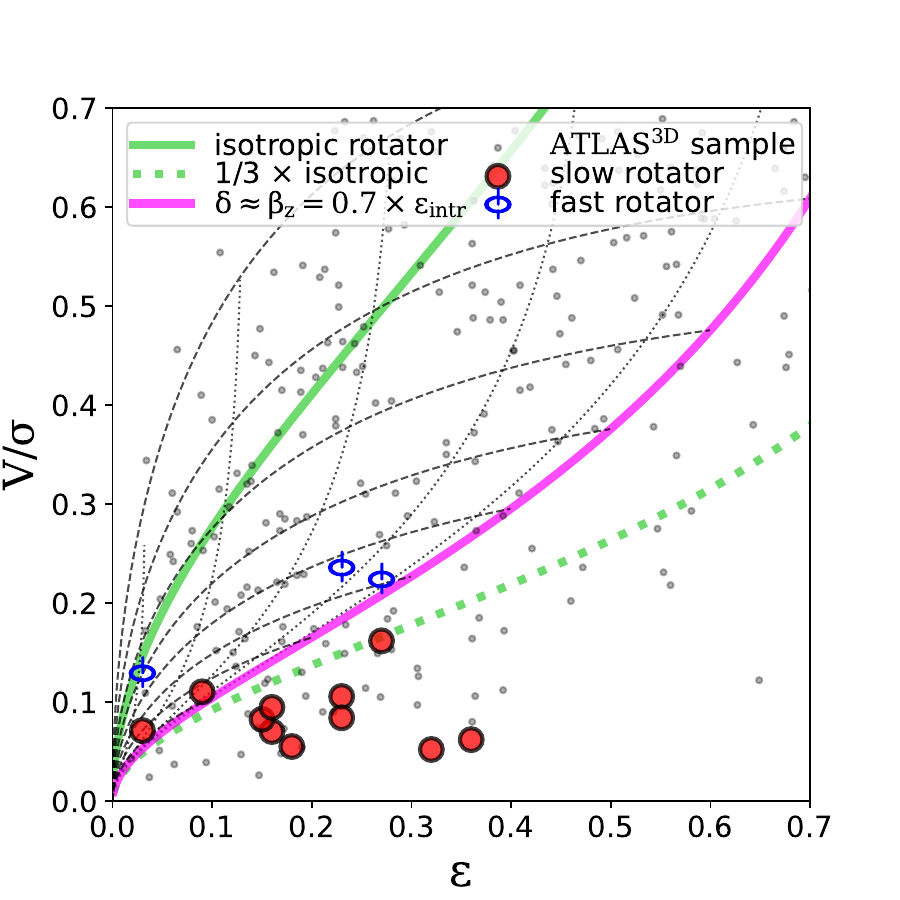}    \includegraphics[width=0.49\textwidth]{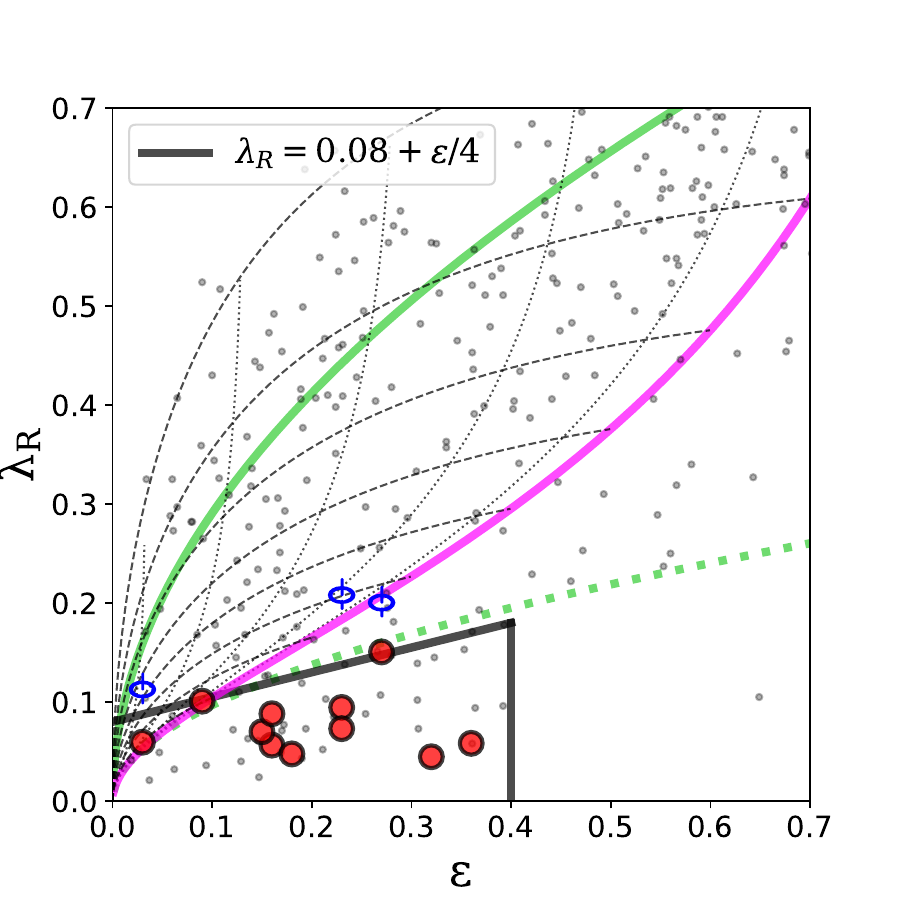}
    \caption{We plot ($\mathrm{V}/\sigma, \ \epsilon$) (\textit{left}) and ($\lambda_{\rm R}, \ \epsilon$) (\textit{right}) calculated for each of our 14 SLACS lenses. See Section \ref{sect:classification}.
    $\epsilon$ is the observed ellipticity, measured from the MGE models at the isophote enclosing half the total luminosity. $\mathrm{V}/\sigma$ and $\lambda_{\rm R}$ are calculated from the kinematic maps integrated within the circularized restframe-corrected V-band effective radius as reported in SLACS-X.
    The solid green curve shows the theoretical values for a perfectly isotropic galaxy seen edge-on \citep{binney05}, and the dotted green line shows 1/3 that value, which has been shown to roughly approximate the separation of fast/slow rotators on the ($\mathrm{V}/\sigma,\epsilon$) diagram for nearby ETGs \citep{Cappellari2016}. The magenta curve shows the edge-on relation $\delta\approx\beta_z=0.7\times\epsilon_{\mathrm{intr}}$ from \cite{cappellari07}. Dotted lines are the same relation projected at different inclinations at intervals of $10^{\circ}$, and dashed lines are contours of equal intrinsic ellipticity for the relation at different projections. Fast rotators are consistent with being randomly oriented oblate axisymmetric galaxies with anisotropy $\beta \leq 0.7$ (i.e., they lie above the magenta curve). In the right plot, the black lines show selection criteria $\lambda_{\rm R} < 0.08 + \epsilon/0.4$ and $\epsilon < 0.4$ \citep[eq. 19 of][]{Cappellari2016}, which has been shown to reliably separate fast and slow rotators as well. Gray scatter markers indicate the ATLAS$^{\mathrm{3D}}$ sample. Blue and red markers indicate quantitative classification as fast and slow rotators as reported in Table \ref{tab:results}.}
    \label{fig:classification}
\end{figure}

\noindent Values of $\mathrm{V}/\sigma$ and $\lambda_{\rm R}$ plotted against $\epsilon$ for the 14 objects are shown in Figure \ref{fig:classification} and listed in Table \ref{tab:results}.

The green curve in both plots of Figure \ref{fig:classification} shows the theoretical $\mathrm{V}/\sigma$ values for a perfectly isotropic system with given ellipticity \citep[see equation 26 in][]{binney05}; here $\alpha=0.15$ as in \cite{cappellari07}. A dividing line between fast and slow rotators can be taken as the curve delineating 1/3 of this value \citep[see, e.g.,][section 3.5.3]{Cappellari2016}. SAURON, ATLAS$^{\mathrm{3D}}$, etc, were analyzed in this way, and the data pointed to a stronger separating feature set \citep{emsellem11}. A dividing region of low rotation and low ellipticities defined by $\lambda_{\rm R} < 0.08 + \epsilon/0.4$ and $\epsilon < 0.4$ \citep[equation 19 in][]{Cappellari2016} reliably separates fast and slow rotators from those datasets. These classifications are useful for characterizing these systems in the context of more detailed studies of nearby galaxies.

We are able to detect rotation in all mean velocity maps. \cite{Czoske12} found evidence for rotation in about a third of their sample of 17 SLACS lenses (many of which are also in our sample), which demonstrates the quality of KCWI data. Despite the detectability of rotational velocity, 11 of the 14 SLACS lenses are quantitively slow/nonregular rotators when plotted in comparison to SAURON and ATLAS$^{\mathrm{3D}}$ samples on the ($\mathrm{V}/\sigma$, $\epsilon$) and ($\lambda_{\rm R}, \ \epsilon$) diagrams (see Figure \ref{fig:classification}). For some objects, rotational velocities that reach $\sim$100 km $\mathrm{s}^{-1}$ are suppressed in the $\mathrm{V}/\sigma$ contest by velocity dispersions in the 300 km $\mathrm{s}^{-1}$ range. This is what we expect for our sample of very large ETGs in the range $\sim10^{11}-10^{12} M_{\rm \rm \odot}$, which are expected to be representative of the population that likely formed from multiple dry mergers. However, several of them have significant enough rotation that they may need to be accounted for during dynamical modeling, in order to constrain the dynamical mass models at the few-percent level needed for precision cosmology. We leave this analysis for follow-up work (Knabel et al. 2024, in prep). When it matters, we rely primarily upon the ($\lambda_{\rm R},\epsilon$) to guide our classification, and we report our overall classifications in Table \ref{tab:results}. 

\begin{figure}
    \centering
    \includegraphics[width=0.5\textwidth]{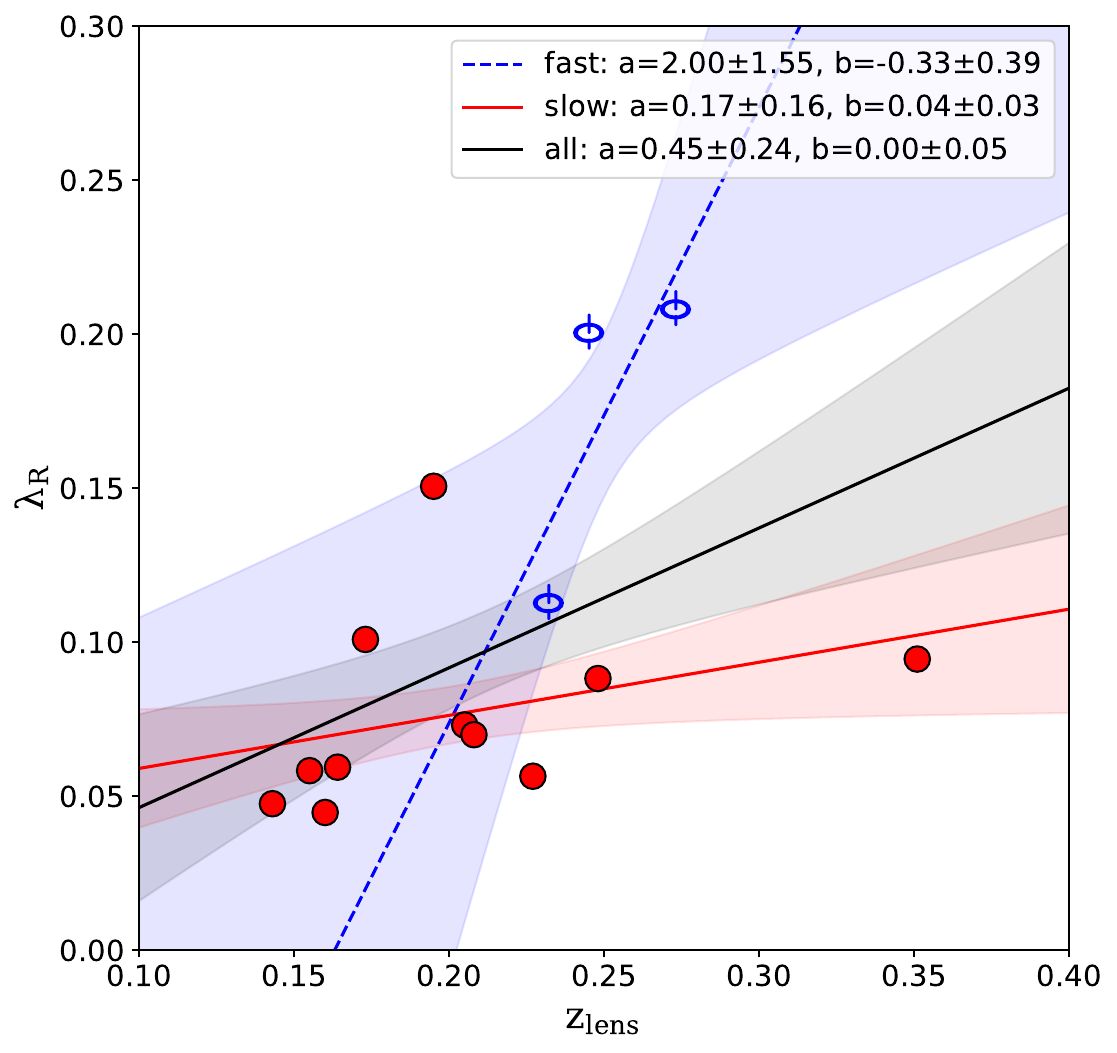}
    \caption{$\lambda_{\rm R}$ as a function of lens redshift. Markers show the objects' classification as fast/slow rotators, as given in Table \ref{tab:results}. Dotted blue line is fit to fast rotators. Solid red line is fit to slow rotators. Solid black line fits all objects.}
    \label{fig:lambda_R_redshift}
\end{figure}

In Figure~\ref{fig:lambda_R_redshift}, we show $\lambda_{\rm R}$ as a function of redshift, distinguishing the fast and slow rotators with blue and red markers as in Figure \ref{fig:classification}. The blue dotted line is fit to the fast rotators, and the red solid line is fit to the slow rotators. The data hints at an increase in $\lambda_{\rm R}$ with redshift for both classes. The extension to higher redshift of $\lambda_{\rm R}$ kinematic classification schemes derived from higher-resolution kinematics of more nearby surveys (e.g., SAURON, ATLAS$^{\mathrm{3D}}$, SAMI \citep{vandesande17_sami}, SDSS MaNGA \citep{Graham18}, CALIFA \citep{Falcon-Barroso19}) is not fool-proof. MaNGA observations \citep{bevacqua22, greene17} have shown a number of visually ``misclassified" fast rotators falling in the slow rotators zone in the ($\lambda_{\rm R},\epsilon$) diagram, which could be an effect of low spatial resolution at higher redshifts than e.g., SAURON observations. Even with the quality of KCWI spatially resolved kinematics, it is difficult to parse out this effect until a larger representative sample is observed. 

The dominance of slow rotators in our sample is expected because the lensing cross-section scales with $\sigma^4$, which favors the inclusion in lens samples of higher masses and velocity dispersions with respect to the general population of ETGs.

\subsection{Integrated Velocity Dispersions, Aperture Effects, and Comparison with SDSS}
\label{sect:aperture}

Velocity dispersions obtained with single-aperture spectroscopy (hereafter ``integrated kinematics") have limitations when compared to spatially resolved kinematics, but they are still extremely useful and will continue to be so for many applications. Very distant and compact sources, large spectroscopic surveys, multi-object spectroscopy \citep[e.g., MOSFIRE,][]{mclean10_mosfire, mclean12_mosfire}, and studies of archival data (e.g., SDSS, etc.) will still provide compelling data. Inferences based on observed integrated kinematics have established and expanded our understanding of galaxy properties and scaling relations. In the context of time-delay cosmography, integrated kinematics are still constraining (though less so) for constraining the Hubble constant. For all of these reasons, it is worth comparing the KCWI kinematic maps with the SLACS-XII SDSS kinematics that are used in the hierarchical inference of TDCOSMO-IV. 

Single-aperture velocity dispersion measurements are influenced by the size of the spectroscopic aperture. This effect cannot be fully decoupled from other instrument-specific effects, but spatially resolved kinematics allow us to examine the effects of integrating the velocity maps within various ``aperture" sizes. The projected velocity dispersion profile of ETGs decreases with angular radius because the intrinsic orbital velocities generally decrease with radius from the center of the gravitational potential. 
This means a smaller aperture will generally measure a higher integrated velocity dispersion, as a result of varying aniostropy and the intrinsic mass density profile.  Inclination effects can also influence integrated velocity dispersions. The interpretation of velocity dispersions reported in the literature is highly dependent on their definition and the specifics of the instrument and situation. This problem has been tested in \cite{bennert15} and shown to introduce up to 30-40\% differences between different definitions (inclination corrections and aperture sizes). Dynamical inferences from integrated kinematics (e.g., virial mass estimates) require relating the aperture to a more physical radius of the galaxy (e.g., the galaxy effective radius or half-thereof).

To extract an integrated velocity dispersion for each object, we define a circular aperture centered on the lens galaxy center and integrate over all spaxels within that radius. We integrate within these apertures at the datacube level before fitting the 1D composite spectra as in the first step of our full kinematic extraction described in Section \ref{sect:analysis}. As an additional test, we also extract a luminosity-weighted integrated velocity from the 2D binned kinematic map within the same radii (as in Section \ref{sect:classification} and \ref{fig:classification}. These two methods are consistent to $1\%$ on average.

\begin{figure}
    \centering    
    \includegraphics[width=0.49\linewidth]{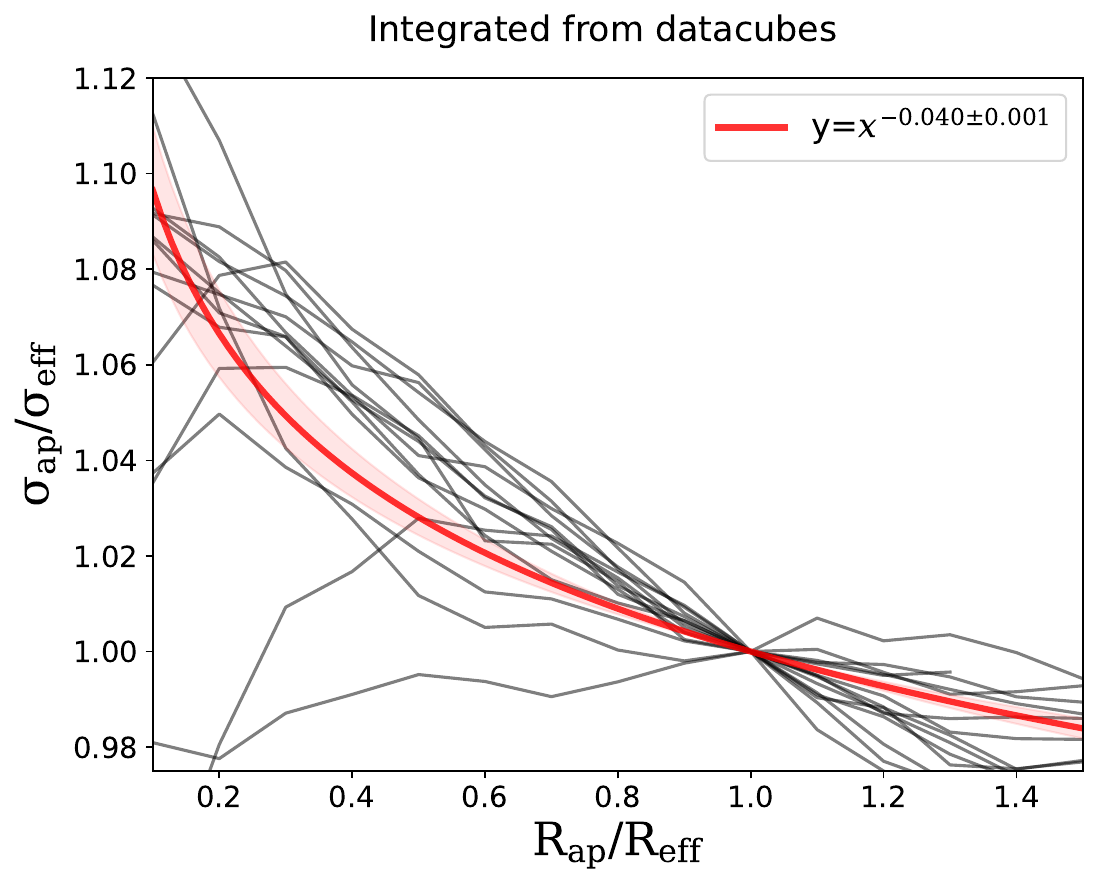} 
    \includegraphics[width=0.49\linewidth]{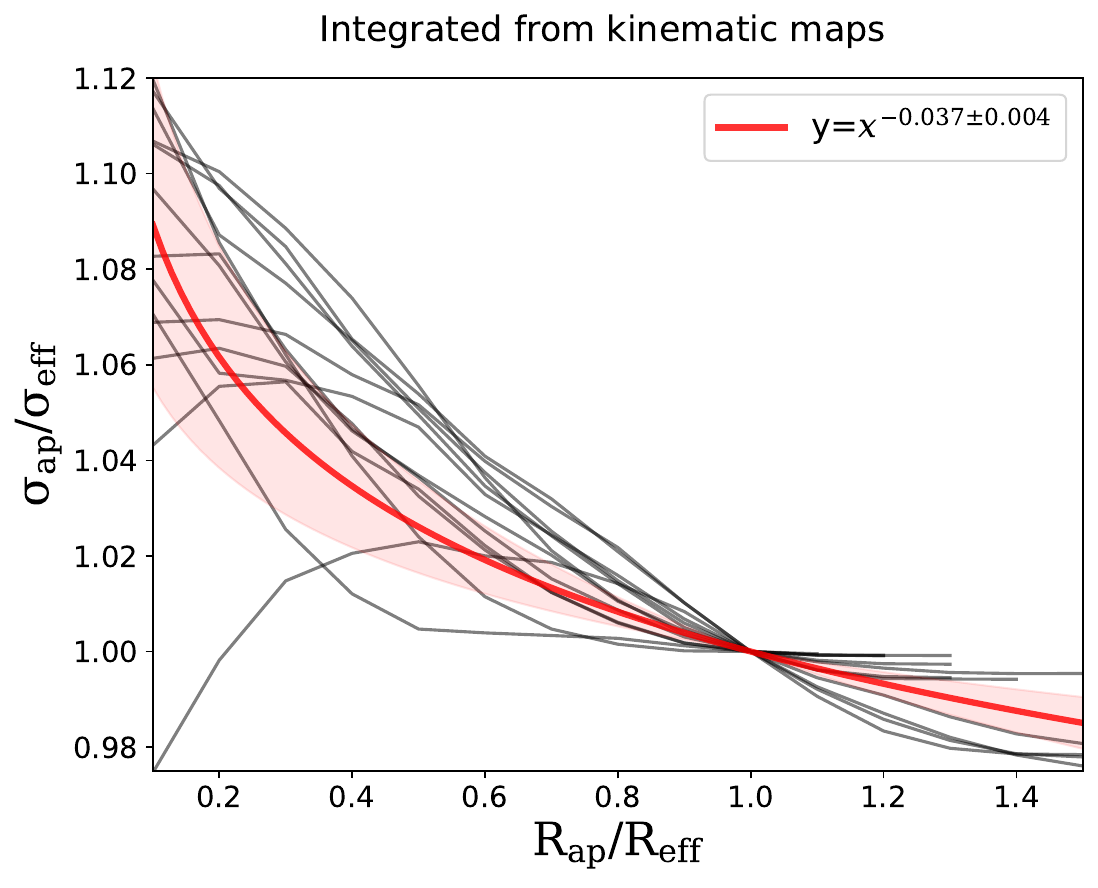} 
    \caption{Tests of aperture-correction formulas. \textit{y-axis}: aperture-integrated velocity dispersions calculated by luminosity-weighted integration of the kinematic maps. \textit{x-axis:}: radius of integration. Both are normalized to the restframe V-band effective radius as in SLACS-X. The red line shows the mean of power law fits to each of the 13 of the 14 objects (shown in gray). SDSSJ2303+1422 has a very large effective radius that is not reached by the kinematic maps. We therefore do not include this object in the average.}
    \label{fig:ap_int_effects_ratio}
\end{figure}

We look further at the effect of the integration aperture radius on the integrated velocity dispersion. A relation of $\sigma_{\rm ap}/\sigma_{\rm eff} \sim \left(R_{\rm ap}/R_{\rm eff}\right)^{\eta}$, with $\eta \sim -0.06-0.04$ has been suggested in the literature for ETGs: e.g., -0.04 \citep{jorgensen96}, -0.06 \citep{mehlert03}, -0.066 \citep{cappellari06}, -0.05 \citep{shu15}. These relations can be utilized to ``correct" for aperture-size effects when comparing data collected within different aperture sizes. Figure \ref{fig:ap_int_effects_ratio} shows aperture-integrated velocity dispersions for each object integrated within radii in steps of 0.1$\times$ the restframe V-band effective radius as reported in SLACS-X, shown as gray curves, normalized by the velocity dispersion integrated within the effective radius. The left panel displays velocity dispersions from spectra integrated at the datacube level, while the right panel shows flux-weighted integrated velocity dispersions from the 2D binned kinematic maps. We fit the curve for each galaxy with a simple power law weighted by the inverse variance of the integrated velocity dispersions. For velocity dispersions integrated from datacubes and kinematic maps, respectively, we find $\eta=-0.040\pm0.001$ and $\eta=-0.037\pm0.004$. A weighted linear regression in log-log space closely agrees. This mean power law fit is the red curve of both panels. We use $\eta=-0.040$ in the analysis of scaling relations in Section \ref{sect:correlations}.

\subsubsection{Our Method Applied to SDSS Spectra}\label{sect:sdss_compare}

% Table of SDSS uncertainties
\begin{table}[]
    \centering
    \begin{tabular}{l|c|c|c|c|c|c|c}%|c|c|c|c}
    
    Sample & $\rm <\delta \bar{\sigma}/\bar{\sigma}>$ & $\rm <\Delta \bar{\sigma}/\bar{\sigma}>$ & $\rm \sqrt{<C_{i,j}\bar{\sigma}/\bar{\sigma_i}\bar{\sigma_j}>_{i\neq j}}$ & $\rm <\Delta_B \bar{\sigma}/\bar{\sigma}>$ & $\rm \sqrt{<C_{B,i,j}\bar{\sigma}/\bar{\sigma_i}\bar{\sigma_j}>_{i\neq j}}$ & Median SNR & \# objects \\
\hline
KCWI &   0.90\%  &  0.79\%  &   0.70\%  &  0.97\%  &  0.86\% & 153.6 & 14 \\
SDSS S/N $>$ 15 & 3.30\% & 1.39\% & 1.15\% & 1.70\% & 1.40\% & 17.3 & 38 \\
SDSS S/N (10, 15] & 5.30\% & 1.88\% & 1.31\% & 2.30\% & 1.61\% & 11.7 & 98 \\
SDSS S/N [5,10] & 7.71\% & 2.70\% & 1.86\% & 3.31\% & 2.28\% & 8.1 & 151 \\

\hline
\end{tabular}
\caption{Uncertainties, as in Table~2 of TDCOSMO-XIX. For each dataset, we report the average statistical error, the average systematic uncertainty associated with the choice of clean template library, the average amplitude of the off-diagonal terms of the covariance matrix between elements of the sample, the Bessel-corrected average systematic uncertainty and off-diagonal covariance matrix. Averages are calculated with equal weights, as in the lower panel of Table~2 of TDCOSMO-XIX. KCWI values are identical to those found by TDCOSMO-XIX. SDSS spectra have been separated into bins of S/N per $\rm \AA$, calculated in the range 4000-5000$\rm \AA$.}
\label{tab:sdss}
\end{table}

% Figure of KCWI and SDSS spectra with S/N for each object
\begin{figure}
    \centering
    \includegraphics[width=0.8\textwidth]{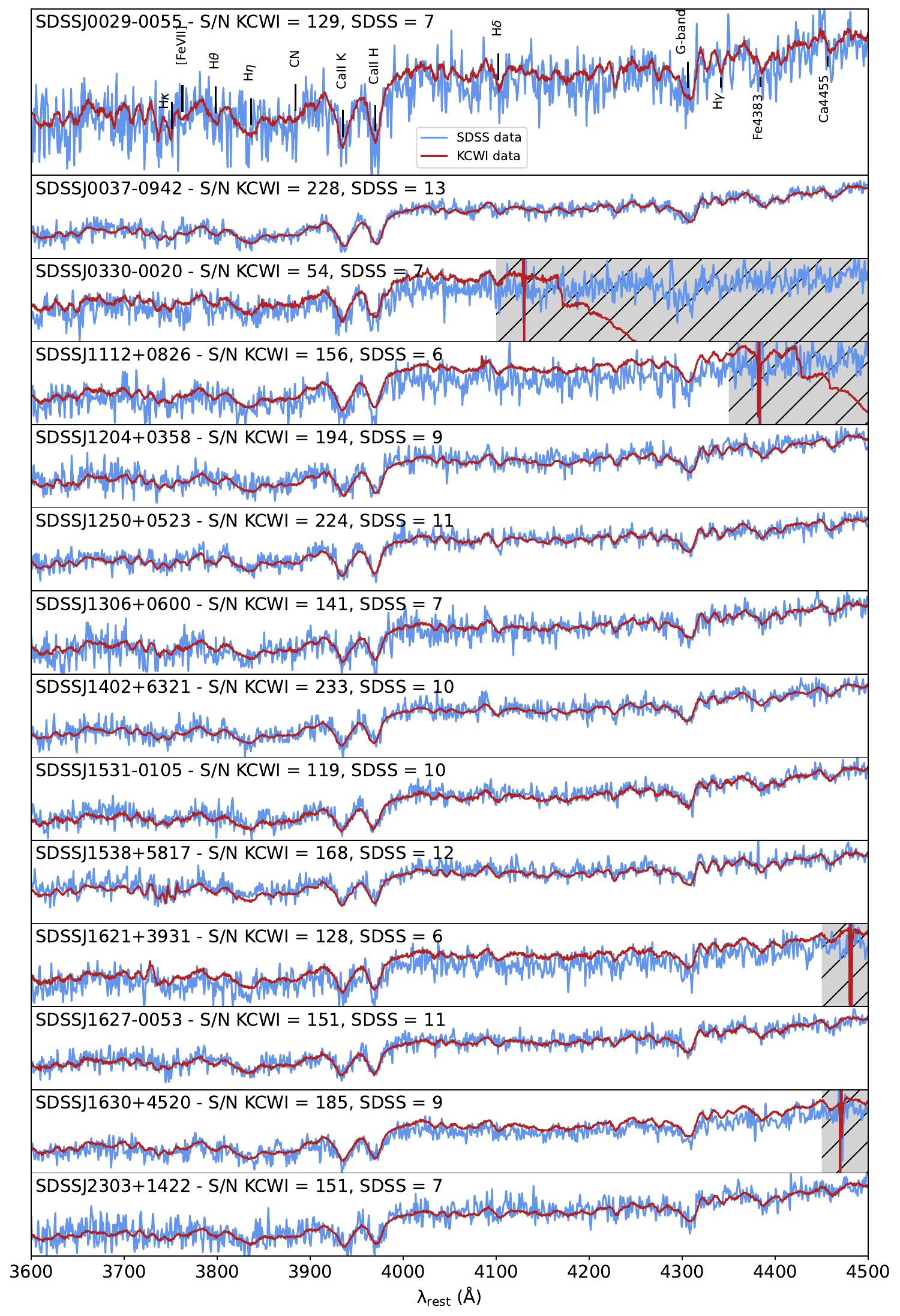} 
    \caption{Restframe spectrum of each deflector galaxy in our sample. Red shows the KCWI data extracted from an aperture of the SDSS fiber size ($1.5''$). Blue shows the SDSS spectrum for the object. We list the S/N per $\rm \AA$ for each spectrum measured in the range $4000-4500\rm \AA$. SDSSJ1538+5817 has two fiber observations that conflict in extracted velocity dispersions. Here we show the fiber measured in SLACS-XII.
    }
    \label{fig:sdss_spectra}
\end{figure}

We use our methods to extract new velocity dispersions from SDSS spectra for the SLACS sample and compare with our KCWI measurements for the 14 galaxies of our sample, finding agreement between the datasets within the uncertainties, which are dominated by the lower S/N of the SDSS spectra. We retrieved the SDSS fiber spectra from SDSS DR17 \citep{abdurrouf22_sdss_dr17} via the SDSS Science Archive Server (SAS) and measured the spectra using {\sc pPXF} for the 341 ETGS from the SLACS candidate parent sample. Where possible, we queried the same fiber spectra used for the SLACS-XII velocity dispersions. TDCOSMO-XIX used only the highest S/N spectra ($>15$/\AA\ from this sample) to test for systematic uncertainties as a result of stellar template library selection, finding systematic uncertainties of less than $2\%$ and a strong preference for the XSL stellar library by BIC evidence. We extend that analysis using the same three cleaned stellar libraries (Indo-US, MILES, and XSL) to lower S/N spectra and measure stellar library systematics for S/N bins of 5-10 and 10-15 per $\rm \AA$. Table~\ref{tab:sdss} shows the average uncertainties over the SDSS S/N bin samples and the KCWI data for reference, as in the lower section of Table 2 in TDCOSMO-XIX. For very low S/N spectra, the fits result in high statistical uncertainties $\rm <\delta \bar{\sigma}/\bar{\sigma}>$, making the data much more sensitive to changes in stellar libraries or other factors, enhancing systematic errors. Furthermore, the BICs for the template libraries become completely indistinguishable. The SDSS data for the 14 KCWI objects has a mean S/N per \AA\ of $\sim9$, and a maximum of 13. In Table~\ref{tab:sdss}, the bottom row has S/N in the range 5-10 with a median of 8.1, and it encompasses 10 of the 14 galaxies. The larger sample of SDSS spectra at this S/N resulted in an average statistical uncertainty of $7.7\%$, which is far above the systematic uncertainties attributable to template libraries that we find for any of the subsamples. For these reasons, we derive the systematic uncertainties for the purposes of this paper with equal weights to the three libraries instead of using BIC-weighting. With a Bessel correction to account for the small number of libraries (see equations 14-16 and Appendix A of TDCOSMO-XIX) the average systematic uncertainties for the bin with S/N in the range 5-10 are $3.31\%$ in the diagonal and $2.28\%$ in the off-diagonal. The off-diagonal represents the correlated effect of the differences in the stellar template libraries across a sample of measured velocity dispersions. In other words, changing from one library to another moves all the velocity dispersions up or down, in a coherent fashion, introducing covariance across the sample. For the bin with S/N in the range 10-15, the systematics are $2.30\%$ and $1.61\%$ in the diagonal and off-diagonal, respectively. 

We extract kinematics from the KCWI datacube integrated within the 1.5 arcsecond SDSS aperture for comparison with the results from SDSS spectra. Typical seeing for our Keck observations is significantly better ($\sim 0.8-0.9''$ FWHM) than that of the SDSS observations ($\sim2''$ FWHM). PSF blurring tends to lower integrated velocity dispersions as it introduces light from outer regions of the galaxy where velocity dispersions are typically lower than the central regions. We simulate the effect of the seeing difference by convolving the datacube with a Gaussian kernel when performing the aperture integration before kinematic extraction. We tested the dependence of the extracted velocity dispersions on the assumed seeing of SDSS observations by fitting with SDSS seeing of 1.5 and 2.5$''$ FWHM, resulting in $0.58\pm0.26\%$ larger and $0.21\pm0.25\%$ smaller velocity dispersions, respectively. These are completely consistent within statistical errors. Because seeing FWHM for the Keck observations is much smaller, the effect is neglible. In Figure \ref{fig:sdss_spectra}, we show the SDSS spectra in comparison with the KCWI spectra integrated within the same $1.5''$ aperture. The average S/N per $\rm \AA$ of the SDSS spectra for the 14 galaxies of this sample is $\sim9$, where the systematics in Table \ref{tab:sdss} are $\sim3\%$. The average S/N of the KCWI sample is more than one order of magnitude higher. 

The KCWI sample shows a strong BIC-preference for Indo-US over MILES (greater than $6\sigma$) and over XSL (at $3\sigma)$. The uncertainty on the difference in BIC between two libraries is estimated by bootstrapping over the 14 galaxies' fits and constructing a distribution of the difference in total BIC values calculated from the combined number of parameters, number of data points, and likelihoods of each bootstrapped sample (see Section 5 of TDCOSMO-XIX for more details). The highest-S/N subsample of SLACS SDSS spectra with S/N $ > $ 15 per \AA\ strongly prefers the XSL library over Indo-US and MILES libaries (both at around $5\sigma$). For all following comparisons, as well as comparisons with other analyses of SDSS data in Section~\ref{sect:previous_sdss}, we take the BIC-preferred library for each dataset (Indo-US for KCWI and XSL for SDSS) as the fiducial model for that dataset, and we report the results of these fiducial fits in Table~\ref{tab:results} as $\rm \sigma^{KCWI}_{1.5}$ and $\rm \sigma^{this \ work}_{SDSS}$. For these BIC-preferred libraries, the SDSS velocity dispersions are on average $1.7\pm2.2\%$ larger than the KCWI values for the same galaxies. Under the assumption of 1.5$''$ (2.5$''$) SDSS seeing, the SDSS velocity dispersions are $1.2\pm2.2\%$ ($1.9\pm2.2\%$) larger than the KCWI values. For fixed stellar template library, the SDSS velocity dispersions are also on larger on average. For Indo-US, SDSS velocity dispersions are $1.2\pm1.8\%$ larger than the corresponding KCWI fit. For MILES, which was strongly downweighted by BIC for both KCWI and SDSS, we measure SDSS velocity dispersions that are $5.2\pm2.2\%$ larger than KCWI. For XSL, SDSS is $2.1\pm2.1\%$ larger.

As an alternative avenue to evaluate our measured SDSS velocity dispersions and their uncertainties in comparison with the higher-quality KCWI data, we carry out a two-parameter inference to examine any bias as well as residual uncertainties in our SDSS fits when including only the formal statistical uncertainties, as in TDCOSMO-IV. This relies upon our confidence in the KCWI velocity dispersions to determine how much additional uncertainty is appropriate to add to the SDSS errors without assuming the source of those systematics. We include a parameter $\beta$ to describe any multiplicative bias and a parameter $\gamma$ to describe underestimated errors in the velocity dispersions extracted from SDSS spectra. The likelihood is then

\begin{equation}\label{eq:sdss_comp1}
    \frac{1}{\sqrt{2 \pi} \delta \sigma_{\rm T}}
    \exp{\left[-\frac{1}{2}\left( \frac{\sigma_{\rm KCWI,1.5}-\beta \sigma_{\rm SDSS}}{\delta \sigma_{\rm T}}\right)^2\right]}
\end{equation}

where 

\begin{equation}\label{eq:sdss_comp2}
    \delta\sigma_{\rm T}^2=\delta\sigma_{\rm KCWI,1.5}^2+(1+\gamma)^2 \delta \sigma_{\rm SDSS}^2
\end{equation}

or

\begin{equation}\label{eq:sdss_comp3}
    \delta\sigma_{\rm T}^2=\delta\sigma_{\rm KCWI,1.5}^2 + \delta \sigma_{\rm SDSS}^2 + \gamma_*^2 \sigma_{\rm SDSS}^2
\end{equation}

\noindent is the total uncertainty of the comparison between our measured velocity dispersion from SDSS spectra and the one estimated based on the KCWI data within the same aperture with mock seeing. $\gamma$ in Equation \ref{eq:sdss_comp2} describes a percent underestimate of the absolute velocity dispersion errors from our SDSS extractions. $\gamma_*$ in Equation \ref{eq:sdss_comp3} describes the underestimate of relative SDSS errors by a percentage of the velocity dispersions and is the parametrization as in TDCOSMO-IV (Equation 57), where the equivalent parameter is called $\mathrm{\sigma_{\sigma^P,sys}}$. $\beta$ will be very close to the same value for both equations. We optimize the likelihood function using Markov Chain Monte Carlo (MCMC) sampling using the software package\footnote{\url{https://pypi.org/project/emcee/}} \textsc{emcee} \citep{Goodman10, Foreman-Mackey13}. We use a Jeffreys prior of $1/\gamma$ and with $\delta\sigma_{\rm T}^2$ as in Equation \ref{eq:sdss_comp2}. In Table~\ref{tab:my_label}, we show the inferred values of $\gamma$, $\gamma_*$, and $\beta$ by comparing our velocity dispersions measured from KCWI and from SDSS spectra for the fiducial, BIC-preferred libraries (Indo-US for KCWI and XSL for SDSS), as well as for fixed stellar template libraries.

\begin{table}[]
    \centering
    \begin{tabular}{cccccc}
	Equation & Parameter & Best-BIC & Indo &	MILES &	XSL \\
    \hline
        \ref{eq:sdss_comp2} & $\gamma$ & $0.097^{+0.189}_{-0.075}$ & $0.041^{+0.094}_{-0.026}$	& $0.152^{+0.199}_{-0.121}$	& $0.090 ^{+0.185}_{-0.069}$ \\
         & $\beta$	& $1.002{\pm0.020}$	& $1.008 \pm 0.019$	& $0.957 \pm 0.020$	& $0.993 \pm 0.020$ \\
         \hline
        \ref{eq:sdss_comp3} & $\gamma_*$	& $0.039 ^{+0.027}_{-0.021}$	& $0.020^{+0.017}_{-0.007}$	& $0.050^{+0.027}_{-0.024}$	& $0.037^{+0.028}_{-0.020}$ \\
        & $\beta$	& $1.002 {\pm0.022}$	& $1.006{\pm0.020}$	& $0.959\pm{0.023}$	& $0.993\pm{0.022}$ \\
    \end{tabular}
    \caption{Estimates of residual uncertainties and biases of our velocity dispersion estimates extracted from SDSS spectra with our methods in comparison with KCWI aperture-integrated velocity dispersions with mock SDSS seeing, as in Equations~\ref{eq:sdss_comp1}-\ref{eq:sdss_comp3}. We compare the best-BIC libraries for each dataset (Indo-US for KCWI and XSL for SDSS) as well as for fixed stellar template libraries.}
    \label{tab:my_label}
\end{table}

For the fiducial models with the best-BIC template libraries, for the parameters in Equation~\ref{eq:sdss_comp2}, we find $ \gamma = 0.097^{+0.189}_{-0.075}$ and $ \beta = 1.002\pm{0.020}$. This again indicates no evidence of bias in the velocity dispersions we measure from SDSS; however, uncertainties are shown to be underestimated by $9.7^{+18.9}_{-7.5}\%$ (with respect to the uncertainty) to be added in quadrature to the estimated SDSS uncertainty. Using $\delta\sigma_{\rm T}^2$ as in Equation \ref{eq:sdss_comp3}, we find $ \gamma_* = 0.039^{+0.027}_{-0.021}$, and $\beta$ is unchanged. The underestimated error could then be accounted for by adding a relative 3.9$^{+2.7}_{-2.1}\%$ error (with respect to the velocity dispersion) in quadrature. This is consistent with our own estimate of the systematic residual error due to the template library selection shown in Table~\ref{tab:sdss}. We find results consistent with this picture when comparing the KCWI and SDSS velocity dispersions when both are fitted with templates from Indo-US or XSL. The MILES library, which was strongly disfavored for both KCWI and SDSS datasets in our fits by BIC evidence, results in a $\sim 4\%$ difference between the the datasets ($\beta=0.957\pm0.020$ for Equation~\ref{eq:sdss_comp2} and $\beta=0.959\pm0.023$ for Equation~\ref{eq:sdss_comp3}). We infer a $5.0^{+2.7}_{-2.4}\%$ (relative to the velocity dispersions) residual error from the MILES fits. These are consistent within $1\sigma$ with our estimated systematic uncertainty from the stellar template library mismatch (see Table~\ref{tab:sdss}). Because the differences between our measurements of the KCWI and SDSS data are within uncertainties on the mean at the sample level, and our SDSS velocity dispersions include both statistical and systematic uncertainties well above these differences, we conclude that there is no measurable bias between the velocity dispersions we measure from SDSS spectra versus those obtained with KCWI data, using the same procedure and cleaned stellar libraries. Our analysis shows that a systematic uncertainty of $\sim3-4\%$, of which the majority is attributable to the stellar template library selection, must be added in quadrature to the errors of our extracted SDSS velocity dispersions. For the ease of interpretation and implementation, we recommend the $\gamma_*$ model be used to calculate the term to be added in quadrature to uncertainties on SDSS velocity dispersions whenever used.

\subsubsection{Comparison with previously published SDSS velocity dispersions}\label{sect:previous_sdss}

We are now in a position to compare our measured velocity dispersions with the SDSS measurements that have been utilized in TDCOSMO-IV (from SLACS-XII). We also compare with earlier SLACS velocity dispersions extracted from SDSS spectra. SLACS-IX and -X explore fundamental plane relations and other correlations in SLACS galaxies through joint lensing-dynamical inference, using velocity dispersions measured with the ELODIE stellar
template library \citep{prugniel_soubrian_2001_elodie}, which we have not used in any of our analyses. We compare the correlations and other inferred quantities with our KCWI data in Section \ref{sect:correlations}. SLACS-XII updated and superseded these velocity dispersions with methodology outlined by \cite{Bolton12a} and \cite{Bolton12}, using the Indo-US library to cover broader wavelength ranges than were available previously, which represents a change in the most significant contributors to systematic error we find in our methods (see \ref{sect:uncertainties}). 

Given the superior quality of our data with respect to SDSS, this comparison gives us an opportunity to assess any bias in prior SDSS measurements and by how much the SDSS uncertainties are underestimated, as found by TDCOSMO-IV. We note that these analyses utilized the same spectra we analyzed, which have S/N $\sim9$ per \AA\ and are likely subject to statistical and systematic uncertainties similar to those reported in the bottom row of Table~\ref{tab:sdss}, $7.71\%$ statistical and $3.31\%$ systematic, as well as $2.28\%$ covariance over the sample due to the correlated effect of the stellar template libraries. With that in mind, any trend of differences in the fitted velocity dispersions between our measurements and previous velocity dispersions measured from SDSS spectra at $3.31\%$ or less do not indicate a significant measurable bias. We note one object (SDSSJ1538+5817) is a strong outlier in our comparisons. We retrieved two observations of this galaxy from SDSS. One yields $180\pm12$ km $\mathrm{s}^{-1}$, which agrees closely with the SLACS-IX and -XII values. The other yeilds $253\pm15$ km $\mathrm{s}^{-1}$. The value we measure over the same aperture with KCWI data and mock seeing $216\pm2$ is at the midpoint between these two values. The difference in mean velocity is $22\pm11$ km $\mathrm{s}^{-1}$, suggesting that the two observations are not both pointed at the center of the galaxy, but at least one of the fibers was misaligned. This may explain the difference in velocity dispersion. Therefore we do not consider this object in the comparison here. 

\begin{figure}
    \centering
    \includegraphics[width=0.8\linewidth]{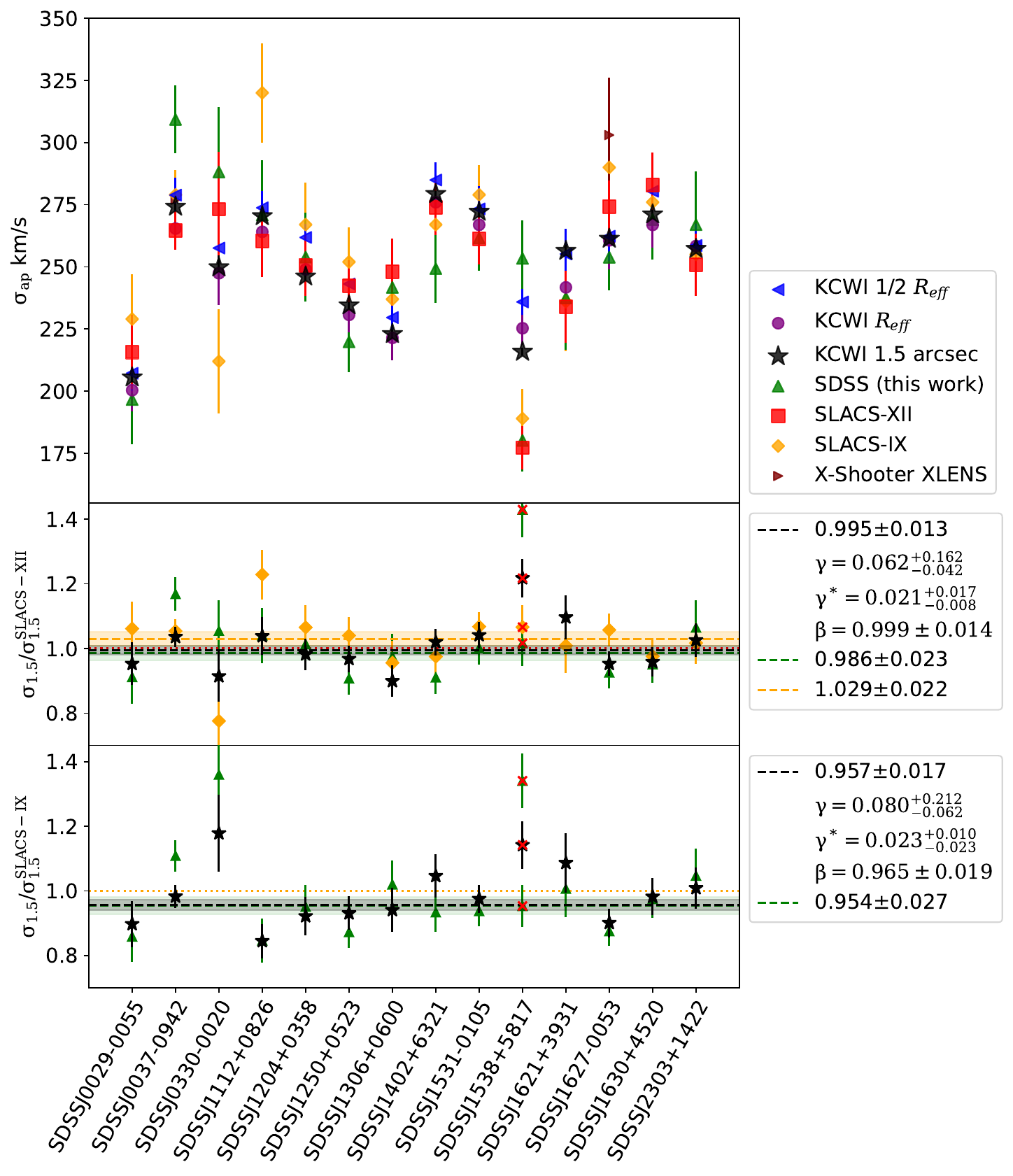}
    \caption{We compare integrated velocity dispersions within various apertures from KCWI datacubes and those extracted from SDSS spectra. The blue left-pointing triangle and purple circle markers indicate the velocity dispersions extracted from the kinematic maps by luminosity-weighted integration. These are the values we use, e.g., for ($\mathrm{V}/\sigma,\epsilon$) and ($\lambda_{\rm R},\epsilon$) calculations and inferences in Section \ref{sect:correlations}. We also integrate KCWI kinematics at the datacube level within the SDSS 1.5-arcsecond aperture radius (black star) for comparison with kinematics from  SDSS-III BOSS fiber spectroscopy: extracted with our methods (green triangle), reported in SLACS-XII (red square), and from SLACS-IX (orange diamond). We also include X-Shooter slit-spectroscopy velocity dispersions from XLENS \citep{Spiniello15} for two objects (maroon right-pointing triangle). In the middle panel of the figure, we plot the ratios of the 1.5 arcsecond aperture-integrated KCWI velocity dispersions, our re-analysis of the SDSS spectra, and the SLACS-IX velocity dispersions against the SLACS-XII velocity dispersions. Markers and colors are identical to the upper panel. We inferred parameters $\gamma$, $\gamma*$, and $\beta$, defined as in Equations \ref{eq:sdss_comp1}-\ref{eq:sdss_comp3} for the comparison between KCWI and SLACS-XII. We remove SDSSJ1538+5817 from these inferences, indicated in the plot with a red X, because there may be an issue with the SDSS target centering. The lower plot shows the agreement between our extractions from both KCWI and SDSS when compared with the kinematics from SLACS-IX.} 
    \label{fig:int_sigma}
\end{figure}

The upper panel of Figure \ref{fig:int_sigma} shows our measured velocity dispersions for KCWI and SDSS among single-aperture stellar velocity dispersions for SLACS lenses based on different datasets and apertures. The middle panel compares three different measurements (the KCWI $1.5''$-aperture velocity dispersions with mock SDSS seeing in black, our velocity dispersions measured from SDSS spectra in green, and the SLACS-IX velocity dispersion in yellow) against the SLACS-XII velocity dispersions. Averaged over the sample, the KCWI data agrees with the SLACS-XII values to $0.5\pm1.3$\%. Our velocity dispersions measured on SDSS spectra agree with SLACS-XII to an average of $1.4\pm2.3\%$. The SLACS-IX values are higher with respect to SLACS-XII on average by $2.9\pm2.2\%$. The lower panel of the figure shows the comparison between the velocity dispersions from this work compared against SLACS-IX. KCWI velocity dispersions are smaller than SLACS-IX values by $4.3\pm1.7\%$, and our measured SDSS velocity dispersions are smaller by $4.5\pm2.7\%$. These differences reflect the systematic uncertainties and covariance we expect for changes in stellar template library given the low S/N of the SDSS spectra, as described in Table~\ref{tab:sdss}.

We compare our KCWI velocity dispersions with the published SDSS values with Equations~\ref{eq:sdss_comp1}-\ref{eq:sdss_comp3} as in Section~\ref{sect:sdss_compare}. These results are also given in the legends for the middle and lower panels of Figure \ref{fig:int_sigma}. For SLACS-XII velocity dispersions, we find $ \gamma = 0.062^{+0.162}_{-0.042}$ and $ \beta = 0.999\pm{0.014}$. Using Equation \ref{eq:sdss_comp3}, we find $ \gamma_* = 0.021^{+0.017}_{-0.008}$, and $\beta$ is unchanged. This again indicates no evidence of bias in the SDSS velocity dispersions measured by SLACS-XII. The uncertainties are shown to be underestimated by $6.2^{+16.2}_{-4.2}\%$ (with respect to the uncertainty) or 2.1$^{+1.7}_{-0.8}\%$ error (with respect to the velocity dispersion) to be added in quadrature to the reported uncertainty. This is consistent with our own estimate of the systematic residual error shown in Table~\ref{tab:sdss}. Our estimate of $\gamma_*$ is somewhat lower than the inference of TDCOSMO-IV, which found $\gamma_* = 0.06\pm0.02$, although it is consistent within the errors. We repeat the comparison of KCWI data with SLACS-IX velocity dispersions in the formalism of Equations \ref{eq:sdss_comp1}-\ref{eq:sdss_comp3}. We find $\gamma = 0.080_{-0.062}^{+0.212}$, $\gamma_* = 0.023_{-0.023}^{+0.010}$, and $\beta = 0.965\pm0.019$, indicating a broad scatter and a $3.5\pm1.9\%$ difference on the mean. We attribute this difference primarily to the update in stellar libraries, but there is no significant bias indicated beyond the level of systematic uncertainty we expect for the SDSS spectra at this S/N. 

In summary, we find no measurable bias between velocity dispersions measured with SDSS spectra when compared with KCWI data for the same galaxies, for previous analyses or for our own. We find the uncertainties for velocity dispersions measured from SDSS spectra for our subsample of 13 galaxies to be consistent with the systematic uncertainties attributable to the stellar template library selection that we estimate for the larger SLACS parent sample with similar S/N. Furthermore, with the exception of the very highest S/N spectra, velocity dispersions of SLACS galaxies measured with SDSS are not suitable for measuring the Hubble constant with precision and accuracy below 2\%. Across our sample and in comparison with previous SLACS studies, the velocity dispersions extracted from SDSS spectra are consistent within the $5\%$ level that is sufficient for the galaxy formation and evolution purposes for which they were intended. However, we find that the relative uncertainties of the SDSS velocity dispersions are underestimated, and need an additional $\sim2-4\%$ depending on the analysis. Furthermore, the $\sim$2\% covariance that we estimate between our velocity dispersions measured from SDSS spectra, as a result of the correlated effect of the different stellar template libraries, is significant for H$_0$.
We therefore conclude that the inference in TDCOSMO-IV
based on the SLACS SDSS velocity dispersions underestimated the overall uncertainty because while it boosted the individual errors it did not include the covariance.
We recommend that future hierarchical cosmological inferences utilize velocity dispersions extracted from higher-S/N spectra with the methods outlined in TDCOSMO-XIX.

Two of our objects were also observed in the XLENS survey \citep{spiniello11_xlens} and extracted with single-slit kinematics from X-Shooter spectroscopy \citep{Spiniello15}. SDSSJ0037-0942 shows excellent agreement between X-Shooter and the integrated KCWI velocity dispersions, well within $1\sigma$ uncertainties. SDSSJ1627-0053, on the other hand, is $2\sigma \sim15\%$ larger for X-Shooter velocity dispersions with respect to KCWI integrated values. Unfortunately, without more overlap between the KCWI SLACS sample and XLENS survey, we are unable to draw significant conclusions from the comparison.

%%%%%%%%%%%%%%%%%%%
%%% Discussion

\section{Discussion}\label{sect:discussion}

%%%%%%%%%%%%%%%%%%%
%%%% Figures from SLACS-X

\subsection{Correlations with Other Observables}
\label{sect:correlations}

With updated spatially resolved kinematics and SDSS extractions, we are able to test some inferred quantities against values inferred from SDSS spectra and previous SLACS studies. We take the most recent lens model results for these objects from \cite{tan23}, utilizing their lensing mass power law slope $\gamma$ and Einstein radius. Following SLACS-X, we focus on the centermost region (within 1/2 the effective radius) of the galaxies.
We calculate $\sigma_{\rm 1/2}$ from our updated SDSS 1.5-arcsecond aperture velocity dispersions by the following equation (taking $\sigma_{\rm ap} = \sigma_{\rm eff}\left(R_{\rm ap}/R_{\rm eff}\right)^{-0.040}$, the power law fit given in Section \ref{sect:aperture} and shown in Figure \ref{fig:ap_int_effects_ratio}):

$$\sigma_{\rm 1/2} = \sigma_{\rm eff}\left(1/2\right)^{-0.040}=\sigma_{\rm 1.5''}\left(\frac{1.5''}{R_{\rm eff}}\right)^{0.040}\left(1/2\right)^{-0.040}=\sigma_{\rm 1.5''}\left(\frac{R_{\rm eff}}{3}\right)^{-0.040}$$

\noindent We take all stellar masses from stellar population models assuming Salpeter IMF and restframe V-band effective radius estimates from SLACS-IX and -X. 
$\sigma_{\rm 1/2}$ for KCWI targets is integrated from the kinematic map within the aperture radius of 1/2 the effective radius in a flux-weighted average, as described in Section \ref{sect:classification} and listed in Table \ref{tab:results}. In the following Figures \ref{fig:vd_to_log_stellar_mass}-\ref{fig:gamma_to_sigma}, blue markers indicate KCWI values, and green markers indicate SDSS values. Fainter gray markers indicate the full SLACS sample values from SLACS-IX and -X. In Section \ref{sect:sdss_compare}, we found these velocity dispersions to be consistent within $1\sigma$ with KCWI observations extracted from the same aperture size.

Everything else (effective radius, stellar mass, power law slope, Einstein radius) is consistent between the blue and green points. Linear fits are performed for all three samples with {\sc numpy.polyfit} and {\sc numpy.curvefit} \citep{harris20_numpy}, and we describe them with $a$ (slope) and $b$ (intercept) in the text and figures. Shaded areas denote 1$\sigma$ confidence levels.

\subsubsection{Correlations to stellar mass}

The Stellar Mass Plane \citep[$M_*$P,][]{hyde_bernardi09a} is an extension of the FP where surface brightness is replaced by stellar mass. In general total mass and central velocity dispersions correlate with stellar mass. SDSS studies \citep{hyde_bernardi09b, bernardi11} revealed a characteristic break in $M_*-\sigma$ in ETGs at a mass of $M_{\rm crit}=2\times10^{11} M_{\rm \odot}$. \cite{cappellari13_atlas3dxx} show this break to describe a transition between fast and slow rotators in ATLAS$^{\mathrm{3D}}$ at masses below and above $M_{\rm crit}$ respectively. Our subsample of 14 SLACS lenses lie above $M_{\rm crit}$. 

Figure \ref{fig:vd_to_log_stellar_mass} shows the log of the velocity dispersion integrated within half the effective radius $\sigma_{\rm 1/2}$ against the log of the stellar mass ($M_*/10^{11} M_{\rm \odot}$) for a Chabrier IMF (compare to Figure 1 in SLACS-X) with $1\sigma$ uncertainties. We note that there is evidence for these massive galaxies to have a heavier IMF, closer to Salpeter IMF \citep[see e.g][]{Treu10, auger10_imf, cappellari12_imf}, which would imply that the true stellar masses may be larger than plotted here by $\sim0.21$ dex. 
The full sample from SLACS-X is shown in the background as faded gray markers. Blue and green fits indicate fits to our 14 objects using KCWI and SDSS velocity dispersions, respectively. 

Our data is consistent with the results found for larger samples of SDSS and ATLAS$^{\mathrm{3D}}$ galaxies and are supportive of previously reported $M_*$P relations. 

As in SLACS-X, we define $M_{\rm 1/2}$ as the total mass contained within a projected cylinder of radius half the effective radius (this is 32\% of the total mass for a \citealt{deVaucouleur48} profile) estimated using the virial relation:

$$M_{\rm 1/2}=\frac{C_V R_{\rm eff} \sigma_{\rm 1/2}^2}{2 G}$$

\noindent where $\log_{\rm 10} C_V=0.53\pm0.09$ is the virial coefficient. SLACS-X shows $M_{\rm 1/2}$ to be tightly linearly correlated with the dynamical mass, and we use $M_{\rm 1/2}$ for consistency with their analysis. Figure \ref{fig:log_stellar_to_log_half_mass} shows the stellar mass as a function of $M_{\rm 1/2}$ (compare to the centermost plot of Figure 3 in SLACS-X). 

The tightness of the correlation shows the power of the $M_*$P and FP. The fits are again consistent with previous studies.

\subsubsection{Correlations to lensing power law slope}

Several strong lens studies have found a remarkably good fit to the central total mass density profile of ETGs within a few effective radii with a nearly isothermal power law \citep[e.g.,][]{rusin_ma01, rusin03, cohn01, munoz01, rusin_kochanek05, shajib21, etherington22, tan23}. X-ray studies find similar results to scales much larger than the effective radius \citep[e.g.,][]{thomas86, kim_fabbiano95, matsushita98, humphrey_buote10}. Using spatially resolved stellar kinematics from ATLAS$^{\mathrm{3D}}$ and SLUGGS surveys \citep{brodie14}, \cite{cappellari15} showed the average power law slope to be very nearly isothermal ($<\gamma> = 2.19 \pm 0.03$) out to four effective radii in nearby galaxies, with slightly shallower values for galaxies below $\sigma=100$ km $\mathrm{s}^{-1}$ and above $M_{\rm crit}=2\times10^{11} M_{\rm \odot}$ \citep[fig.~22c]{Cappellari2016}. 
Since our subsample of 14 SLACS lenses do not span across these breaks, we expect relations involving the power law slope $\gamma$ to agree with previous SLACS findings. \cite{Koopmans06, koopmans09} showed through lensing and dynamical analysis that the total mass density profiles of SLACS lenses are well described by a single nearly isothermal power law, and that these do not correlate strongly with other galaxy observables. SLACS-X explored possible correlations between $\gamma$ and aperture kinematics measurements and found some tentative trends that showed the need for further analysis. 
The situation has dramatically improved with dynamical analysis with spatially resolved kinematics of large samples of ETGs. \cite{zhu23} show a steepening of the average slope of the total mass density within the effective radius with increasing velocity dispersion (also within 1$R_\mathrm{eff}$) and with age for their MaNGA sample (see their Figure 8). This trend flattens above $\sigma\sim200$ km $\mathrm{s}^{-1}$ where our subsample of SLACS lenses lies, which indicates that we should find negligible trends between power law slope and kinematics observables if the SLACS subsample is consistent with the MaNGA population.

We discuss possible correlations between lensing power law slope and kinematics in Figures \ref{fig:gamma_to_f_sie}-\ref{fig:gamma_to_sigma} as analogs to the similar discussions in SLACS-X Figures 4 and 5. An important distinction is the difference in  definitions of the power law slope $\gamma$ between \cite{tan23} and SLACS-X. The power law slope $\gamma$ in SLACS-X (as in ATLAS$^{\mathrm{3D}}$, MaNGA, and earlier SLACS analyses) is defined as the average slope within a certain radius (in SLACS-X, the Einstein radius), 
while the $\gamma$ value we take from \cite{tan23} is defined \textit{at} the Einstein radius. These plots are comparable to those from SLACS-X only on the surface-level, so we do not show SLACS-X data as in Figures \ref{fig:vd_to_log_stellar_mass} and \ref{fig:log_stellar_to_log_half_mass}. Compared with Figures~\ref{fig:vd_to_log_stellar_mass} and \ref{fig:log_stellar_to_log_half_mass}, the relations we see between KCWI kinematics observables and $\gamma$ show much more scatter, and correlations are significantly less convincing. \cite{tan23} rigorously estimate systematic uncertainties in the lens model power law slopes through a comparison with previous lensing and joint lensing/dynamical studies of the same strong lenses with different codes and assumptions. This could be why the scatter is so high for correlations involving our small subsample of 14 objects.
At similar velocity dispersions to SLACS, MaNGA galaxies have nearly constant slopes $\gamma\sim2.2$ with a $1\sigma$ scatter of 0.2. The lens model power law slopes $\gamma$ for our sample are all within $2\sigma$ of the MaNGA scatter, as one would expect for a normal population.

In the past, kinematics and lensing data have been compared in the context of less detailed observations. A useful parameter is the deviation of the central velocity dispersion within half the effective radius to the theoretical velocity dispersion for a singular isothermal ellipse (SIE) mass profile model. A value greater than one indicates a mass density profile steeper than isothermal, while a value less than 1 indicates a flatter slope. \cite{treu09} found this ratio to be strongly correlated to the mass density power law slope determined from joint lensing and dynamical models. This relation has been used as a proxy to estimate the power law slope from kinematics data for which Jeans modeling is unfeasible. SLACS-X found further evidence to support this correlation. We calculate SIE model velocity dispersions from the Einstein radius from lens models from \cite{tan23}:

$$\sigma_{\rm SIE}=\frac{c}{2}\sqrt{\frac{\theta_E}{\pi}\frac{D_{\rm LS}}{D_S}}$$

\noindent  We plot the ratio $f_{\rm SIE}=\sigma_{\rm 1/2}/\sigma_{\rm SIE}$ against the power law slope $\gamma$ from \cite{tan23} lens models in Figure \ref{fig:gamma_to_f_sie}.  

We find a milder correlation than the one reported in SLACS-X, which measured a slope of $2.67\pm0.15$. SDSS and KCWI slopes are $0.81\pm0.33$ and $0.97\pm0.37$ respectively. This supports the application of a proxy relation like $f_{\rm SIE}$ as a rough estimate in lieu of the possibility for more complex dynamical analysis. However, as further evidenced in Figures~\ref{fig:gamma_to_surface_mass_density} and \ref{fig:gamma_to_sigma}, the scatter in the power law slope $\gamma$ is not resolved by these approximations with integrated kinematics.

SLACS-X Figure 5 showed another correlation between $\gamma$ and the central surface mass density $\Sigma_0 \equiv M_{\rm 1/2}/R_{\rm eff}^2$. Our Figure \ref{fig:gamma_to_surface_mass_density} shows on the vertical axis the power law slope $\gamma$, and the horizontal axis is $\Sigma_0$ calculated from the virial estimate of projected total central mass $M_{\rm 1/2}$ as in Figure~\ref{fig:log_stellar_to_log_half_mass}. The SDSS and KCWI data for our subsample of 14 objects agree on a neglible correlation. We find another negligible correlation when we look at the direct relation between lensing power law slope $\gamma$ and the integrated velocity dispersion within half the effective radius $\sigma_{\rm 1/2}$ in Figure \ref{fig:gamma_to_sigma}, in agreement with SLACS-X. When compared with the MaNGA DynPop sample in \cite{zhu23}, we see that our SLACS subsample lies in the region where $\gamma$ is nearly constant, indicating consistent populations in the observed mass range.

% insert correlations plots
% shows correlation plots

\def\plotw{0.5\textwidth}

\begin{figure}
        \centering
        \includegraphics[width=\plotw]{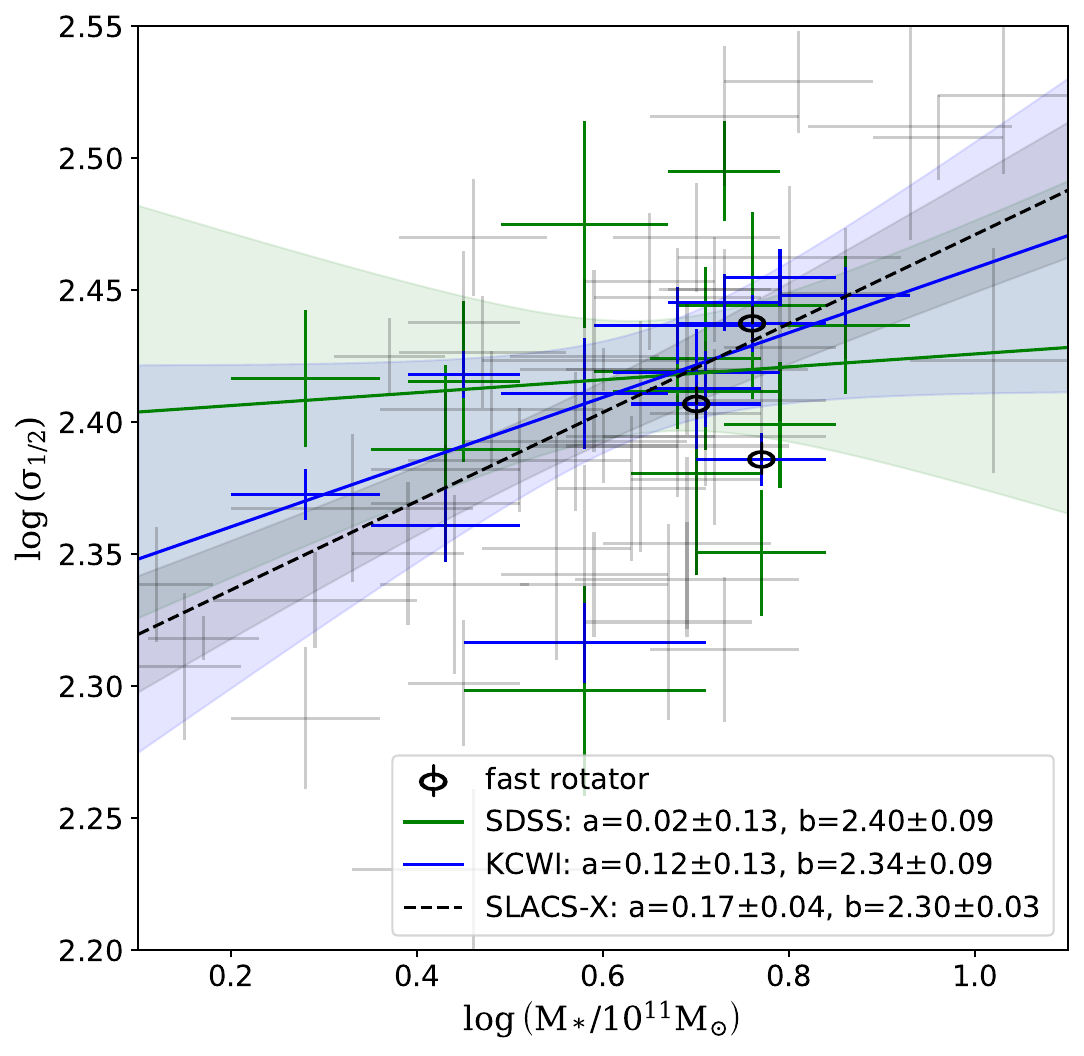}
        \caption{\textit{y-axis}: $\log_{10}$ of the velocity dispersion within half the effective radius. \textit{x-axis}: $\log_{10}$ of the stellar mass from stellar population synthesis in units of $10^{11} M_{\odot}$. KCWI (blue) and SDSS (green) quantities are derived from kinematic extractions in this work for each object in our subsample of 14 SLACS lenses. Gray markers and dotted black lines indicate SLACS-X data and fits. Fast rotators are marked as in Figure \ref{fig:classification}. 
        %Black triangle markers show KCWI and SDSS data for SDSSJ1538+5817, in which we have identified a kinematically decoupled core (KDC). See Section \ref{sect:aperture} for discussion regarding the disagreement between the measured KCWI and SDSS kinematics for this object. 
        Uncertainties on fits and errorbars are $1\sigma$. Fit parameters $a$  and $b$ are slope and intercept.}
        \label{fig:vd_to_log_stellar_mass}
\end{figure}%
\begin{figure}
        \centering
        \includegraphics[width=\plotw]{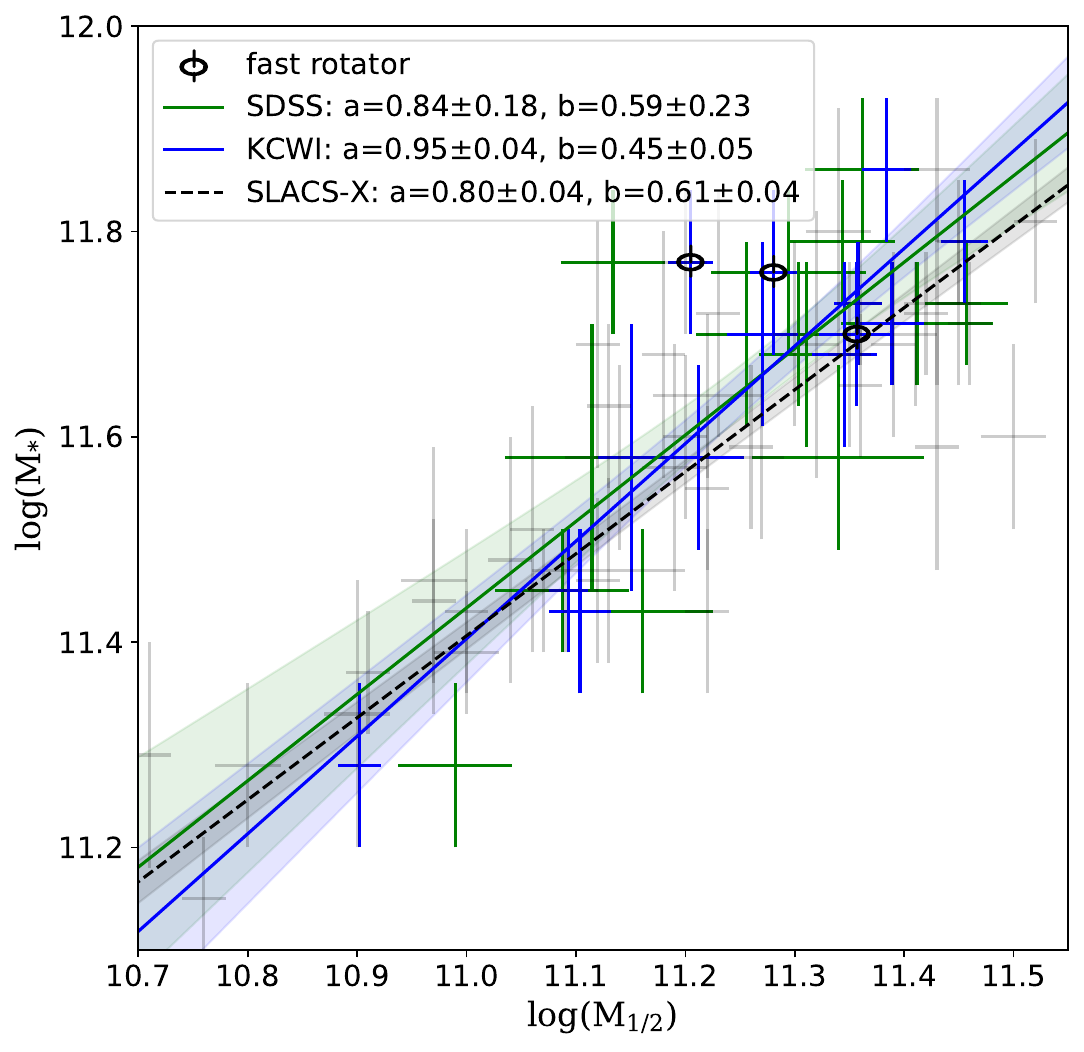}
        \caption{\textit{y-axis}: $\log_{10}$ of the stellar mass from stellar population synthesis. \textit{x-axis}: $\log_{10}$ of the virial estimate of the total projected mass within a cylinder of radius half the effective radius. Markers and lines are as defined in Figure \ref{fig:vd_to_log_stellar_mass}.}  
        \label{fig:log_stellar_to_log_half_mass}
\end{figure}%
\begin{figure}
        \centering
        \includegraphics[width=\plotw]{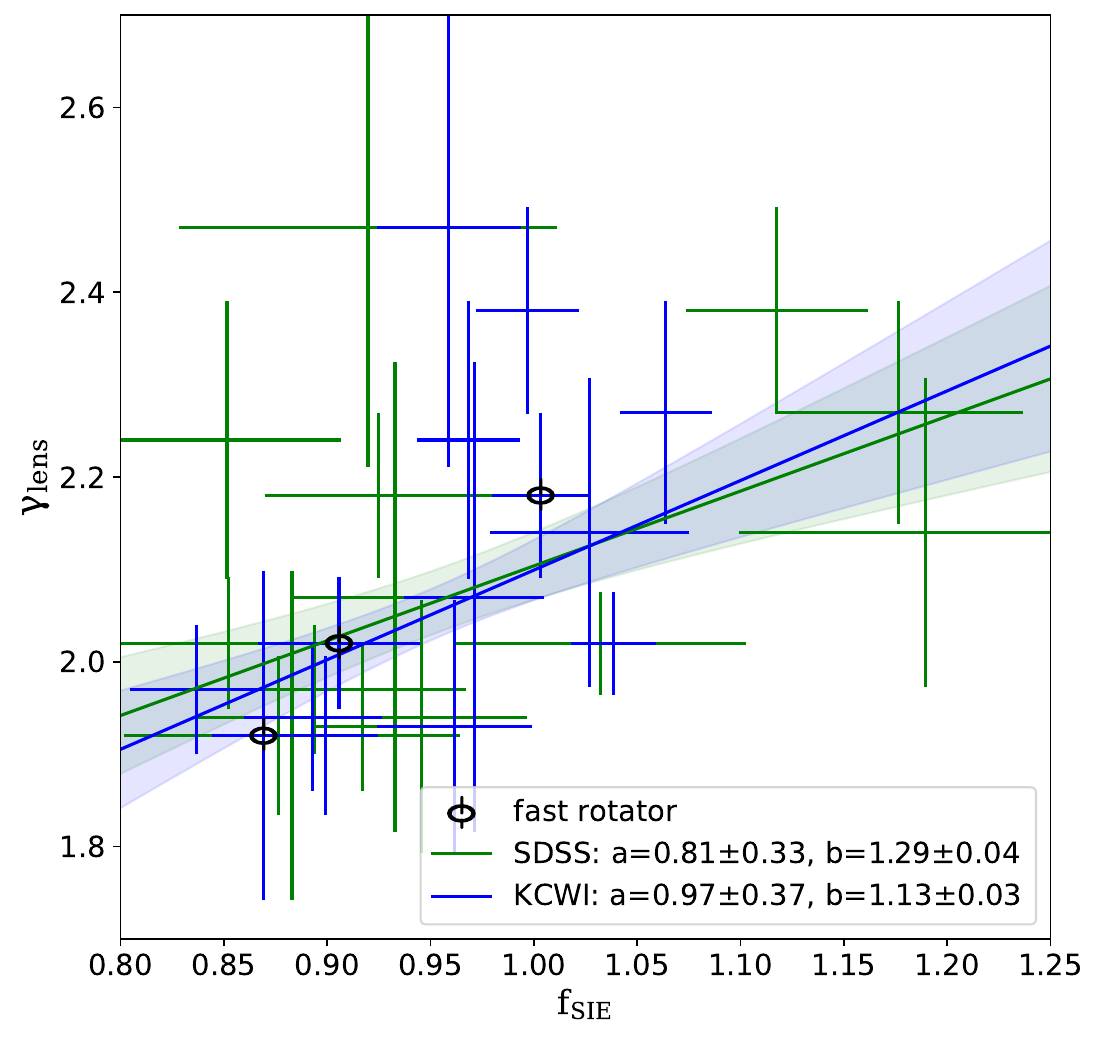}
        \caption{\textit{y-axis}: slope of the lensing power law mass density profile at the Einstein radius. \textit{x-axis}: ratio of the velocity dispersion within half the effective radius to the velocity dispersion calculated from the lens model Einstein radius assuming a singular isothermal ellipse, $f_{\mathrm{SIE}}=\sigma_{1/2}/\sigma_{\mathrm{SIE}}$. We do not show SLACS-X data here because the power law slope is not defined in the same way as in \cite{tan23}. Markers and lines are as defined in Figure \ref{fig:vd_to_log_stellar_mass}.}
        \label{fig:gamma_to_f_sie}
\end{figure}%
% Should I remove this one?
%\begin{figure}
%    \centering
    %\includegraphics[width=\plotw]{f_sie_vs_redshif}
    %\caption{\textit{y-axis}: ratio of the velocity dispersion within half the effective radius to the velocity dispersion calculated from the lens model Einstein radius assuming a singular isothermal ellipse, $f_{SIE}=\sigma_{1/2}/\sigma_{SIE}$. \textit{x-axis}: lens redshift. Markers for fast (open marker) and slow (closed marker) rotators are plotted on top of only the KCWI data. }
    %\label{fig:fsie_to_redshift}
%\end{figure}
\begin{figure}
        \centering
        \includegraphics[width=\plotw]{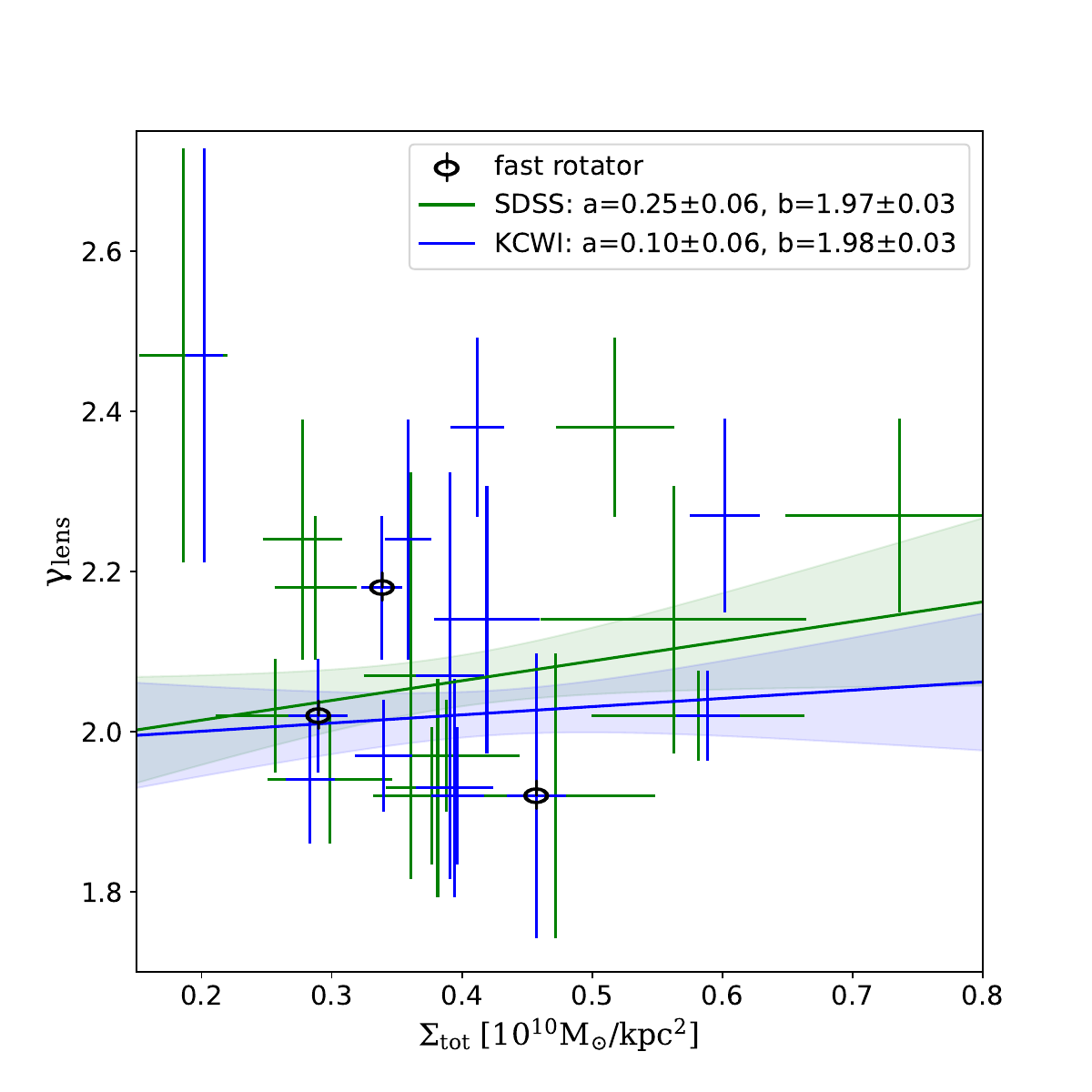}
        \caption{\textit{y-axis}:  slope of the lensing power law mass density profile at the Einstein radius. \textit{x-axis}: central surface mass density calculated from dynamical mass within half the effective radius, $\Sigma_0 \equiv M_{1/2}/R_{eff}^2$. Markers and lines are as defined in Figure \ref{fig:vd_to_log_stellar_mass}.}
        \label{fig:gamma_to_surface_mass_density}
\end{figure}%
\begin{figure}
        \centering
        \includegraphics[width=\plotw]{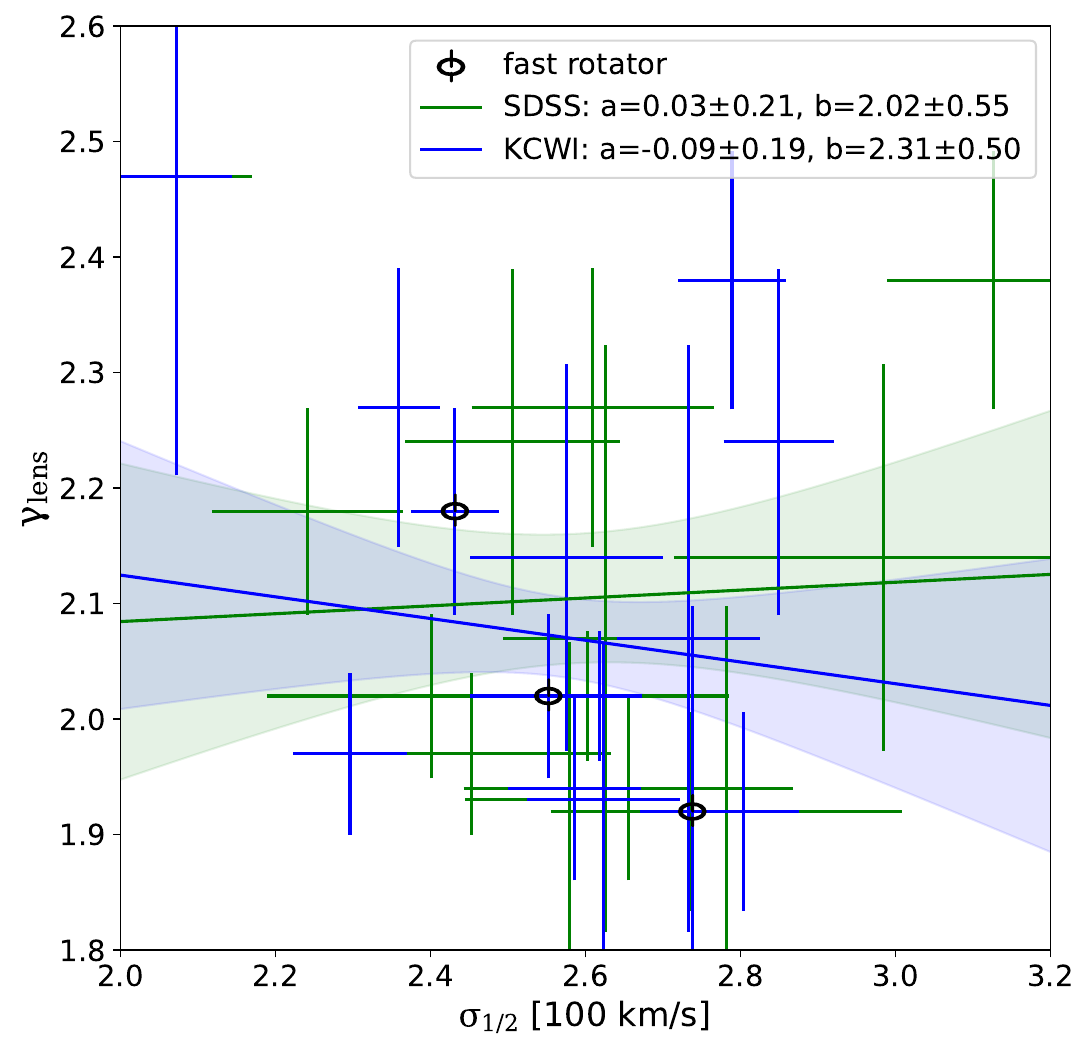}
        \caption{\textit{y-axis}:  slope of the lensing power law mass density profile at the Einstein radius. \textit{x-axis}: velocity dispersion within half the effective radius. Markers and lines are as defined in Figure \ref{fig:vd_to_log_stellar_mass}.}
        \label{fig:gamma_to_sigma}
\end{figure}

%\subsection{Oblate vs Prolate}
\subsection{Axisymmetric vs Triaxial}

\begin{figure}
    \centering
    \includegraphics[width=0.49\textwidth]{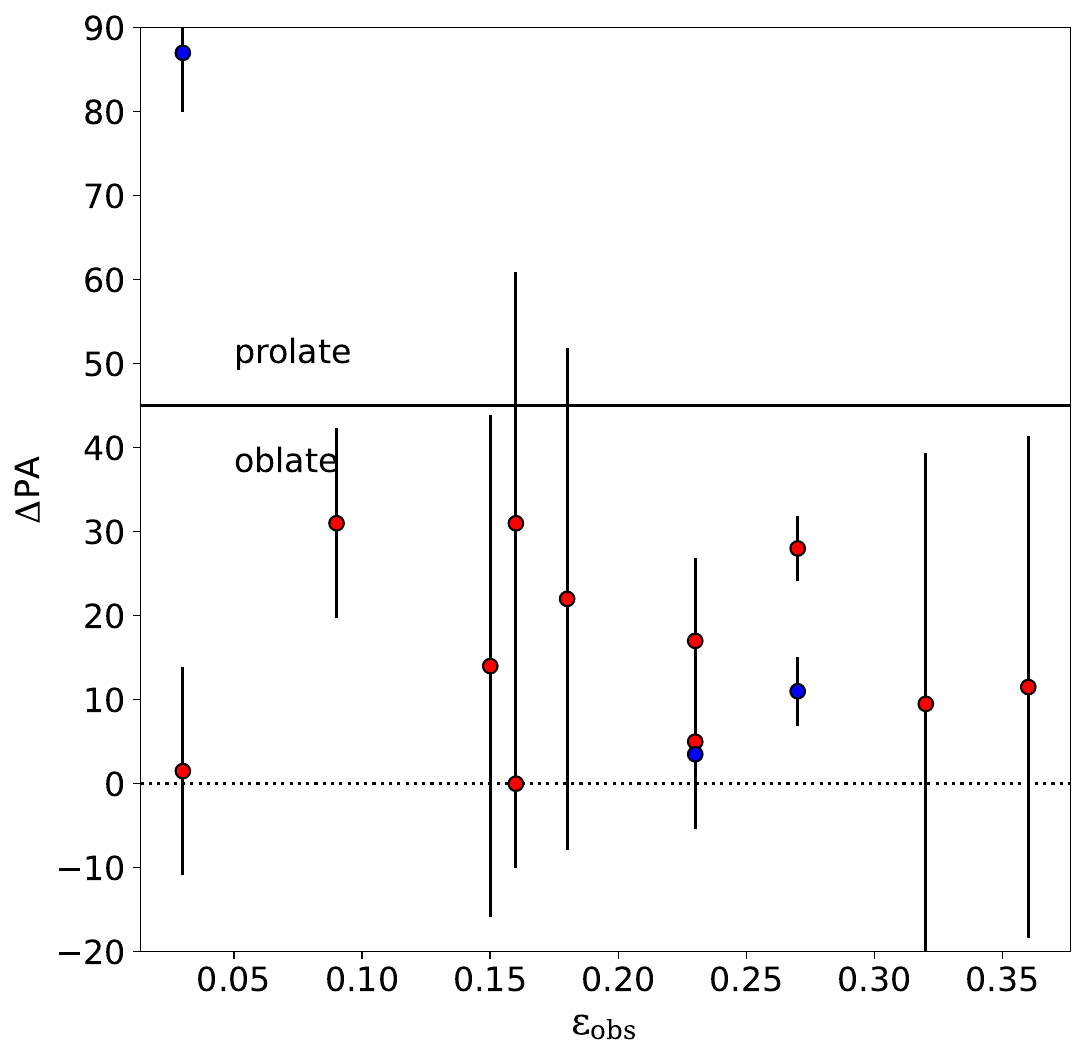}
    \caption{\textit{y-axis}: misalignment of photometric and kinematic axes for each object. \textit{x-axis}: ellipticity of the galaxy measured from the MGE models at the isophote enclosing half the total luminosity. Blue and red markers indicate fast rotators and slow rotators, respectively (see Section \ref{sect:classification} and Table \ref{tab:results}). The kinematic major axis has been measured from mean velocity maps, and the photometric axis is from SLACS-IX. The solid black horizontal line shows the designation between oblate and prolate cases at 45$^{\circ}$. All objects are oblate, with the exception of SDSSJ1250+0523, which is very round $\epsilon\sim0.03$ and likely not truly prolate.}
    \label{fig:oblate_prolate}
\end{figure}

An important factor for the dynamical interpretation of the kinematics is the alignment of photometric and kinematic axes. The kinematic major axis has been measured from mean velocity maps using the \texttt{PaFit} package\footnote{\url{https://pypi.org/project/pafit/}} \citep{Krajnovic2006}, and the photometric position angle is from \textit{HST} photometry as reported in \cite{slacs5}. In Figure \ref{fig:oblate_prolate}, we show the angle between misalignment of these axes $\mathrm{\Delta{PA}}$ for each object as a function of observed ellipticity. Observed ellipticity is measured from the MGE models at the isophote enclosing half the total luminosity, using the procedure \texttt{mge\_half\_light\_isophote} from the JAM software package\footnote{\url{https://pypi.org/project/jampy/}}. Blue markers indicate objects we identify as fast rotators 
(see Section \ref{sect:classification} and Table \ref{tab:results}), which have significantly better constrained kinematic axes than the objects we identify as slow rotators (shown here in red markers). 
The dominant uncertainty in $\mathrm{\Delta PA}$ comes from the kinematic major axis, which is poorly defined for slow rotators that do not show any regular rotation.  Photometric uncertainty is not published in the SLACS papers, so it cannot be added. However, based on experience and data quality it is much smaller than that of the kinematic axis and therefore negligible for this comparison. Misalignments $\mathrm{0<|\Delta PA|<45}$ are considered most likely axisymmetric and oblate, and $45<|\mathrm{\Delta PA}|<90$ are most likely triaxial or prolate. The solid black horizontal line shows this separation. We note one fast rotator (SDSSJ1250+0523) that appears to show prolate rotation due to its nearly 90$^{\circ}$ measured misalignment. This object is very round with an observed ellipticity $\epsilon_{\rm obs}\sim0.03$, so the photometric axis may be highly uncertain.

\subsection{Time-Delay Cosmography}

The kinematic maps and analysis in this work illustrate that the KCWI data is of sufficient quality to achieve strong constraints on the mass profiles and anisotropy of lensing ETGs. Correlations also highlight the need for this quality of data and more detailed dynamical analysis, which will be presented in the upcoming companion paper. The 1$\%$ off-diagonal term of the covariance matrix at the sample level indicates that 2$\%$ precision on $\mathrm{H_0}$ is achievable given an expanded sample of non-time-delay lenses of the same level of quality. Improved datasets with an expanded SLACS sample (e.g. with the new KCRM red arm of KCWI) will improve constraints from the non-time-delay data beyond this 2$\%$ level. Complementary to parallel ongoing kinematics measurements with KCWI of time-delay lenses \citep[e.g.,][]{shajib23}, we project that the combined sample of our measurements with time-delay lenses will help to achieve precision of H$_0$ that is independent of the best current determinations of H$_0$ based on the local distance ladder \citep{freedman21,riess22} and comparably precise (1.9\%, using the same methodology as \cite{birrer_treu21} and updated based on ongoing and scheduled observations). It will also provide a key verification of the TDCOSMO 2\% precision measurement \citep{millon20b} achieved under mass density profile assumptions. 

%%%%%%%%%%%%%%%%%%%

\section{Conclusions}\label{sect:conclusions}

We present kinematic data and analysis for 14 SLACS strong gravitational lens deflector galaxies, from which we determine the following main conclusions:

\begin{enumerate}
    \item Most (11) of our 14 SLACS lenses are consistent in terms of ($\mathrm{V}/\sigma, \ \epsilon$) and ($\lambda_{\rm R}, \ \epsilon$) with the round ``slow" non-regular rotators of the local universe (as in ATLAS$^{\mathrm{3D}}$), with the exception of 3 that have significant rotational support and are close to the boundary between ``slow" and ``fast" regular rotators.
    \item 
    We quantify systematic error at the spatial bin level for each object and find $1-1.4\%$ systematic error, dominated by template library choice. At the sample level, we find a positive covariance of $<1\%$ that indicates that with current kinematics the SLACS sample is sufficient to achieve hierarchical inference of $\mathrm{H_0}$ to $2\%$ precision. 
    \item  
    Systematics between velocity dispersions extracted from SDSS spectra of the SLACS galaxies are at the level of 2-4\% with positive sample covariance of $\sim2\%$ depending on the S/N of the spectra, primarily due to stellar template library selection. Relative uncertainties of previously published SDSS velocity dispersions for these 14 galaxies are underestimated by $\sim2-3\%$. We therefore conclude that they are sufficiently accurate for galaxy studies, but that their systematics and covariance are significant for accurate time delay cosmography. Case in point, the TDCOSMO+SLACS measurement of H$_0=67.4_{-3.2}^{+4.1}$ km s$^{-1}$ Mpc$^{-1}$ reported by \citet[][TDCOSMO-IV]{birrer20_tdcosmo_iv} based on SDSS velocity dispersions neglects the covariance among the SDSS velocity dispersion measurements and thus overestimates the constraining power of the SDSS sample, resulting in underestimated uncertainties and a potential bias in the inferred value of H$_0$. We recommend that future hierarchical cosmological inferences utilize velocity dispersions extracted from higher-S/N spectra with the methods outlined by \citet[][TDCOSMO-XIX]{TDCOSMO19}.
    \item ETG scaling relations are overall consistent with previous SLACS studies. In particular:
        \begin{enumerate}
            \item Stellar mass correlates positively with velocity dispersion and central projected total mass, agreeing within 1$\sigma$ uncertainty with the correlations published in SLACS-X,
            \item Lensing power law slope correlates mildly with $f_{\rm \rm SIE}$ (deviation from SIE velocity dispersion) with a slope of $0.97\pm0.37$,
            \item Lensing power law slope shows no correlation with surface mass density or central velocity dispersion.
        \end{enumerate}
\end{enumerate}

These kinematic data are prepared for dynamical fitting with Jeans Anisotropic Modeling \citep[JAM]{cappellari08, cappellari20_jeans_axis}. The result of this approach will produce constraints on the mass profiles and anisotropy of ETGs, which will in turn help to constrain the Hubble constant through time-delay cosmography of distant lensed quasars.

%%%%%%%%%%%%%%%%%%%

\begin{acknowledgments}

We acknowledge support by the National Science Foundation in the form of a graduate fellowship to SK, and grants NSF-AST-2407277, NSF-AST-1906976, NSF-AST-1836016, and NSF-AST-1909297. Note that findings and conclusions do not necessarily represent views of
the NSF.
We also acknowledge support by the Gordon and Betty Moore Foundation through a grant to TT.

We thank A.S.~Bolton and M.Bernardi for useful discussions regarding SDSS velocity dispersions. 

Some of the data presented herein were obtained at Keck Observatory, which is a private 501(c)3 non-profit organization operated as a scientific partnership among the California Institute of Technology, the University of California, and the National Aeronautics and Space Administration. The Observatory was made possible by the generous financial support of the W. M. Keck Foundation. 
The authors wish to recognize and acknowledge the very significant cultural role and reverence that the summit of Maunakea has always had within the Native Hawaiian community. We are most fortunate to have the opportunity to conduct observations from this mountain.

This research is based in part on observations made with the NASA/ESA Hubble Space Telescope obtained from the Space Telescope Science Institute, which is operated by the Association of Universities for Research in Astronomy, Inc., under NASA contract NAS 5–26555. These observations are associated with program(s) 10174 (L. Koopmans), 10886 (A. Bolton), 10494 (L. Koopmans), and 10587 (A. Bolton).

\end{acknowledgments}

%%%%%%%%%%%%%%%%%%%

\bibliographystyle{aasjournal}
\bibliography{shawn_bibliography, bibliography2}

%%%%%%%%%%%%%%%%%%%

\end{document}